\documentclass[english,12pt]{article}
\pdfoutput=1
\usepackage[T1]{fontenc}
\usepackage[latin9]{inputenc}
\usepackage{amstext}
\usepackage{babel}
\usepackage{verbatim}
\usepackage{amsfonts}
\usepackage{mathrsfs}
\usepackage{amsmath}
\usepackage{amssymb}
\usepackage{bm}
\usepackage{extarrows}
\usepackage{slashed}
\usepackage{isodateo}
\usepackage{xcolor}
\usepackage{graphicx}
\usepackage[low-sup]{subdepth}
\usepackage{float}
\usepackage{isodateo}
\usepackage{tensor}
\usepackage{multirow}
\usepackage{enumitem}
\usepackage{ytableau}
\usepackage[linktocpage=true]{hyperref}
\hypersetup{
colorlinks=true,
citecolor=blue,
linkcolor=blue,
urlcolor=blue,
pdfauthor={},
pdftitle={},
pdfsubject={}
}
\makeatletter
\def\simgt{\mathrel{\lower2.5pt\vbox{\lineskip=0pt\baselineskip=0pt
\hbox{$>$}\hbox{$\sim$}}}}
\def\simlt{\mathrel{\lower2.5pt\vbox{\lineskip=0pt\baselineskip=0pt
\hbox{$<$}\hbox{$\sim$}}}}
\numberwithin{equation}{section}

\usepackage{titlesec}
\titleformat*{\section}{\large\bfseries}
\titleformat*{\subsection}{\normalsize\bfseries}
\titleformat*{\subsubsection}{\small\bfseries}

\makeatletter
\newsavebox\myboxA
\newsavebox\myboxB
\newlength\mylenA
\newcommand*\xoverline[2][0.7]{%
\sbox{\myboxA}{$\m@th#2$}%
\setbox\myboxB\null
\ht\myboxB=\ht\myboxA%
\dp\myboxB=\dp\myboxA%
\wd\myboxB=#1\wd\myboxA
\sbox\myboxB{$\m@th\overline{\copy\myboxB}$}
\setlength\mylenA{\the\wd\myboxA}
\addtolength\mylenA{-\the\wd\myboxB}%
\ifdim\wd\myboxB<\wd\myboxA%
\rlap{\hskip 0.8\mylenA\usebox\myboxB}{\usebox\myboxA}%
\else
\hskip -0.5\mylenA\rlap{\usebox\myboxA}{\hskip 0.5\mylenA\usebox\myboxB}%
\fi}

\definecolor{Green}{RGB}{199,238,206}

\newcommand{\eqrefe}{Eq.\eqref}
\newcommand{\beq}{\begin{equation}}
\newcommand{\eeq}{\end{equation}}
\newcommand{\ba}{\begin{array}}
\newcommand{\ea}{\end{array}}
\newcommand{\beqa}{\begin{eqnarray}}
\newcommand{\eeqa}{\end{eqnarray}}
\newcommand{\beqs}{\begin{subequations}}
\newcommand{\eeqs}{\end{subequations}}

\newcommand{\la}{\langle}
\newcommand{\ra}{\rangle}
\newcommand{\fr}[2]{\mbox{$\frac{\,{#1}\,}{#2}$}}
\renewcommand{\rm}{\mathrm}
\def\Re{\mathfrak{Re}}
\def\Im{\mathfrak{Im}}

\def\leqq{\leqslant}
\def\geqq{\geqslant}
\def\({\left(}
\def\){\right)}
\def\[{\left[\,}
\def\]{\,\right]}
\def\LB{\left\{}
\def\RB{\right\}}
\def\nn{\nonumber}
\def\pd{\partial}
\def\pp{\prime}
\def\to{\rightarrow}
\def\ito{\!\rightarrow\!}

\def\over{\overline}

\def\tr{\text{tr}}
\def\diag{\text{diag}}
\def\MP{M_{\text{Pl}}^{}}

\def\ba{\bar{a}}

\def\bfA{\mathbf{A}}

\def\CC{\mathcal{C}}
\def\D{\mathcal{D}}
\def\td{\text{d}}
\def\bE{\bar{E}}
\def\EE{\mathcal{E}}

\def\bfF{\bold{F}}

\def\FF{\mathcal{F}}

\def\ii{\text{i}}
\def\iii{\text{i}\hspace*{0.3mm}}

\def\La{\mathcal{L}}
\def\bM{\xoverline{\mathcal{M}}}
\def\M{\mathcal{M}}

\def\NN{\mathcal{N}}

\def\mO{\mathcal{O}}

\def\bs{\bar{s}}

\def\Sb{\mathbb{S}}

\def\TT{\mathcal{T}}

\def\vt{\tilde{v}}
\def\VV{\mathcal{V}}

\def\al{\alpha}

\def\be{\beta}
\def\ga{\gamma}
\def\Ga{\Gamma}
\def\ka{\kappa}
\def\ab{\alpha\beta}

\def\mn{\mu\nu}

\def\mnr{\mu\nu\rho}

\def\ep{\epsilon}
\def\vep{\varepsilon}
\def\lam{\lambda}
\def\tGa{\tensor{\Gamma}}

\def\si{\sigma}
\def\slp{\slashed{p}}
\def\ct{c_\theta^{}}
\def\st{s_\theta^{}}
\def\cht{c_{\theta/2}^{}}
\def\sht{s_{\theta/2}^{}}
\def\cct{c_\theta^2}

\def\ctt{c_{2\theta}^{}}
\def\cttt{c_{3\theta}^{}}
\def\ctf{c_{4\theta}^{}}
\def\ctfv{c_{5\theta}^{}}

\def\stt{s_{2\theta}^{}}
\def\sttt{s_{3\theta}^{}}

\def\sz{s_0^{}}
\def\tz{t_0^{}}
\def\uz{u_0^{}}
\def\bsz{\bar{s}_0}

\def\sp{\mathfrak{s}}
\def\epT{\epsilon_{\text{T}}}
\def\epP{\epsilon_{\text{P}}}
\def\epL{\epsilon_{\text{L}}}
\def\epS{\epsilon_{\text{S}}}
\def\epX{\epsilon_{\text{X}}}
\def\tGa{\tensor{\Gamma}}
\def\mt{\widetilde{m}}
\def\P{\text{P}}
\def\L{\text{L}}
\def\T{\text{T}}

\def\APP{A_{\text{P}}^{}}
\def\ATT{A_{\text{T}}^{}}
\def\AP{A^a_{\text{P}}}
\def\AL{A^a_{\text{L}}}
\def\AT{A^a_{\text{T}}}
\def\AS{A^a_{\text{S}}}
\def\ASS{A^{}_{\text{S}}}
\def\AX{A^a_{\text{X}}}
\def\Ap{A_{\text{P}}}
\def\At{A_{\text{T}}}
\def\tdA{\tilde{A}}
\def\tdAT{\tilde{A}^a_{\text{T}}}
\def\tdATT{\tilde{A}^{}_{\text{T}}}
\def\hP{h_{\text{P}^{}}}
\def\ZZ{\mathbb{Z}}
\def\hs{\hspace*{0.3mm}}
\def\hsx{\hspace*{0.5mm}}
\def\hsm{\hspace*{-0.3mm}}
\def\hsmx{\hspace*{-0.5mm}}
\def\hf{\frac{1}{2}}
\def\End{\end{document}}
\allowdisplaybreaks
\begin{document}

\interfootnotelinepenalty=10000
\baselineskip=17pt

\vspace*{-2cm}
\thispagestyle{empty}

\begin{center}
{\Large\bf 
Structure of Chern-Simons Scattering Amplitudes from \\[1mm] 
Topological Equivalence Theorem and Double-Copy}
	
\vspace*{8mm}
	
{\sc Yan-Feng Hang}\,$^{a}$\footnote{Email: {yfhang@sjtu.edu.cn}},
~~ 
{\sc Hong-Jian He}\,$^{a,b,c}$\hs\footnote{Email: {hjhe@sjtu.edu.cn}},
~~
{{\sc Cong Shen}\,$^a$}
	
\vspace*{3mm}
$^a$\,Tsung-Dao~Lee Institute \& 
School of Physics and Astronomy, 
\\[-0.7mm]
Key Laboratory for Particle Astrophysics and Cosmology (MOE),
\\[-0.7mm]
Shanghai Key Laboratory for Particle Physics and Cosmology,
\\[-0.7mm]
Shanghai Jiao Tong University, Shanghai 200240, China
\\[0.5mm]
$^b$\,Institute of Modern Physics and Department of Physics,
\\[-0.7mm]
Tsinghua University, Beijing 100084, China
\\[0.5mm]
$^c$\,Center for High Energy Physics, Peking University, Beijing 100871, China
	
\vspace{12mm}
	
\end{center}

\begin{abstract}
\baselineskip 16pt
\noindent
We study the mechanism of topological mass-generation for 3d Chern-Simons (CS) gauge theories, where the CS term can retain the gauge symmetry and make gauge boson topologically massive. Without CS term the 3d massless gauge boson has a single physical transverse polarization state, while adding the CS term converts it into a massive physical polarization state and conserves the total physical degrees of freedom.
We newly formulate the mechanism of topological mass-generation at $S$-matrix level. For this, we propose and prove a new Topological Equivalence Theorem (TET) which connects the $N$-point scattering amplitude of the gauge boson's physical polarization states ($\AP$) to that of the transverse 
polarization states ($\AT$) under high energy expansion.\ 
We present a general 3d power counting method on the leading energy dependence of $N$-point scattering amplitudes in both topologically massive Yang-Mills (TMYM) and topologically massive gravity (TMG) theories. 
With these, we uncover {\it a general energy cancellation mechanism 
for $N$-gauge boson scattering amplitudes} which predicts the
cancellation $\hs E^4\hsm\ito E^{4-N}$ at tree level. 
Then, we compute the four-point amplitudes of 
$\AP$'s and of $\AT$'s, with which we explicitly 
demonstrate the TET and establish such energy cancellations. 
We further extend the double-copy approach 
and construct the four-point massive graviton amplitude of the 
TMG theory from the massive gauge boson amplitude of the TMYM theory. 
With these, we newly uncover {\it striking large energy cancellations 
$E^{12}\!\ito E^1$ in the four-graviton amplitude of the TMG,}\,
and establish 
{\it its new correspondence to the leading energy cancellations 
$E^4 \ito E^0\hs$ in the four-gauge boson amplitude of the TMYM.}
\\[2mm]
$[\hsx$arXiv:2110.05399 [hep-th]$\hs ]$
\end{abstract}

\setcounter{footnote}{0}

\newpage

\tableofcontents    

\newpage

\baselineskip=17.5pt

\section{\hspace*{-3mm}Introduction}
\label{sec:1}

Chern-Simons gauge theories in (2+1)-dimensional (3d)  
spacetime play an important role 
in studying modern quantum field theories for particle physics and 
condensed matter 
physics\,\cite{Deser:1981wh}\cite{Dunne:1998}\cite{Tong:2016kpv}.
Such 3d theories can always contain gauge-invariant mass terms
of gauge bosons through the topological mass-generation 
\`{a} la Chern-Simons (CS)\,\cite{CS}, 
without invoking the conventional 
Higgs mechanism\,\cite{Higgs}\footnote{The conventional Higgs mechanism\,\cite{Higgs} is sometimes also called 
Brout-Englert-Higgs (BEH) mechanism or Anderson-Higgs mechanism\,\cite{Zee} in the literature.} 
in the 4d standard model (SM). 

\vspace*{1mm}

In this work, we study the dynamics of 
3d topological mass-generation for the (Abelian and non-Abelian) 
gauge bosons $A^a_{\mu}\hsx$.\
A spin-1 massless gauge boson in 3d contains only one 
physical degree of freedom (DoF) which is the transversely 
polarized state $\AT\hsx$. Including the gauge-invariant
topological CS term converts this massless transverse 
polarization state $\AT$ into a massive physical state
$\AP$\,.
We newly formulate this 3d topological mass-generation mechanism
at $S$-matrix level. 
For this, we propose and prove a new
Topological Equivalence Theorem (TET) 
which connects the $N$-point scattering amplitudes of the 
physical polarization states of massive gauge bosons ($\AP$) 
to the scattering amplitudes of the corresponding 
transversely polarized gauge boson states ($\AT$)
under high energy expansion. 
This differs essentially from the conventional equivalence theorem (ET)
\cite{ET-Rev} in the 4d SM
because the 3d gauge bosons acquire gauge-invariant topological mass-term without invoking the conventional 
Higgs mechanism\,\cite{Higgs}.
We note that the Kaluza-Klein ET 
(KK-ET)\,\cite{5DYM2002}\cite{5DYM2002b}\cite{5DYM2004} 
was formulated for 
the compactified 5d Yang-Mills theories which realize a geometric
Higgs mechanism with the 5th component of 5d gauge field converted
to the longitudinal component of the corresponding 4d massive KK 
gauge boson. But our TET also has essential difference from the KK-ET,
because the 5d gauge symmetry is spontaneously broken 
by compactification down to the 4d residual gauge symmetry 
of the massless zero-modes  
and the induced KK gauge boson mass-term is not 
gauge-invariant. In contrast, the 3d CS term
for the topological mass-generation of gauge bosons 
can be manifestly gauge-invariant and the 3d gauge symmetry is
unchanged before and after including the CS term.

\vspace*{1mm}

We present a general 3d power counting method 
to count the leading energy-power dependence 
of the $N$-point scattering amplitudes in 
both topologically massive Yang-Mills (TMYM) theory 
and topologically massive gravity (TMG) theory. 
Using the TET and power counting method for the 3d TMYM theory, 
we uncover that despite the individual diagrams 
in a given $N$-particle scattering amplitude 
of on-shell physical gauge bosons ($N\!\!\geqq\! 4$) 
having leading energy dependence of $E^4$ at tree level, 
they have to cancel down to $E^{4-N}$ 
in the full tree-level amplitude. 
We will prove that 
{\it the TET provides a general theoretical mechanism to 
guarantee such nontrivial energy cancellations:
$\!E^4\!\!\to\!E^{4-N}\hsm$,}$\hs$
without invoking any conventional Higgs boson.
For the massive $4$-gauge boson scattering amplitudes at tree level, 
we will demonstrate explicitly the large energy cancellations of
$\,E^4\!\to\! E^0\,$ under high energy expansion. 

\vspace*{1mm}

Furthermore, using the scattering amplitude of topologically massive 
gauge bosons in the 3d TMYM theory, we will reconstruct the 
topologically massive graviton scattering amplitude 
of the 3d TMG theory, by extending 
the conventional double-copy method of 
Bern-Carrasco-Johansson (BCJ)\,\cite{BCJ}\cite{BCJ-Rev}
for massless gauge/gravity theories  
to the current 3d topologically massive gauge/gravity theories.\ 
The BCJ method was inspired by
the Kawai-Lewellen-Tye (KLT)\,\cite{KLT} relation which connects
the product of open string amplitudes to that of the closed string 
at tree level. Analyzing the properties of the
heterotic string and open string amplitudes can prove and refine
parts of the BCJ conjecture\,\cite{Tye-2010}.
Many studies appeared in the literature to test the double-copy
conjecture in massless gauge/gravity field theories\,\cite{BCJ-Rev}, 
and some recent works attempted to extend the double-copy method to 
the 4d massive YM theory versus Fierz-Pauli-like 
massive gravity\,\cite{dRGT}\cite{DC-4dx1}\cite{DC-4dx2}, 
to the KK-inspired effective gauge theory with extra 
global U(1)\,\cite{DC-5dx}, 
and to the compactified 5d KK gauge/gravity 
theories and KK string theories\,\cite{Hang:2021fmp}\cite{Li:2021yfk}.
Double-copies of three- and two-algebra gauge theories were considered 
previously for the 3d supersymmetric theories\,\cite{Agarwal:2008pu}\cite{songhe}\cite{ythuang}, 
and some double-copy analyses for the amplitudes with matter fields
in 3d CS gauge theory as well as the study of 3d covariant 
color-kinematics duality appeared very recently\,\cite{DC-3dx}\cite{DC-3dx2}\cite{Moynihan:2021rwh}.

\vspace*{1mm} 

We stress that the topological mass-generation for
gauge bosons and gravitons in the 3d TMYM and TMG theories 
can be realized in a fully gauge-invariant way 
under the path integral formulation,
which is important for the successful double-copy construction
in the massive gauge/gravity theories.
In this work, we will use an extended double-copy approach 
to construct the massive four-graviton amplitude of the TMG theory
from the corresponding massive four-gauge boson amplitude 
of the TMYM theory with properly improved kinematic numerators.
Our findings newly demonstrate a series of {\it strikingly large energy 
cancellations, $\,E^{12}\!\!\to\!E^1\hsm$,\, 
in the massive four-graviton amplitude}
under high energy expansion. With these we establish  
{\it a new correspondence between the two types of 
leading energy cancellations in the massive scattering amplitudes:
$E^4 \ito E^0\hs$ in the TMYM theory and
$\,E^{12}\!\to\!E^1$ in the TMG theory.}

\vspace*{1mm}

This paper is organized as follows.
In section\,\ref{sec:2}, 
we study the mechanism of the topological mass-generation
in the 3d CS gauge theories at the Lagrangian level 
via path integral formulation.\ 
We identify the conversion of transverse polarization state $\AT$ 
in the massless theory into the massive physical polarization state
$\AP$ under the topological mass-generation of the CS gauge theories.
In section\,\ref{sec:3.1}, we propose and prove the new TET 
which connects the $N$-point 
$\AP$-amplitudes to the corresponding $\AT$-amplitudes 
under high energy expansion.\ 
Using the TET, we newly formulate the mechanism of topological 
mass-generation at $S$-matrix level.\ 
Then, in section\,\ref{sec:3.2},
we present the general 3d power counting rules 
on the leading energy dependence of 
the $N$-point scattering amplitudes in 
both the CS gauge theories and the TMG theory. 
Using the TET and the power counting rule, 
we prove in section\,\ref{sec:3.3}
a general energy cancellation mechanism 
for the $N$-gauge boson scattering amplitudes which predicts
the cancellation 
$E^4\ito E^{4-N}$ at tree level.\ 
For sections\,\ref{sec:4.1}-\ref{sec:4.2}, 
we first compute the four-point matter-induced gauge boson amplitudes,
and then compute the pure four-gauge boson amplitudes for the 
$\AP$-states and $\AT$-states 
in the Abelian and non-Abelian CS gauge theories.\ 
These analyses explicitly demonstrate the TET for the first time,
and newly establish the energy cancellations 
$E^2\hsm\ito E^0$ of the four-point amplitudes
with just two gauge bosons 
(in either Abelian or non-Abelian CS theories)
and the energy cancellations $E^4\hsm\ito E^0\hs$ 
of the four-gauge boson amplitudes (in TMYM theories). 
In section\,\ref{sec:4.3}, we analyze the perturbative unitarity bounds 
for both the TMYM theory and the TMG theory. 
We demonstrate that their partial wave amplitudes can exhibit good 
high energy behaviors. 
In section\,\ref{sec:5}, we further extend the double-copy approach 
and construct the massive four-graviton amplitude of the TMG from  
the massive four-gauge boson amplitude of the TMYM. 
With these, we newly uncover strikingly large energy cancellations 
in the four-graviton amplitude:\ $E^{12} \ito E^1$,\,
and establish its new correspondence to the leading energy cancellation 
$E^4 \ito E^0\hs$ in the massive four-gauge boson amplitude of the TMYM. 
We conclude in section\,\ref{sec:6}.
Finally, we provide more derivations and formulas in
Appendices\,\ref{app:A}-\ref{app:E} which are used 
for the analyses in the main text.

\section{\hspace*{-3mm}Topological Mass Generation in Chern-Simons Gauge Theories}
\label{sec:2}

We consider the 3d topological massive gauge theories including 
the Chern-Simons (CS) Lagrangian 
with Abelian or non-Abelian gauge symmetry, 
where the former may be denoted as 
Topologically Massive QED (TMQED) 
and the latter as Topologically Massive Yang-Mills (TMYM) theory. 
In either case, the CS Lagrangian provides a gauge-invariant 
topological mass-term for the 3d gauge bosons.
The 3d TMQED and TMYM Lagrangians have their gauge sectors 
take the following forms:
\beqs
\label{eq:L-MCS-YMCS}
\begin{align}
\label{eq:L-MCS}
\hspace*{-2mm}
\La_{\rm{TMQED}}^{} 
&=-\frac{1}{4} F_{\mn}^2 \!+\! 
\frac{1}{2}\mt\,\vep^{\mnr} A_{\mu} \pd_{\nu} A_{\rho} \,,
\\[2mm]
\hspace*{-3.5mm}
\label{eq:L-YMCS}
\hspace*{-3.5mm}
\La_{\rm{TMYM}} &=  -\frac{1}{2} \tr\bfF_{\mn}^2
\!+ \mt\,\vep^{\mnr} \tr\!\(\!\bfA_{\mu}  
\pd_{\nu}\bfA_{\rho}  \!-\! 
\frac{\,\ii\hspace*{0.3mm}2g\,}{3} 
\bfA_{\mu}\bfA_{\nu}\bfA_{\rho} \!\) \!,
\end{align}
\eeqs
where the non-Abelian gauge field
$\,\bfA_\mu \!\!=\! A^a_\mu T^a$,
and its field strength
$\,\bfF_{\mn} \!=\! F^a_{\mn}T^a$\, with 
$\bfF_{\mn}\!=\!\pd_{\mu} \bfA_{\nu } \!\!-\! \pd_{\nu } \bfA_{\mu} \!-\! \ii g[\bfA_{\mu},\bfA_{\nu}]$\,
and $\,T^a$ denotes the generator of 
the non-Abelian group SU($N$).
The gauge coupling $g$ has mass-dimension
$\fr{1}{2}$.
The gauge boson acquires a topological mass
$\,m\!=\!|\mt|$\, from the CS term, and the ratio
$\,\mathfrak{s}={\mt}/{m}=\pm 1$\,
corresponds to its spin projection\,\cite{Deser:1981wh}\cite{Dunne:1998}.
The mass parameter $\mt$ is related to the CS level
$\,n =4\pi\mt/g^{2} \in \mathbb{Z}\,$ 
\cite{Dunne:1998}\cite{Tong:2016kpv}.
The CS terms in Eq.\eqref{eq:L-MCS-YMCS} violate
the discrete symmetries $P$, $T$ and $CP$.

\vspace*{1mm} 

For the TMQED \eqref{eq:L-MCS}, the action
$\int\!\hspace*{-0.25mm}\td^3x\hspace*{0.35mm}\La_{\rm{TMQED}}^{}\,$ 
is gauge-invariant up to a total derivative 
which vanishes at the boundary for trivial topology.
For the TMYM theory \eqref{eq:L-YMCS}, under the
gauge transformation
$\,\bfA_{\mu}^{}\hsmx \to\hs\bfA_{\mu}' \hsm =\hs   
U^{-1} \bfA_{\mu} U \!+\! \fr{\ii}{g}U^{-1} \pd_{\mu} U$,\, 
the action changes by
\beqs
\begin{align}
\label{eq:DeltaS}
&\Delta S_{\rm{TMYM}} \,=\, 2 \pi nw +\! \int \! \td^{3} x \! 
\[\ii\hspace*{0.35mm}\mt g^{-1} \vep^{\mnr} \pd_{\nu}
\hspace*{0.3mm} 
\tr \!\(\pd_{\mu} U U^{-1} A_{\rho} \) \!\] \!,
\\[2mm]
&  w\,=\,\frac{1}{\,24 \pi^2\,} \!\!\int \! \td^{3} x \!
\[\! \vep^{\mnr} \,\tr(U^{-1} \pd_{\mu} UU^{-1} \pd_{\nu}
UU^{-1} \pd_{\rho}U ) \!\] \!,
\end{align}
\eeqs
where $\,w\,$ is the winding number which follows from the homotopy group $\Pi_{3}[\rm{SU}(N)] \!\cong\! \ZZ$ 
\cite{Tong:2016kpv}.
Hence, \eqrefe{eq:DeltaS} will not contribute to the path integral
since $\,e^{\ii 2\pi n w}=1\,$,\, and the second term is a 
total derivative (similar to the Abelian case).

\vspace{1mm}

With the path integral formulation, we can add the covariant 
gauge-fixing term and the Faddeev-Popov ghost term:
\beqs 
\label{eq:LGF-FP} 
\begin{align}
\label{eq:LGF}
\La_{\rm{GF}}^{} &= -\frac{1}{\,2\xi\,}(\FF^a)^2 ,
~~~~~ \FF^a\!= \pd^\mu \!A_\mu^a\,,
\\[2mm]
\label{eq:LFP}
\La_{\rm{FP}}^{} &=
\bar{c}^a\pd^\mu \!\(\delta^{ab} \pd_\mu -gC^{abc}A_\mu^c\)\!c^b,
\end{align}	
\eeqs
where $C^{abc}$ is the gauge group structure constant and
$(c^a,\,\bar{c}^a)$ denote the Faddeev-Popov ghost and anti-ghost fields.
Eq.\eqref{eq:LGF-FP} can be reduced to the Abelian case by
simply setting $\,C^{abc}$$=0\,$ and $\,A^a_{\mu}\!=\! A_\mu$.
So, hereafter we need not to specify the Abelian case unless needed.
The quantized CS action 
$\,\int\!\td^{3}x(\La +\La_{\rm{GF}}^{}+\La_{\rm{FP}}^{})\,$
is BRST-invariant (Becchi-Rouet-Stora-Tyutin), with which we
can derive the relevant BRST identities.

\vspace{1mm}

The equation of motion (EOM) for the massive gauge boson $A^a_\mu$ 
can be derived from the quadratic part of the CS action,
\begin{equation}
\label{eq:EOM-A}
\[\eta^{\mu\nu}\pd^2 
+(\xi^{-1}\! -\!1)\pd^\mu\pd^\nu\!
+\mt\,\vep^{\mu\rho\nu}\pd_{\rho}^{}
\]\!A^{a}_{\nu} \,=\,0\,,
\end{equation}
which describes the propagation of the free field $A^a_\mu$\,.
For the on-shell wave solution  
$\,A^a_{\mu} \!\sim\!\ep_\mu(p)e^{-\ii p \cdot x}\,$
with $\,p^\mu\!A_\mu^a\!=0$\,, 
the polarization vector should satisfy the 
equation    
\begin{equation}
\label{eq:PolEOM}
( m \eta^{\mn}-\ii\hspace*{0.2mm}\sp\hspace*{0.2mm}
\vep^{\mu\rho\nu}p_{\rho})\,\ep_{\nu}^{}(p) 
\,=\, 0 \,,
\end{equation}
under the on-shell condition $\,p^2 \!=\! -m^2$\, and with
$\,\mathfrak{s}\!=\!{\mt}/{m}\!=\!\pm 1$\,. 
The 3d Poincar\'e group ISO(2,1)
contains the proper Lorentz group 
SO(2,1) and translations. The little group 
is $\,\ZZ_2^{}\otimes\mathbb{R}$ for massless particles and 
SO(2) for massive particles\,\cite{UIR3}.
The Poincar\'e algebra is characterized by two Casimir operators
$(P^2,\,W)=(P_\mu^{}P^\mu,\,P_\mu^{}J^\mu)$,\,
where $W$ is the Pauli-Lubanski pseudoscalar and the angular
momentum $J^\mu$ can be generally expressed as\,\cite{Jackiw:1991}:
\begin{equation}
\label{eq:J}
J^\mu \,=\, -\ii\hspace*{0.3mm}\vep^{\mn\al}p_\nu^{}
\frac{\!\partial}{\partial p_\al^{}}
-\sp\frac{\,\,p^\mu\!+\!\eta^\mu m\,}{\,p\!\cdot\!\eta -m\,}\,,
\end{equation}
with $\,\eta^\mu \!=\!(1,0,0)\,$. 
Thus, in the rest frame it gives
\,$W\!=P\!\cdot\! J=-\sp\hspace*{0.3mm}m\,$.
We see that the spin is a pseudoscalar 
and takes the values $\,\sp\! =\pm 1$\, 
for gauge fields $A^a_\mu$\,. 
The polarization state with either
$\,\sp\!=\!+1\,$ or $\,\sp\!=\!-1\,$ 
is physically equivalent.
(More discussions on the gauge boson polarization vector 
are given in Appendix\,\ref{app:A}.)

\vspace*{1mm}

Note that the 3d massless gauge field can be viewed as a 
scalar field of spin-0 with one physical degree of 
freedom (DoF)\,\cite{UIR3}\cite{Jackiw:1991}.
Including the CS term does not add any new field,
and the total physical DoF remains as one because 
the physical DoF of $A^a_{\mu}$
should be conserved\,\cite{Jackiw:1991}\cite{Pisarski:1985}.
For the on-shell one-particle state, 
the 3d massless gauge boson has a single 
(transverse) physical polarization state 
$\,\AT =\hsm\epT^\mu A_\mu^a\hs$.
As the physical DoF is conserved, 
the CS term could only convert the massless $\AT$ state   
into a massive physical polarization state 
$\AP\!=\!\epP^\mu A_\mu^a$\,.

\vspace{1mm}

For the on-shell gauge boson in the rest frame with momentum 
$\,p^\mu\! =\!(m,\,0,\,0)\equiv \bar{p}^\mu$,\,
the physical polarization vector $\epP^\mu(\bar{p})$
can be solved from \eqrefe{eq:PolEOM}:
\begin{equation}
\epP^\mu(\bar{p}) \,=\,
\fr{1}{\sqrt{2\,}\,}(0,\,1,\,-\ii\hspace*{0.2mm}\sp )\,,
\end{equation}
in agreement with \cite{Pisarski:1985}\cite{Banerjee:2000gc}.
Then, by making a Lorentz boost we can express $\hs\epP^\mu(p)\hs$
in the moving frame:
\begin{align}
\label{eq:Pol-Physical}  
\hspace*{-3mm}
\epP^\mu(p) \,= \,\frac{1}{\sqrt{2\,}\,} \!\!\(\!
\frac{\,\ii p_1^{} \!\!+\!\sp p_2^{}}{m},\, 
\ii\!+\!  
\frac{\,p_1^{}(\ii p_1^{} \!\!+\! \sp p_2^{})}{m(m\!-\! p_0^{})},\,  
\sp\!+\!  \frac{\,p_2^{}(\ii p_1 \!\!+\! \sp p_2)}
{m(m\!-\! p_0^{})}\!\) \!,
\end{align}
which agrees with \cite{Banerjee:2000gc}
up to an overall factor $\ii\,$.
The on-shell physical polarization vector 
$\epP^\mu(p)$ obeys the conditions
$\,\epP^{\mu} \ep^{\,*}_{\text{P}\mu} \!\!=\! 1$\, 
and $\,p_\mu \ep^\mu_P \!\!=\!0$\,.
We can express the general momentum $p^\mu$ 
in a familiar form
$\,p^\mu \!=\!E(1,\,\be\st,\,\be\ct)\,$,
where the notations $\,p^0\!=\!-p_0^{}\!=\!E\hs$,
$\hs (\st,\,\ct)\!=\!(\sin\theta,\,\cos\theta)$, 
$\be\!=\!\hsm\sqrt{1\!-\!m^2\hsm /E^2\,}$,\,
and $\,\theta\,$ denotes the angle between the moving direction
and $y$-axis. With these, we can rewrite the 
polarization vector \eqref{eq:Pol-Physical} as follows:
\begin{equation}
\label{eq:Pol-epsionP}
\epP^\mu(p) = \fr{1}{\sqrt{2\,}\,} 
( \bE\be,\, \bE\st \!+\! 
\ii\hspace*{0.2mm}\mathfrak{s}\hspace*{0.2mm}\ct,\, 
\bE\ct \!-\! \ii\hspace*{0.2mm}\sp \st )
\,,
\end{equation}
where \,$\bE\!=\! E/m$\, 
and we have removed an irrelevant overall phase factor. 
Inspecting the structure of the physical
polarization vector \eqref{eq:Pol-epsionP}, we derive the
following general decomposition:
\begin{equation}
\label{eq:epP=epT+epL}
\epP^\mu = \fr{1}{\sqrt{2\,}\,}\!\(\epT^\mu +\epL^\mu\),
\end{equation}
which contains the transverse and longitudinal polarization vectors
$(\epT^\mu ,\,\epL^\mu)$ 
of the massive gauge boson $A^a_{\mu}$\,,
\begin{equation}
\epT^\mu = (0,\,\ii\hspace*{0.2mm}\sp\hspace*{0.2mm}\ct,\,
-\ii\hspace*{0.2mm}\sp\hspace*{0.2mm}\st),~~~~
\epL^\mu = \bE (\be ,\,\st,\, \ct)\,.
\end{equation}
Thus, we have the relation for the on-shell polarization states
of $A^a_\mu\hs$:
\begin{equation}
\label{eq:Ap}
\AP = \fr{1}{\sqrt{2\,}\,}\!\(\AT +\AL\),
\end{equation}
where  
$(\AP,\,\AT,\,\AL)\!=\!(\epP^\mu,\,\epT^\mu,\,\epL^\mu)A_\mu^a\,$.
The gauge boson $A_\mu^a$ also has an unphysical scalar 
polarization state $\,\AS\!=\!\epS^\mu A_\mu^a\,$ with
$\,\epS^\mu\!=\!p^\mu\!/m\,$.\
It is important to note that the polarization vectors
$(\epP^\mu,\,\epL^\mu,\,\epS^\mu)$ 
are all enhanced by energy and scale
as $\mO(E/m)$ under the high energy expansion.
The 3d gauge boson $A^a_\mu$ has 3 possible polarization states
in total, including 1 physical polarization and 2 unphysical
polarizations. In the massless case ($m\!=\!0$), 
$A^a_{\mu}$ contains 1 physical transverse polarization state 
$\,\AT\!=\!\epT^\mu A^a_\mu$\,
and 2 unphysical (longitudinal,~scalar) polarization states 
$(\AL,\AS)\!=(\epL^\mu A_\mu^a,\,\epS^\mu A_\mu^a)$
with $\,\epL^\mu\! +\epS^\mu\propto\! p^\mu\,$. 
On the other hand, 
for the massive case with CS term ($m\!\neq\! 0$), 
$A^a_\mu\,$ includes 1 physical polarization state $\AP$ 
as in Eq.\eqref{eq:Ap} 
and 2 orthogonal unphysical polarization states:
\beqs 
\begin{align}
\label{eq:Ax}
\AX &\,=\, \epX^\mu A_\mu^a 
= \fr{1}{\sqrt{2\,}\,}\!\(\AT -\AL\),
\\
\AS &\,=\,  \epS^\mu A_\mu^a \,,
\end{align}
\eeqs 
where
$\,\ep_{\rm{X}}^\mu \!=(\epT^\mu\!-\epL^\mu)/\!\sqrt{2\,}\,$
and $\,\epS^\mu\!=p^\mu\!/m\,$.
The three polarization vectors obey the orthogonal conditions  
$\,\epP^{}\!\cdot\epX^{\,*}=\epP^{}\!\cdot\epS^{\,*}
=\epX^{}\!\cdot\epS^{\,*} =0\,$. 
We see that adding the CS term for gauge boson $A^a_{\mu}$ 
dynamically generates a new physical polarization state 
$\AP$ of spin-1 (which has mass $m$ and is composed of
$\,\AT\!+\!\AL\,$), and converts its orthogonal combination
$\,\AX\!\propto\!(\AT\!-\!\AL)\,$ into the unphysical state,
while the scalar-polarization state $\AS=\epS^\mu A_\mu^a$
(with $\epS^\mu\!\propto\!p^\mu$) remains unphysical
as constrained by the gauge-fixing function 
$\,\FF^a\!=\!-\ii p^\mu\! A^a_\mu$\,
in Eq.\eqref{eq:LGF}.

\vspace{1mm}

The above mechanism of 3d topological mass-generation 
might be called a ``topological Higgs mechanism''
to resemble the dynamical {\it conversion} 
of $\,(\AT+\!\AL)$\, into the massive physical state $\AP$ of the 
gauge field $A^a_\mu$\,,\, while making the orthogonal combination 
$\,\AX\!\hsmx\propto\!\hsmx (\AT\hsmx -\!\AL)\,$ be an unphysical 
``Goldstone boson'' state.\ 
However, for the reasons given below, 
the ``topological Higgs mechanism'' is not the most appropriate name
for the 3d topological mass-generation.
We stress that the mechanism of topological mass-generation 
of gauge bosons {\it differs} from the
conventional Higgs mechanism\,\cite{Higgs} 
in essential ways:
(i)\,topological CS mass-term automatically holds the 
exact gauge symmetry in the path integral formulation, 
without invoking any spontaneous gauge symmetry breaking 
by the vacuum of Higgs potential; 
(ii)\,before including the CS term, the transverse $\AT$ 
is the physical polarization state and 
is exactly massless as ensured by the gauge symmetry; 
while after including the CS term, $\AT$ combines with $\AL$
to form the massive physical state $\AP$ and makes its orthogonal
combination $\AX$ become unphysical; hence there is no 
spontaneous symmetry breaking invoked to generate massless 
Goldstone boson, nor is there  
any extra physical Higgs boson component; 
(iii)\,the massive physical gauge boson state $\AP$ is converted
from the massless transverse polarization state $\AT$
combined with the longitudinal polarization state $\AL$
via Eqs.\eqref{eq:Ap} and \eqref{eq:Ax}. 
The single physical degree of freedom of $A^a_\mu$ 
is conserved before and after adding the CS term,
through the topological conversion
$\AT\!\to\!\AP$\,.
Taking the massless limit $m\!\to 0\,$,
we see that the massive state $\AP$ disappears and the massless
state $\AT$ is released 
to be the physical transverse polarization, while
the longitudinal state $\AL$ becomes fully unphysical again.\
As we will demonstrate shortly, in the high energy limit
the scattering amplitudes of the physical polarization states
($\AP$) equal the corresponding amplitudes of the
transverse polarization states ($\AT$),
which means that the $\AP$ state remembers its origin
of $\AT$ state under the limit $\,m/E\!\to\!0$\,.

\section{\hspace*{-3mm}Topological Equivalence Theorem for  
Chern-Simons Gauge Theories}
\label{sec:3}

As shown above, for the 3d topological gauge theories
\eqref{eq:L-MCS-YMCS}, 
the Chern-Simons (CS) Lagrangian generates 
a topological mass for gauge boson
$A_\mu^a$ by converting the massless transverse polarization state
$\AT$ (combined with the longitudinal polarization state $\AL$)
into the massive physical polarization state $\AP$\,.
In this section, we formulate the mechanism of 
topological mass-generation at the $S$-matrix level
by newly proposing and proving a general
Topological Equivalence Theorem (TET),
which quantitatively connects the $N$-point 
scattering amplitudes of $\AP$'s
to the corresponding amplitudes of the $\AT$'s
in the high energy limit $\,m/E\!\to\!0$\,.

\subsection{\hspace*{-3mm}Topological Equivalence Theorem for 
Topological Mass Generation}
\label{sec:3.1}
 
Inspecting the quantized CS Lagrangians 
\eqref{eq:L-MCS-YMCS} and \eqref{eq:LGF-FP} 
and following the method in Refs.\,\cite{ET94}
\cite{ET-Rev},
we can derive the following Slavnov-Taylor-type identity 
in momentum space:
\begin{equation}
\label{eq:F-ID}
\hspace*{-3mm}
\la 0|
\FF^{a_1}(p_1^{})\FF^{a_2}(p_2^{})\cdots \FF^{a_N}(p_N^{})\Phi
|0 \ra \,=\, 0\,,~~
\end{equation}
which is based on the 3d gauge symmetry, 
where $\,\FF^a$ 
is the gauge-fixing function defined in Eq.\eqref{eq:LGF}, 
and the symbol $\Phi$ denotes any other on-shell physical fields after the Lehmann-Symanzik-Zimmermann (LSZ) amputation.
Since the function $\,\FF^a$ contains only 
a single gauge field $A_\mu^a$
having no mixing with any other field, 
it is straightforward to amputate each external 
$\,\FF^a$ line by the LSZ reduction. 
We impose the on-shell condition $\,p_j^2\!=\!-m^2$\, 
for each external momentum.
In the momentum space, we can express the gauge-fixing function 
$\FF^a\!=\!-\ii p^\mu A_\mu^a\!=\!-\ii m \AS$\,. 
We also deduce 
$\,v^\mu\!\equiv \epL^\mu\!-\epS^\mu=\mO(m/E)\,$.
With Eq.\eqref{eq:epP=epT+epL}, we can express the scalar 
polarization vector $\,\epS^\mu $\, as
\begin{equation}
\label{eq:epS}
\epS^\mu \,=\,\sqrt{2\,}\epP^\mu - (\epT^\mu\!+\!v^\mu)\,.
\end{equation}
Thus, we derive the following formula 
for the gauge-fixing function:
\beqs 
\beqa
\FF^a  &\!\!=\!\!\!& 
-\ii \sqrt{2\,} m\hspace*{0.3mm} 
(\AP \!-\! \Omega^a)\hsx , 
\\[1mm]
\Omega^a &\!\!=\!\!\!& \fr{1}{\sqrt{2\,}\,}(\AT \!+\! v^a) 
\hsx ,
\eeqa
\eeqs 
where $(\AP,\,\AT)\!=\!(\epP^\mu,\,\epT^\mu)A_\mu^a\,$
and $\,v^a \!=\! v^\mu A^a_\mu$\,
with $\,v^\mu\!=\!\epL^\mu\!-\!\epS^\mu =\! \mO(m/E)$\,.
With these and Eq.\eqref{eq:F-ID} after the LSZ reduction, 
we can derive the following TET identity
which connects two scattering amplitudes:
\begin{align}
\label{eq:TET-ID}
& \TT [A_{\P}^{a_1^{}},\!\cdots\!,A_{\P}^{a_N^{}},\Phi]
\,=\, \TT [\Omega^{a_1^{}} ,\!\cdots\!,\Omega^{a_N^{}} \!,\Phi] 
\,,
\end{align}
where we have made use of the fact that an amplitude including 
one or more external $\FF$ lines plus any other external on-shell physical fields (such as $\AP$ and/or $\Phi$) must vanish 
due to the identity \eqref{eq:F-ID}.
Thus, we can expand the TET identity as follows:
\beqs 
\label{eq:TET-ID2}
\begin{align}
\label{eq:TET-ID2a}
\hspace*{-1mm}
& \TT [A_{\P}^{a_1^{}},\!\cdots\!,A_{\P}^{a_N^{}},\Phi]
\,=\, \TT [\tdA_{\T}^{a_1^{}} ,\!\cdots\!,\tdA_{\T}^{a_N^{}} \!,\Phi]\,+\,\TT_{v}^{}\,,
\hspace*{3mm}
\\[-0.5mm]
\label{eq:Tv}
\hspace*{-1mm}
& \TT_{v}^{} \,=\,  \sum_{j=1}^N
\TT[\vt^{a_1} \!,\!\cdots\!,\vt^{a_{j}}\!,\tdA_{\T}^{a_{j+1}}
\!,\!\cdots\!,\tdA_{\T}^{a_N} \!,\Phi] \,,
\end{align}
\eeqs 
where for convenience we have adopted the notations 
$\,\tdA_{\T}^a\!=\!\fr{1}{\sqrt{2\,}\,}\AT\,$ and
$\,\vt^a\!=\!\fr{1}{\sqrt{2\,}\,}v^a$.
Under the high energy expansion, the residual term behaves as 
$\,\TT_v^{}=\mO (m/E)\!\ll\! 1\,$
due to the suppression factor $v^\mu$. 
Thus, we can derive the Topological Equivalence Theorem
(TET):
\begin{equation}
\label{eq:TET}
\TT [A_{\P}^{a_1^{}},\!\cdots\!,A_{\P}^{a_N^{}},\Phi]
\,=\,\TT [\tdA_{\T}^{a_1^{}} ,\!\cdots\!,\tdA_{\T}^{a_N^{}} \!,\Phi]
+\mO\!\(\frac{m}{E}\)\hsm .
\end{equation}
The TET \eqref{eq:TET} states that any $\AP$-scattering amplitude 
equals the corresponding $\AT$-scattering amplitude 
in the high energy limit.
We note that different from the conventional equivalence theorem 
(ET)\,\cite{ET94}\cite{ET96}\footnote{%
The 4d ET in the presence of the Higgs-gravity interactions
was established in Ref.\,\cite{GET4d}
which can be applied to studying cosmological models (such as the
Higgs inflation\,\cite{HiggsInfx}\cite{GET4d}\cite{HiggsInf})
or to testing self-interactions of weak gauge bosons
and Higgs bosons\,\cite{GET4d}\cite{HRY}.}
for the case of the SM Higgs mechanism, the right-hand-side (RHS) of 
Eq.\eqref{eq:TET-ID2} or Eq.\eqref{eq:TET} 
receives no multiplicative modification factor
at loop level. This is because in the present case both
$\AP$ and $\AT$ belong to the same gauge field $A^a_\mu$
and the LSZ reduction on the left-hand-side (LHS) 
of Eq.\eqref{eq:F-ID} becomes much simpler.

Finally, we note that our present formulation of the 
TET \eqref{eq:TET} in the 3d CS gauge theories differs 
essentially from the conventional ET \cite{ET-Rev} in the 4d SM
because the 3d gauge bosons acquire gauge-invariant topological mass-term without invoking the conventional 
Higgs mechanism\,\cite{Higgs}.
We also note that the KK-ET \cite{5DYM2002}\cite{5DYM2002b}\cite{5DYM2004} 
for the compactified 5d Yang-Mills theories 
formulates the geometric Higgs mechanism at $S$-matrix level
where the 5th component of 5d gauge field is converted
to the longitudinal component of the corresponding 4d massive KK 
gauge boson. But our TET has essential difference from the KK-ET
because the 5d gauge symmetry is spontaneously broken 
down to the 4d residual gauge symmetry of zero-modes 
by the boundary conditions of 
compactification and the induced KK gauge boson mass-term is not 
gauge-invariant. On the contrary, the 3d CS term
for the topological mass-generation of gauge bosons 
can be manifestly gauge-invariant, and the inclusion of CS term 
does not change the gauge symmetry of the 3d theory.

\subsection{\hspace*{-3mm}Power Counting Method
for 3d Chern-Simons Theories}
\label{sec:3.2}

In this subsection, we develop a general energy power counting method 
for the scattering amplitudes
in the 3d topologically massive gauge and gravity theories.  
We also present a general energy power counting rule on
the $d$-dimensional scattering amplitudes
in Appendix\,\ref{app:B}.

\vspace*{1mm}

We note that Weinberg proposed a power counting method
for the 4d ungauged nonlinear $\sigma$-model as an 
effective theory of low energy QCD\,\cite{weinbergPC}.
The extensions of Weinberg's power counting method to
the compactified 5d Kaluza-Klein (KK) gauge theories 
and 5d KK gravity theory were recently given 
in Ref.\,\cite{Hang:2021fmp}.\footnote{%
Weinberg's power counting rule
was also extended previously\,\cite{ET-Rev}\cite{ETPC-97}
to the 4d gauge theories including the SM,
the SM effective theory (SMEFT), and the electroweak chiral Lagrangian.}
Weinberg's power counting method includes the following key points:
{(i).}\,For an $S$-matrix element $\,\mathbb{S}\,$, its 
total mass-dimension $D_{\mathbb{S}}^{}$ is determined 
by the number of external states ($\mathcal{E}$)
and the number of spacetime dimensions,\ 
$\,D_{\mathbb{S}}^{}\!=\hsmx 4-\mathcal{E}\,$, 
in the 4d field theories.\
{(ii).}\,Consider the scattering amplitude $\,{\mathbb{S}}^{}$
having scattering energy $E$ much larger than all the masses 
of the internal propagators as well as the masses of the external states.
Thus, for the $E$-independent coupling constants contained in the amplitude $\hs\mathbb{S}\hs$, their total mass-dimension $D_C^{}$
can be counted directly according to the type of vertices therein.
Based on these, the total energy-power dependence
$D_E^{}$ of the amplitude $\,\mathbb{S}\,$ is given by
$\,D_E^{}= D_{\mathbb{S}}^{} -D_C^{}\,$.
We note that for our following derivation in 3d spacetime (or
the general derivation in $d$-dimensional spacetime 
in Appendix\,\ref{app:B}), we should modify the formula of
$\,D_{\mathbb{S}}^{}\,$ in point\,(i) accordingly.
As for the point\,(ii), it should hold for any high energy scattering
with energy $E$ much larger than the involved particle masses.
The nontrivial energy-dependence from the polarization vectors (tensors)
of the gauge bosons (gravitons) can be taken into account accordingly.
Keeping these in mind, we will construct the new power counting rules 
for the 3d topologically massive gauge and gravity theories.

\vspace*{1mm}

Consider a general scattering $S$-matrix element $\,\mathbb{S}\,$ having
$\,\EE\,$ external states and $L$ loops ($L\!\geqq\! 0$)
in the (2+1)d spacetime.
Thus, we can deduce that the amplitude $\,\Sb\,$ 
has a mass-dimension:
%
\begin{equation}
\label{eq:DS}
D_{\mathbb{S}}^{} \,=\, 3 - \fr{1}{2}\EE \,,
\end{equation}
where the number of external states
$\,\EE \!=\EE_B^{}+\EE_F^{}\,$,
with $\,\EE_B^{}\,(\EE_F^{})\,$ being the number of
external bosonic (fermionic) states. 
We note that the above Eq.\eqref{eq:DS} agrees to the
$\hs d\!=\!3\hs$ case of our general formula \eqref{eq:DS-d}
in Appendix\,\ref{app:B}.
For the fermions, we only
consider the SM fermions whose masses are much smaller than the
scattering energy $E$\,.\,
We denote the number of vertices of type-$j$ as $\VV_j^{}$\,.
Each vertex of type-$j$ contains $\,d_j^{}\,$ derivatives,
$\,b_j^{}\,$ bosonic lines, and $\,f_j^{}\,$ fermionic lines.
Then, the energy-independent effective coupling constant in
the amplitude $\,\mathbb{S}\,$ has its total mass-dimension given by
\begin{equation}
\label{eq:DC}
D_C^{} \,=\, \sum_j \VV_j^{}\!
\(3-d_j^{}\!- \fr{1}{2} b_j^{}\!- f_j^{}\) \!.
\end{equation}
For each Feynman diagram in the scattering amplitude
\,$\mathbb{S}$\,,\,
we denote the number of the internal lines as
$\,I=I_B^{}+I_F^{}\,$ with
$\,I_B^{}$ ($\,I_F^{}\,$) being the number of the internal
bosonic (fermionic) lines. Thus, we have the following general
relations:
\begin{equation}
\label{eq:L-V-I}
L \!= \! 1+I-\VV\,, \quad~~
\sum_j \VV_j^{}b_j^{} = 2I_B^{}+\EE_B^{}\,, \quad~~
\sum_j \VV_j^{}f_j^{} = 2I_F^{}+\EE_F^{}\,,
\end{equation}
where $\,\VV=\sum_j\!\VV_j^{}\,$
is the total number of vertices in a given Feynman diagram.
With these, we can derive the following leading energy dependence
$\,D_E^{}\!=\! D_{\mathbb{S}}^{}\! -\!D_C^{}\,$
from Eqs.\eqref{eq:DS}-\eqref{eq:L-V-I}:
%
\begin{equation}
\label{eq:DE0}
D_E^{} \,=\, 2(1-\VV)+
L+\sum_j  \VV_j^{}\!
\(d_j^{}\!+\!\fr{1}{2}f_j^{}\) .
\end{equation}
\\[-6mm]
Furthermore, 
we have the following relations:
\begin{align}
\label{eq:V-1}
\sum_j \VV_j^{}d_j^{} &=\VV_d^{}\,, \quad
\sum_j \VV_j^{}f_j^{} = 2\VV_F^{}\,, \quad
\VV \!=\!\sum_j\VV_j^{} = \VV_3^{} \!+\! \VV_4^{}\,, \quad
\VV_3^{}\!=\! \VV_d^{} \!+\! \VV_F^{}\!+\! \over{\VV}_3^{}\,,
\end{align}
where $\,\VV_d^{}\,$ denotes the number of all cubic vertices 
including one partial derivative
and $\,\over{\VV}_3^{}\,$ 
denotes the number of bosonic cubic vertices
having no partial derivative. 

\vspace*{1mm}

Then, we consider the topologically massive CS gauge theories.
In such gauge theories, we have the relation
$\,2I+\EE \!=3\VV_3^{}+4\VV_4^{}\,$.
With these, we can derive the following power counting rule on the
leading energy-power dependence of a general scattering amplitude:
\begin{equation}
\label{eq:DE}
D_E^{} \,=\, (\EE_{\!\APP}^{} \!-\EE_{v}^{})+
(4 - \EE - \over{\VV}_{\!3}^{}) -L  \,,
\end{equation}
where $\,\EE_{\!\APP}^{}$ is the number of external
gauge boson states with physical polarizations
($\AP\!=\epP^\mu A_\mu^a$),
and $\,\EE_{v}^{}\,$ denotes the number of external gauge bosons
$\,v^a\!=\!v_\mu^{}A^{a\mu}\,$.
In Eq.\eqref{eq:DE}, the terms 
$(\EE_{\!\APP}^{} \!\!-\EE_{v}^{})\,$ 
arise from the high energy behaviors
$\,\epP^\mu=\mO(E/m)\,$ and
$\,v^\mu \!=\!\epL^\mu\!-\epS^\mu\!=\mO(m/E)$\,.

\vspace*{1mm}

For the sake of later applications, 
we further consider the 3d topologically massive gravity (TMG)
and derive the energy power counting rule for general 
scattering amplitudes of massive gravitons. 
The graviton self-interaction vertices from the 
gravitational CS term 
\eqref{eq:S-TMG} (cf.\ Sec.\,\ref{sec:5})
always contain 3 partial derivatives 
and contribute to the leading energy dependence
of the graviton scattering amplitudes, 
which correspond to 
$\,d_j^{} \!=\!3\,$ and $\,f_j^{} \!= 0\,$ in \eqrefe{eq:DE0}. 
Thus, we have 
$\,\sum_j\!\VV_j^{}d_j^{}\!=3\VV_{d3}^{}\,$
and $\,\VV\!=\!\VV_{\!d3}^{}\,$ 
in such leading diagrams, where
$\VV_{\!d3}^{}$ denotes the number of vertices containing
3 partial derivatives. 
Hence, the leading energy dependence of the pure graviton 
scattering amplitudes in (2+1)d arise from the Feynman diagrams 
containing the CS graviton vertices with 3 derivatives,
and can be derived as follows: 
%
\begin{equation}
\label{eq:DE-TMG}
D_E^{} \,=\, 2\EE_{\hP}^{} \!+ (2 + \VV_{\!d3}^{} +L) \,,
\end{equation}
where $\,\EE_{\hP}^{}$ denotes the number of external
graviton states with physical polarizations
($\hs\hP \!\!=\epP^{\mn}h_{\mn}^{}\hs$) and the physical graviton
polarization tensor scales as
$\,\epP^{\mn}\!= \mO(E^2\!/m^2)$.
For the leading tree-level diagrams composed of the
cubic CS vertices with 
$\,d_j^{}\!=\!3\,$,\, 
we derive a relation
$\,\EE_{\hP}^{} \!= 2+\VV_{\!d3}^{}\,$.\, 
Hence, using \eqrefe{eq:DE-TMG}, we can deduce the
leading energy dependence of such tree-level diagrams:
\begin{equation}
\label{eq:DE0-TMG}
D_E^{0} \,=\, 3\hspace*{0.3mm} \EE_{h_P} \,.
\end{equation}
For instance, the leading four-graviton scattering amplitudes
of the TMG theory contain individual leading energy terms
of $\,E^{12}\,$ at the tree level. We will analyze these further
in section\,\ref{sec:5}.

\subsection{\hspace*{-3mm}Energy Cancellations for Topological
Scattering Amplitudes}
\label{sec:3.3}

In this subsection, we will apply our power counting rule
\eqref{eq:DE} to analyze the leading energy dependence 
of the pure gauge boson scattering amplitudes in the
3d topological massive CS gauge theory.
We also note that 
because the 3d CS theory is super-renormalizable,
the leading energy dependence of a given amplitude is 
always given by the diagrams having $L\!=0$ (tree level) and
 $\,\over{\VV}_3^{}\!=0\,$. 
Thus, given the external states of an amplitude, its maximal
energy dependence is realized at tree level:
\begin{equation}
\label{eq:DE-max}
D_E^{\max} \,=\, (\EE_{\!\APP}^{} \!-\EE_{v}^{})+
(4 - \EE)  \,,
\end{equation}
with $L\!=0$ and $\,\over{\VV}_3^{}\!=0\,$. 

\vspace*{1mm}

According to Eq.\eqref{eq:DE-max},
the scattering amplitudes of pure gauge bosons ($\AP$)
with the number of external states
$\,\EE\!=\EE_{\APP}^{}\!\!=\!N\,$ and $\,\EE_v\!=0\,$
can receive leading individual contributions
of $\mO(E^4)$ at the tree level.
For the pure $\AT$-amplitudes with 
$\,\EE\!=\EE_{\ATT}^{}\!\!=\!N\,$ and 
$\,\EE_{\APP}^{}\!\!=\EE_v\!=0\,$,\, 
its individual leading contributions scale like
$\,\mO(E^{4-N})$\, at the tree level.
With these, we find that 
our TET identity \eqref{eq:TET-ID2a} guarantees the energy cancellation in the $N$-gauge boson ($\AP$) scattering amplitude 
on its LHS:
\begin{equation}
\label{eq:DE-cancel}
E^4 \,\to\, E^{4-N}\,.
\end{equation}
This is because on the RHS of Eq.\eqref{eq:TET-ID2a} 
the corresponding pure $N$-gauge boson
($\AT$) amplitude scales as 
$\,\mO(E^{\,4-N})$ 
and the residual term $\,\TT_v\,$  
(with $\EE_v\!\geqq\! 1$) scales no more than 
$\,\mO(E^{\,3-N})$\,.
We can readily generalize this result to up to $L$-loop
level and deduce the following {\it energy cancellations} 
based on \eqrefe{eq:TET-ID2a} and \eqrefe{eq:DE}:
\begin{equation}
\label{eq:DE-cancel2}
\Delta D_E \,=\, D_E[N\!\AP] - D_E[N\!\AT] \,=\, N \,.
\end{equation}
Hence, the TET identity \eqref{eq:TET-ID2} [or the TET
\eqref{eq:TET}] provides a general mechanism
which guarantees the nontrivial energy cancellations
in Eq.\eqref{eq:DE-cancel} or Eq.\eqref{eq:DE-cancel2}.

\vspace*{1mm} 

Before concluding the current section\,\ref{sec:3}, we discuss further   
the conversion of physical degrees of freedom during the 3d topological
mass-generation, in comparison with that realized during the 5d 
geometric mass-generation under the Kaluza-Klein (KK) compactification.  
For the 3d gauge theories, before including the CS term,
the massless gauge boson $A_\mu^a$ has only 1 physical transverse
polarization state $\AT$; 
while after including the CS term, the gauge boson
acquires a topological mass and generates a single 
physical polarization state $\AP$
(by absorbing the massless state $\AT$ combined with the
longitudinal state $\AL$),
without invoking the conventional 
spontaneous gauge symmetry breaking.
Hence, this topological mass-generation mechanism leads
to the conversion of the physical states:\
$\AT\!\to\!\AP$\,,\,  
which conserves the physical degree of freedom:\
$1\!=\!1\,$,\,
as we explained earlier. 
In consequence, we observe that both 
the Lagrangians \eqref{eq:L-MCS}-\eqref{eq:L-YMCS} 
and the gauge boson propagator 
\eqref{eq:A-propagator} indeed have a 
{\it smooth massless limit}
$\,m\ito 0\,$, 
which is similar to the massive KK gauge 
theories\,\cite{5DYM2002}-\cite{5DYM2004}.
Based upon this mechanism of the topological mass-generation, 
we have newly established the TET \eqref{eq:TET}
which connects a given $\AP$-amplitude to
the corresponding $\AT$-amplitude 
under the high energy expansion.

\vspace*{1mm}

In comparison, we note that the 5d geometric mass-generation
for the KK gauge bosons $A_n^{a\mu}$ 
is realized by absorbing (``eating'') the corresponding 
5th components $A_n^{a5}$ of the 5d gauge fields 
$\widehat A^{a}_{M}$ 
at each KK level-$n$
\cite{5DYM2002}\cite{5DYM2002b}. 
The 5th components $A_n^{a 5}$ may be regarded as a kind of 
``geometric Goldstone bosons'' due to the KK compactification,
although they do not arise from a separate scalar Higgs potential
and differs essentially from the conventional Higgs 
mechanism\,\cite{Higgs}.
The 5d massless gauge boson $\widehat{A}^{a}_{M}$ has 3 physical 
transverse polarizations and after KK compactification
each 4d massive KK gauge boson $A_n^{a\mu}$ has 
2 transverse polarizations plus 1 longitudinal polarization
(from absorbing $A_n^{a5}$). So, the physical degrees of freedom
are conserved before and after the KK mass generation:
$\,3=2+1\,$; and this corresponds to the conversion of one
physical degree of freedom at each KK level-$n$: 
$A_5^{an}\!\to\! A_{\L}^{an}$.
This geometric mass generation of KK gauge bosons leads to
the KK Equivalence Theorem (KK-ET) 
which connects the high-energy scattering amplitudes of the
longitudinal KK gauge bosons $A_{\L}^{an}$ to that of the
corresponding KK Goldstone bosons $A_n^{a 5}$
\cite{5DYM2002}\cite{5DYM2004}.\footnote{%
Besides, the study of the geometric mass-generation of 5d KK gravitons 
and its gravitational equivalence theorem (GRET) were presented 
recently in Ref.\,\cite{Hang:2021fmp}, where the KK graviton field 
$\hs h_n^{\mn}\hsm$ becomes massive by absorbing the scalar-component
$h_n^{55}$ and vector-component $h_n^{\mu 5}$ from compactification
of the 5d graviton field $\hat{h}^{MN}$. 
Note that before compactification the massless 5d graviton 
$\hat{h}^{MN}$ has 5 physical degrees of freedom and 
after compactification the massive KK graviton $h_n^{\mn}$ contains
the physical states with helicities 
$\hs\lambda \!=\hsm \pm 2,\pm 1, 0\hs$. 
We see that the physical degrees of freedom
are conserved before and after the KK graviton mass-generation:
$\,5=2+\hsm 2\hsm +1\,$.}

\vspace*{2mm}
\section{\hspace*{-3mm}Topological Scattering Amplitudes and Energy Cancellations}
\label{sec:4}

In this section, we present explicit calculations of the
four-particle scattering amplitudes in the topologically massive
gauge theories including the Abelian QED \eqref{eq:L-MCS} 
and the non-Abelian TMYM theory \eqref{eq:L-YMCS}. 
With these, we newly demonstrate the energy cancellation
of $\,E^2\!\to\!E^0\,$ for the $A_\P^{}$-amplitudes 
in the TMQED and the energy cancellation
of $\,E^4\!\to\!E^0\,$ for the pure $\AP$-amplitudes 
in the TMYM theory, under high energy expansion. 
Then, we verify for the first time that the TET \eqref{eq:TET} 
holds for both the Abelian and non-Abelian CS gauge theories.

\vspace*{1.5mm}
\subsection{\hspace*{-3mm}Topologically Massive QED and 
                          Scattering Amplitudes}
\label{sec:4.1}

In this subsection, we consider two realizations
of the topologically massive QED, namely,
the topologically massive scalar QED (TMSQED)
and the topologically massive spinor QED (TMQED).
We will compute the scattering amplitudes in these
two models and uncover the nontrivial
energy cancellations in these amplitudes.
Then, we will demonstrate explicitly
that the TET \eqref{eq:TET} holds in each model.

\vspace*{1mm}

\subsubsection{\hspace*{-3mm}Scattering Amplitudes
of Topologically Massive Scalar QED}
\label{sec:4.1.1}

We first consider the TMSQED, which is composed by the
scalar QED plus the Chern-Simons term \eqref{eq:L-MCS}.
The Lagrangian contains a scalar sector:
\begin{equation}
\La_{\text{S}}^{} \,=\,  -(D_\mu \phi)^*(D^\mu \phi)
- m_{\phi}^2\hspace*{0.3mm}|\phi|^2
- \lambda |\phi|^4\,,
\end{equation}
where we choose the metric tensor
$\eta_{\mn}\!=\! \eta^{\mn}\!=\! \diag(-1,1,1)$
and denotes the complex scalar field by
$\phi\,$. The covariant derivative is defined as
$\,D_\mu\!\!=\! \pd_\mu^{}\! + \ii e A_\mu^{}$\,.
In the charge eigenstates, we have
$(\phi^-,\,\phi^+)=(\phi,\,\phi^*)$\,,
with $\phi^-$\,($\phi^+$) denoting the scalar electron
(scalar positron).

\vspace*{1mm}

\begin{figure}[t]
\vspace*{-3mm}
\centering
\includegraphics[height=5cm,  width=9.cm]{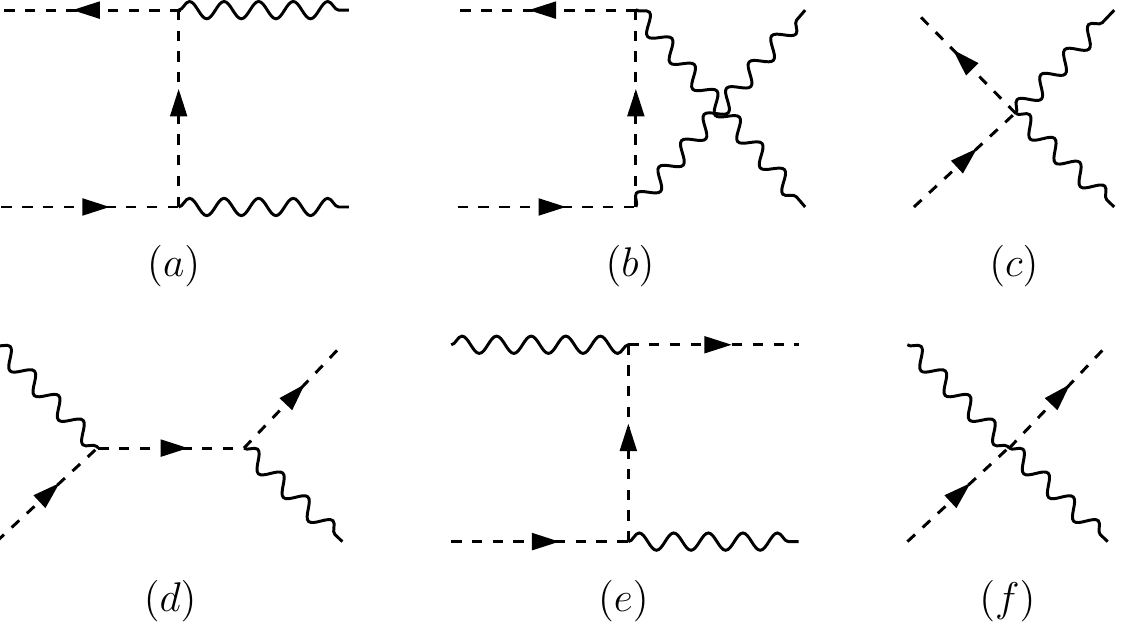}
\vspace*{-2mm}
\caption{\small{Feynman diagrams for the scattering processes
$\,\phi^- \phi^+\!\ito \APP\APP$
$(\phi^-\phi^+\!\ito \ATT\ATT)$\, and
\,$\phi^-\!\APP \ito \phi^-\!\APP$\,
$(\phi^-\!\ATT \to \phi^-\!\ATT)$\,
in 3d topological massive scalar QED.}}
\label{fig:1}
\vspace*{3mm}
\end{figure}

In the following, we compute and analyze
two types of scattering processes
$\,\phi^- \phi^+\!\ito \APP\APP$
$(\phi^-\phi^+\!\ito \ATT\ATT)$\, and
\,$\phi^-\!\APP \ito \phi^-\!\APP$\,
$(\phi^-\!\ATT \ito \phi^-\!\ATT)$
at tree level, where the relevant Feynman diagrams
are shown in Fig.\,\ref{fig:1}.

%

For the annihilation processes
$\,\phi^- \phi^+\!\ito \APP\APP$ and
$\phi^-\phi^+\!\ito \ATT\ATT$\,,
we find that under the high energy expansion and by using
the power counting rule \eqref{eq:DE},
the scattering amplitude
$\,\TT[\phi^-\phi^+\!\ito \APP\APP]$
scales as $E^2$, while the scattering amplitude
$\,\TT[\phi^-\phi^+\!\ito \ATT\ATT]$
scales as $E^0$.
Thus, we can make high energy expansions for
both amplitudes as follows:
\beqs
\label{eq:T[2phi-AA]}
\begin{align}
\label{eq:T[2phi-ApAp]}
\TT[\phi^-\phi^+\!\ito \APP\APP] &\,=\,
\TT^{(2)}_{\P\P}\bE^2 + \TT^{(1)}_{\P\P}\bE^1
+ \TT^{(0)}_{\P\P}\bE^0 + \mO(\bE^{-1})\,,
\\[1mm]
\label{eq:T[2phi-AtAt]}
\TT[\phi^-\phi^+\!\ito \ATT\ATT] &\,=\,
\TT^{(0)}_{\T\T}\bE^0 + \mO(\bE^{-1})\,,
\end{align}
\eeqs
where $\,\bE\!=E/m\,$ and $E$ denotes the energy of the
scalar electron (positron). For simplicity, we set the
scalar mass $\,m_\phi^{}\!\simeq 0\,$.
The amplitude $\TT[\phi^-\phi^+\!\ito \APP\APP]$
contains the contributions of the Feynman diagrams
$(a)$-$(c)$ of Fig.\,\ref{fig:1}.
According to Eq.\eqref{eq:T[2phi-ApAp]}, we compute the amplitude
at each order of the high energy expansion, which is given by
the sum of the three diagrams $(a)$-$(c)$. As shown in
Table\,\ref{tab:1}, we demonstrate explicitly
that the sum of diagrams $(a)$-$(c)$
vanishes at the $\mO(\bE^2)$ and $\mO(\bE^1)$:
\beqs
\label{eq:T[2phi-AA]-cancel}
\begin{align}
\TT_{\P\P}^{(2)}[(a)+(b)+(c)] \,&=\, 0\,,
\\[0.5mm]
\TT_{\P\P}^{(1)}[(a)+(b)+(c)] \,&=\, 0\,.
\end{align}
\eeqs
Furthermore, we compute both amplitudes
$\TT[\phi^-\phi^+\!\ito\! \APP\APP]$ and
$\TT[\phi^-\phi^+\!\ito\! \ATT\ATT]$
at the $\mO(\bE^0)$ and obtain:
\begin{equation}
\label{eq:E0-ApAp=AtAt}
\TT^{(0)}_{\P\P}[\phi^-\phi^+\!\ito \APP\APP]
\,=\,
\fr{1}{2}\TT^{(0)}_{\T\T}[\phi^-\phi^+\!\ito \ATT\ATT]
\,=\, e^2 \,.
\end{equation}
Without losing generality, we set the spin 
$\,\sp\!=\!\mt/m\!=\!+1$\, 
in the above calculations and afterwards.

\linespread{1.5}
\begin{table}[t]
\centering
\begin{tabular}{c||c|c||c|c|c}
\hline\hline
Amplitude
& $\times\bE^2$    & $\times\bE^1$  &  Amplitude
& $\times\bE^2$    & $\times\bE^1$
\\ \hline
$\TT_{\P\P}^{}[(a)]$
&  $-e^2(1\!+\! \ct)$
&  $~-\,\ii\hspace*{0.3mm}2e^2\st$
&  $\TT_{\phi\P}^{}[(d)]$
&  $-2e^2$
&   0
\\ \hline
$\TT_{\P\P}^{}[(b)]$
&  $-e^2(1\!-\! \ct)$
&  $ \ii\hspace*{0.3mm}2e^2\st$
&  $\TT_{\phi\P}^{}[(e)]$
&  $e^2(1\!+\!\ct)$
&  $\ii\hspace*{0.3mm}2e^2\st$
\\ \hline
$\TT_{\P\P}^{}[(c)]$
&  $2e^2~$
&  $0$
&  $\TT_{\phi\P}^{}[(f)]$
&  $e^2(1\!-\!\ct)$
&  $~~-\ii\hspace*{0.3mm}2e^2\st$
\\ \hline\hline
  Sum   & ~0      & 0
& Sum   &  0      & 0
\\ 	
\hline\hline
\end{tabular}
\caption{\small\baselineskip 15pt
Energy cancellations in the scattering amplitudes of 
3d topologically massive scalar QED,
$\,\TT[\phi^-\phi^+\!\ito\! \AP \AP]
\!=\!\TT_{\P\P}^{}[(a)]\!+\!\TT_{\P\P}^{}[(b)]
\!+\!\TT_{\P\P}^{}[(c)]$\,
and
$\,\TT[\phi^-\AP \ito \phi^-\AP]
\!=\!\TT_{\phi\P}^{}[(d)]\!+\!\TT_{\phi\P}^{}[(e)]
\!+\!\TT_{\phi\P}^{}[(f)]$\,,\,
where $\bE \!=\! E/m$\, and \,$(\st,\,\ct) \!=\!
(\sin\theta, \cos\theta)$ with $\theta$
being the scattering angle.
Each full amplitude equals the sum of individual diagrams
$(a)\!+\!(b)\!+\!(c)$ and $(c)\!+\!(d)\!+\!(e)$, respectively,
as shown in Fig.\,\ref{fig:1}.}
\label{tab:1}
\vspace*{2mm}
\end{table}

\vspace*{1mm}

Similarly, we compute the Compton scattering amplitudes
\,$\TT[\phi^-\!\!\APP\! \ito \phi^-\!\APP]$\, and
$\TT[\phi^-\!\ATT \ito \phi^-\!\ATT]$.
Then, we make the following high energy expansions for
both amplitudes:
\beqs
\label{eq:T[phi-A]}
\begin{align}
\label{eq:T[phi-Ap]}
\TT[\phi^-\!\APP \ito \phi^-\!\APP] &\,=\,
\TT^{(2)}_{\phi\P}\bE^2 + \TT^{(1)}_{\phi\P}\bE^1
+ \TT^{(0)}_{\phi\P}\bE^0 + \mO(\bE^{-1})\,,
\\[1mm]
\label{eq:T[phi-At]}
\TT[\phi^-\!\ATT \ito \phi^-\!\ATT] &\,=\,
\TT^{(0)}_{\phi\T}\bE^0 + \mO(\bE^{-1})\,.
\end{align}
\eeqs
As shown in
Table\,\ref{tab:1}, we demonstrate explicitly
that the sum of the three diagrams $(d)$-$(f)$ of Fig.\,\ref{fig:1}
vanishes at the $\mO(\bE^2)$ and $\mO(\bE^1)$:
\beqs
\label{eq:T[phi-Ap]-cancel}  
\begin{align}
\TT_{\phi\P}^{(2)}[(d)+(e)+(f)] \,&=\, 0\,,
\\[0.5mm]
\TT_{\phi\P}^{(1)}[(d)+(e)+(f)] \,&=\, 0\,.
\end{align}
\eeqs
Moreover, we find that both amplitudes
$\TT[\phi^-\!\APP \ito \phi^-\!\APP]$ and
$\TT[\phi^-\!\ATT \ito \phi^-\!\ATT]$
are nonzero and equal at the $\mO(\bE^0)$:
\begin{equation}
\label{eq:E0-phiAp=phiAt}
\TT^{(0)}_{\phi\P}[\phi^-\!\APP \ito \phi^-\!\APP]
\,=\,
\fr{1}{2}\TT^{(0)}_{\phi\T}[\phi^-\!\ATT \ito \phi^-\!\ATT]
\,=\, -\, e^2 \,.
\end{equation}

Finally, from
Eqs.\eqref{eq:T[2phi-AA]-cancel} and \eqref{eq:T[phi-Ap]-cancel}
together with
Eqs.\eqref{eq:E0-ApAp=AtAt} and \eqref{eq:E0-phiAp=phiAt},
we derive
\beqs
\label{eq:TET-SQED}
\begin{align}
\label{eq:TET-ApAp}
\TT[\phi^-\phi^+\!\ito \APP\APP]
&\,=\, \fr{1}{2}
\TT[\phi^-\phi^+\!\ito \ATT\ATT]+\mO\!\(\frac{m}{E}\)\!,
\\[1mm]
\label{eq:TET-phiAp}
\TT[\phi^-\!\APP \ito \phi^-\!\APP]
&\,=\, \fr{1}{2}
\TT[\phi^-\!\ATT \ito \phi^-\!\ATT] +\mO\!\(\frac{m}{E}\)\!,
\end{align}
\eeqs
which explicitly verify the TET \eqref{eq:TET}
for the topologically massive scalar QED.

\subsubsection{\hspace*{-3mm}Scattering Amplitudes of Topologically Massive Spinor QED}
\label{sec:4.1.2}

In this subsection, we consider the topologically massive QED
(TMQED) which includes the gauge sector Lagrangian \eqref{eq:L-MCS}
(with Chern-Simons term)
and the following matter Lagrangian,
\begin{equation}
\La_f^{} \,=\,  \bar{\psi}( \ga^\mu D_\mu\! -m_f^{})\psi \,, \quad
\end{equation}
where the covariant derivative is defined as
$\,D_\mu^{}\!= \pd_\mu^{} + \ii e A_\mu^{}\,$
and the gamma matrices are given by  $(\ga^0\!,\,\ga^1\!,\,\ga^2)=(\ii\si^2\!,\,\si^1\!,\,\si^3)$. 
We define the 3d Dirac spinors and solve the 3d Dirac equation 
in Appendix\,\ref{app:C}.

\vspace*{1mm}

\begin{figure}[t]
\centering
\includegraphics[height=2.5cm, width=14cm]{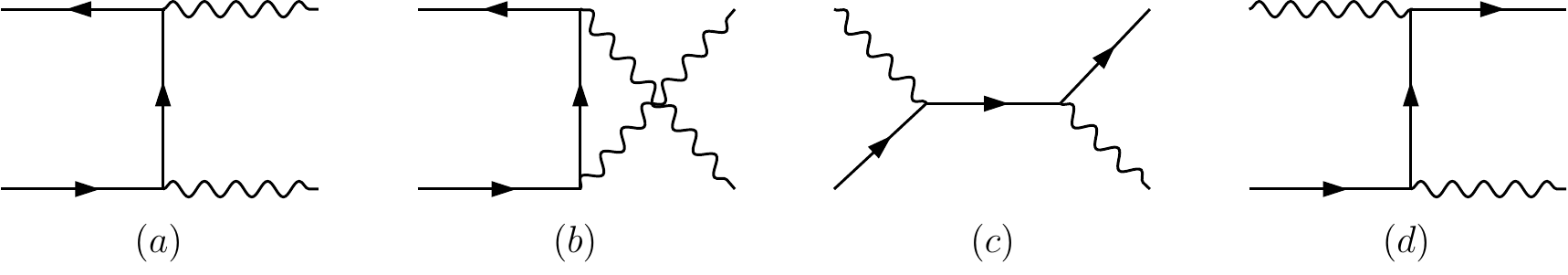}
\vspace*{-2mm}
\caption{\small\baselineskip 15pt
Scattering processes
$\,e^+e^-\!\!\ito\!\APP\APP$\,
($e^+e^- \!\!\ito\! \ATT\ATT$)
via Feynman diagrams $(a)$-$(b)$ and
$\,e^-\!\APP \ito e^-\!\APP$
($\hs e^-\!\ATT \ito e^-\!\ATT\hs$)
via Feynman diagrams $(c)$-$(d)$
in 3d topologically massive spinor QED.}
\label{fig:2}
\vspace*{2mm} 
\end{figure}

Then, we analyze the amplitudes of
the annihilation process
$\,e^+e^-\!\!\to\! \APP\APP$
($e^+e^-\!\!\to\! \ATT\ATT$)
and the Compton scattering
$\,e^-\!\APP\!\!\to\! e^-\!\APP$\,
($e^-\!\ATT\!\!\to\! e^-\!\ATT$).
The relevant Feynman diagrams at tree level are shown
in Fig.\,\ref{fig:2}.
Using the power counting rule \eqref{eq:DE}, we find that
the scattering amplitudes
$\,\TT[e^+e^-\!\!\to\! \APP\APP]$\, and
$\,\TT[e^-\!\APP\!\!\to\! e^-\!\APP]$\, have leading contributions
scale as $E^2$, while the scattering amplitudes
$\,\TT[e^+e^-\!\!\to\! \ATT\ATT]$\, and
$\,\TT[e^-\!\ATT\!\!\to\! e^-\!\ATT]$\, have leading contributions
scale as $E^0$.
Thus, we can make the following high energy expansions:
\beqs
\label{eq:T[ee-AA]}
\begin{align}
\label{eq:T[ee-ApAp]}
\TT[e^-e^+\!\ito \APP\APP] &~=\,
\TT^{(2)}_{\P\P}\bE^2 + \TT^{(1)}_{\P\P}\bE^1
+ \TT^{(0)}_{\P\P}\bE^0 + \mO(\bE^{-1})\,,
\\[1mm]
\label{eq:T[ee-AtAt]}
\TT[e^-e^+\!\ito \ATT\ATT] &~=\,
\TT^{(0)}_{\T\T}\bE^0 + \mO(\bE^{-1})\,,
\\[1mm]
\label{eq:T[eAp-eAp]}
\TT[e^-\!\APP\!\ito e^-\!\APP] &~=\,
\TT^{(2)}_{e\P}\bE^2 + \TT^{(1)}_{e\P}\bE^1
+ \TT^{(0)}_{e\P}\bE^0 + \mO(\bE^{-1})\,,
\\[1mm]
\label{eq:T[eAt-eAt]}
\TT[e^-\ATT\!\ito e^-\!\ATT] &~=\,
\TT^{(0)}_{e\T}\bE^0 + \mO(\bE^{-1})\,,
\end{align}
\eeqs
where $\,\bE\!=E/m\,$ and $E$ denotes the energy of the
incoming electron (positron). For simplicity, we set the
electron mass $\,m_e^{}\!\simeq 0\,$.

\vspace*{1mm}

Then, we explicitly compute the above scattering amplitudes.
We find that all the $\mO(E^2)$ and $\mO(E^1)$ terms cancel exactly
in each amplitude and the final results actually behave as
$\mO(E^0)$. We present these cancellations explicitly in
Table\,\ref{tab:2}. Hence, we have 
\beqs  
\label{eq:T[eeApAp-eAp-E2]-cancel}
\begin{alignat}{2}
\label{eq:T[ee-ApAp-E2]-cancel}
& \TT_{\P\P}^{(2)}[(a)+(b)] \,=\, 0\,,
\hspace*{10mm}
& \TT_{\P\P}^{(1)}[(a)+(b)] \,=\, 0\,;
\\[1mm]
\label{eq:T[eAp-E2]-cancel}
& \TT_{e\P}^{(2)}[(c)+(d)] \,=\, 0\,,
\hspace*{10mm}
& \TT_{e\P}^{(1)}[(c)+(d)] \,=\, 0\,.
\end{alignat}
\eeqs

Finally, we derive the remaining
amplitudes of $\mO(E^0)$ as follows:
\beqs
\label{eq:Amp-CSQED}
\begin{align}
\label{eq:Amp-ee-ApAp}
\TT_{\P\P}^{(0)}[e^-e^+ \!\ito \APP \APP ]
&\,=\, \fr{1}{2} \TT_{\T\T}^{(0)} [e^-e^+ \!\ito\ATT\ATT ]
\,=\, \ii\hspace*{0.3mm}e^2 \cot\hspace*{-0.3mm}\theta  \,,
\\[1mm]
\label{eq:Amp-eAp-eAp}
\TT_{e\P}^{(0)}[e^-\!\APP \ito e^-\!\APP ]
&\,=\, \fr{1}{2}\TT_{e\T}^{(0)}[e^-\!\ATT \ito e^-\!\ATT ]
\,=\,\ii\, e^2
\frac{\,(3 \!+\! \ct) (1 \!+\! \ct \!+\! \st)\,}
{\,4(1 \!+\! \ct)(1\!+\! \st)^{\hf}\,} \,.
\end{align}
\eeqs
For completeness, we also summarize in Appendix\,\ref{app:D}
the full tree-level amplitudes (without high energy expansion) 
for the scattering processes discussed above and 
in sections\,\ref{sec:4.1} and \ref{sec:4.2.1}. 
These exact formulas can provide self-consistency checks for the
corresponding expanded scattering amplitudes given in the main text 
and will also be useful for future studies.

\vspace*{1mm}

From the above Eqs.\eqref{eq:Amp-ee-ApAp}-\eqref{eq:Amp-eAp-eAp}, 
we deduce the following relations under the high energy expansion:
\beqs
\label{eq:Amp-CSQED-AT} 
\begin{align}
\label{eq:Amp-ee-ATAT}
\hspace*{-3mm}
\TT [e^-e^+ \!\ito \APP\APP ]
&\,=\, \fr{1}{2}\TT [e^-e^+ \!\ito \ATT\ATT]
+\mO\!\(\frac{m}{E} \) \!,
\\[1mm]
\label{eq:Amp-eAT-eAT}
\hspace*{-3mm}
\TT [e^-\!\APP \ito e^-\!\APP ]
&\,=\, \fr{1}{2}\TT [e^-\!\ATT \ito e^-\!\ATT]
+\mO\!\(\frac{m}{E} \) \!.
\end{align}
\eeqs
These verify explicitly that the TET
\eqref{eq:TET} does hold, as expected from our general formulation
of the TET in section\,\ref{sec:3.1}.
We observe that the TET identity \eqref{eq:TET-ID2}
[or the TET \eqref{eq:TET}]
provides a {\it general mechanism}
which guarantees the exact energy cancellations
of the $\mO(E^2)$ and $\mO(E^1)$ contributions in the
$\AP$-amplitude and matches the corresponding
$\AT$-amplitude of $\mO(E^0)$\,.
%

\linespread{1.5}
\begin{table}[t]
\centering
\begin{tabular}{c||c|c||c|c|c}
\hline\hline
Amplitude
& $\times \bE^2$
& $\times \bE^1$
& Amplitude
& $\times \bE^2$
& $\times \bE^1$
\\
\hline\hline
   $\TT_{\P\P}^{}[(a)]$
&  $\ii\hspace*{0.3mm}e^2\st$
&  $2e^2\ct $
&  $\TT_{e\P}^{}[(c)]$
&  $\frac{\,\ii 2 e^2 (1+\ct) (1+\st)^{1/2}\,}{1 +  \ct + \st}  $
&  $-\frac{\, 2 e^2 \st (1+\st)^{1/2}\,}{1 +  \ct + \st}$
\\ \hline
   $\TT_{\P\P}^{}[(b)]$
&  $-\ii\hspace*{0.3mm}e^2\st  $
&  $-2 e^2\ct $
&  $\TT_{e\P}^{}[(d)]$
&  $-\frac{\,\ii  2e^2 (1+\ct) (1+\st)^{1/2}\,}{1 +  \ct + \st}  $
&  $\frac{\,2 e^2 \st (1+\st)^{1/2}\,}{1 +  \ct + \st}$
\\ \hline\hline
Sum   & 0      & 0   & Sum & 0      & 0
\\
\hline\hline
\end{tabular}
\caption{\small\baselineskip 15pt
Energy cancellations in the scattering amplitudes of 3d 
topologically massive spinor QED,
$\,\TT[e^+e^-\!\!\to\! \APP\APP]\!=\!
\TT_{\P\P}^{}[(a)]\!+\!\TT_{\P\P}^{}[(b)]$\,  and
$\,\TT[e^-\!\APP\!\!\to\! e^-\!\APP]\!=\!
\TT_{e\P}^{}[(c)]\!+\!\TT_{e\P}^{}[(d)]$\,,\,
where the notations are defined as
$\,\bE \!=\! E/m$\, and
\,$(\st,\,\ct) \!=\!
(\sin\theta,\, \cos\theta)$\, with $\,\theta\,$
denoting the scattering angle.
Each full amplitude equals the sum of individual diagrams
$(a)\!+\!(b)$ and $(c)\!+\!(d)$, respectively,
as shown in Fig.\,\ref{fig:2}.}
\label{tab:2}
\end{table}

\vspace*{1mm}

Before concluding this subsection, 
we further present an exact verification
of the TET identity \eqref{eq:TET-ID} or \eqref{eq:TET-ID2a} without
taking the high energy limit and by considering the simplest case
of $N\!=\hsmx 1\hs$. For the scattering process 
$e^-e^+ \!\ito \APP\APP$, 
we apply the TET identity \eqref{eq:TET-ID2}
to just one external state of $\APP$:
\beqa
\label{eq:TET-N=1}
\TT[e^-e^+ \!\ito \APP \APP ] \,=\,
\TT[e^-e^+ \!\ito \tdATT \APP ]+
\TT[e^-e^+ \!\ito \tilde{v} \APP ]\,,
\eeqa 
where $\,\tdATT\!=\!\fr{1}{\sqrt{2\,}\,}\ATT\,$ and
$\,\vt^{}\!=\!\fr{1}{\sqrt{2\,}\,}v\!=
 \!\fr{1}{\sqrt{2\,}\,}v^\mu\!A_\mu^{}\hs$.
Using the basic relation of polarization vectors in Eq.\eqref{eq:epS},
we can rewrite the above TET identy \eqref{eq:TET-N=1} as follows:
\beqa
\label{eq:TET-N1-AS}
\TT[e^-e^+ \!\ito \ASS \APP ] \,=\, 0\,,
\eeqa 
where $\ASS \!=\epS^\mu A_\mu^{}$ is the unphysical 
scalar polarization state of the photon.
As we explained above Eq.\eqref{eq:epS},
the gauge-fixing function in momentum space can be expressed as
$\hs\FF\!=\!-\iii m\ASS\hs$.
Thus, the above TET identity \eqref{eq:TET-N1-AS} is equivalent to
\beqa
\label{eq:TET-N1-F}
\TT[e^-e^+ \!\ito \FF \APP ] \,=\, 0\,,
\eeqa 
which is just the simplest $N\!=\hsmx 1\hs$ case of
the Slavnov-Taylor-type identity \eqref{eq:F-ID}.
Hence, to verify the TET identity \eqref{eq:TET-ID2}
in the case of $N\!=\hsmx 1\hs$,
we only need to prove explicitly that 
the identity \eqref{eq:TET-N1-AS} holds at the tree level.

\vspace*{1mm}

The tree-level scattering process 
$\hs e^-e^+ \!\ito \ASS\APP\hs$
contains the same type of diagrams $(a)$-$(b)$
via $(t,u)$-channels, as shown in Fig.\,\ref{fig:2}.
Then, we compute directly the contributions of the
$(t,u)$-channels as follows:
\beqa
\TT_t\hs [e^-e^+ \!\hsm\ito\hsmx \ASS\APP] \,=\,
-\TT_u\hs [e^-e^+ \!\hsm\ito\hsmx\ASS\APP]=
\sqrt{2} e^2 \! \(\iii \bE^2 \st + \bE \ct\) \!,
\label{eq:TeeASAP}
\eeqa
which ensures that the full amplitude vanishes:  
\begin{equation}
\TT\hs [e^-e^+ \!\hsm\ito\hsmx \ASS\APP]  \,=\,
\TT_t + \TT_u \,=\, 0\,. 
\end{equation}
This explicitly verifies the TET identity \eqref{eq:TET-N1-AS}
[and thus the TET identity \eqref{eq:TET-N=1}] 
for the case of $N\!=\hsmx 1\hs$,
without taking the high energy limit.

\subsection{\hspace*{-3mm}Topologically Massive QCD and Scattering Amplitudes}
\label{sec:4.2}

In this subsection, we study four-point scattering amplitudes in the 
3d topologically massive QCD (TMQCD) with non-Abelian 
gauge group SU($N$). 
This is also called the topologically massive YM (TMYM) theory 
in Sec.\,\ref{sec:2}
for the pure gauge sector without matter fields. 
We will not discriminate these two terminologies hereafter.
The Lagrangian of the TMQCD can be written as follows:
\begin{equation}
\label{eq:LaCSYMCom}
\La =-\frac{1}{4} (F^{a}_{\mn})^{2} + \frac{m}{2} \vep^{\mnr} A^{a}_{\mu} \pd_{\nu} A^{a}_{\rho} 
+ \frac{\,g\hspace*{0.3mm}m\,}{6}C^{a b c} 
\vep^{\mnr}A^a_\mu A^b_\nu A^c_\rho
+\sum_{i,j=1}^{N}\bar{\psi}_i
(\gamma_\mu^{}{D}_{ij}^\mu-m_q\delta_{ij}) \psi_j   \,,
\end{equation}
where 
$D_{ij}^\mu = \delta_{ij}^{}\pd^\mu -\ii g A^{a\mu}T^a_{ij}$\, 
and $(i,j)$ denote the color indices of the quarks.
We will compute the scattering amplitudes 
of the quark-antiquark annihilation 
and the pure gauge boson scattering,
from which we uncover the nontrivial
energy cancellations.
Then, we will demonstrate explicitly
that the TET \eqref{eq:TET} holds for the non-Abelian TMQCD.

\subsubsection{\hspace*{-3mm}Scattering Amplitudes of 
Quark-Antiquark Annihilation}
\label{sec:4.2.1}

\begin{figure}[t]
\centering
\includegraphics[height=2.7cm, width=11.5cm]{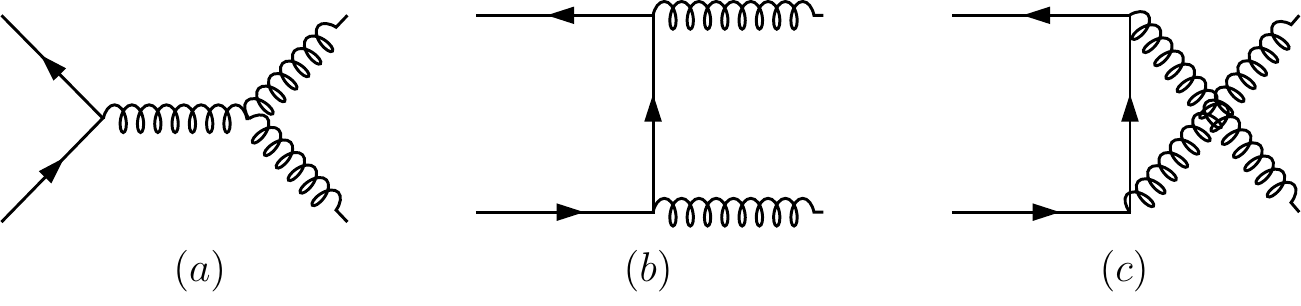}
\vspace*{-4mm}
\caption{\small\baselineskip 15pt
Scattering processes of 
quark-antiquark annihilation into two gluons, 
$q_i^{}\bar{q}_j^{}\!\ito\! \Ap^a\Ap^b$
($q_i^{}\bar{q}_j^{}\!\ito\! \At^a\At^b$)
in the 3d topologically massive QCD.}
\label{fig:3}
\end{figure}

In this subsection,
we analyze the scattering amplitudes of 
quark-antiquark annihilation processes
$\,q\bar{q}\ito \Ap^a\Ap^b$ and $\,q\bar{q}\ito \At^a\At^b$\,,
which include three Feynman diagrams as shown
in Fig.\,\ref{fig:3}. The non-Abelian cubic gluon vertex
generates the $s$-channel diagram of Fig.\,\ref{fig:3}$(a)$ 
which is absent in the $e^-e^+$ annihilation process 
of the TMQED as shown Fig.\,\ref{fig:2}$(a)$-$(b)$.

Applying the power counting rule \eqref{eq:DE}, we find that
the high-energy scattering amplitude 
$\TT[q_i\bar{q}_j\hsmx\ito\hsmx\Ap^a\Ap^b]$ 
has leading contributions scale as $E^2$,\,
while the amplitude 
$\TT[q_i^{}\bar{q}_j^{}\hsm\ito\hsm\At^a\At^b]$\,
scales as $E^0$.
Thus, we can make the following high energy expansions:
\beqs
\label{eq:T[qq-AA]}
\begin{align}
\label{eq:T[qq-ApAp]}
\TT[q_i^{}\bar{q}_j^{}\ito \Ap^a\Ap^b] &~=\,
\TT^{(2)}_{\P\P}\bE^2 + \TT^{(1)}_{\P\P}\bE^1
+ \TT^{(0)}_{\P\P}\bE^0 + \mO(\bE^{-1})\,,
\\[1mm]
\label{eq:T[qq-AtAt]}
\TT[q_i^{}\bar{q}_j^{}\ito \At^a\At^b] &~=\,
\TT^{(0)}_{\T\T}\bE^0 + \mO(\bE^{-1})\,,
\end{align}
\eeqs
where $\,\bE\!=E/m\,$ and $E$ denotes the energy of the
incoming quark (anti-quark). For simplicity, we set the
quark mass $\,m_q^{}\!\simeq 0\,$.
Then, we explicitly compute these scattering amplitudes,
and find that the summed contributions in each amplitude
cancel exactly at $\mO(E^2)$ and $\mO(E^1)$, respectively.  
The final net results could only behave as $\mO(E^0)$. 
We present these cancellations explicitly in
Table\,\ref{tab:3}. From these, we deduce 
\beqs
\label{eq:T[qq-ApAp-E]-cancel}
\begin{align}
\label{eq:T[qq-ApAp-E2]-cancel}
\TT_{\P\P}^{(2)}[(a)+(b)+(c)] \,&=\, 0\,,
\\[0.5mm]
\label{eq:T[qq-ApAp-E1]-cancel}
\TT_{\P\P}^{(1)}[(a)+(b)+(c)] \,&=\, 0\,,
\end{align}
\eeqs
where we have applied the commutation relation
$\,[T^a\hspace*{-0.3mm}, T^b]\!=\!\iii C^{abc}T^c$\,
to the sum of the diagrams $(b)\!+\!(c)$, 
which further cancels the contribution of the
diagram $(a)$ at $\mO(E^2)$ and $\mO(E^1)$ respectively.

\linespread{1.5}
\begin{table}[t]
\centering
\begin{tabular}{c||c|c|c|c}
\hline\hline
Amplitude
& $\TT_{\P\P}^{}[(a)]$
& $\TT_{\P\P}^{}[(b)]$
& $\TT_{\P\P}^{}[(c)]$
&Sum
\\
\hline\hline
   $\times \bE^2$
&  $ g^2\st \hs C^{abc}\hs T^c $
&  $ \iii g^2\st \hs T^a T^b$
&  $ -\iii g^2\st \hs T^b T^a$
& 0
\\ \hline
   $\times \bE^1$
&  $-\iii 2 g^2\ct \hs C^{abc}\hs T^c$
&  $2 g^2\ct \hs T^a T^b$
&  $ -2 g^2\ct \hs T^b T^a$
& 0
\\
\hline\hline
\end{tabular}
\caption{\small\baselineskip 15pt
Energy cancellations in the scattering amplitude 
$\TT [q_i^{}\bar{q}_j^{}\!\ito\! \Ap^a\Ap^b]=\!
\TT_{\P\P}^{}[(a)]\!+\TT_{\P\P}^{}[(b)]\!+\TT_{\P\P}^{}[(c)]\,$
of 3d topologically massive QCD, where the relevant
Feynman diagrams $(a)$-$(c)$ are shown in Fig.\,\ref{fig:3}.}
\label{tab:3}
\end{table}

\vspace*{1mm}

Next, we compute the remaining $q\bar{q}$ annihilation 
amplitudes at $\mO(E^0)$ and derive the following results: 
\begin{equation}
\label{eq:E0-qq-AA}
\hspace*{-8mm}
\TT^{(0)}_{\P\P}[q_i^{}\bar{q}_j^{}\!\ito\!\Ap^a\Ap^b] 
= \fr{1}{2}\TT^{(0)}_{\T\T}[q_i^{}\bar{q}_j^{}\!\ito\!\At^a\At^b] 
= \frac{~\ii\hspace*{0.3mm}g^2\,}{4} \!\!
\[\!\frac{\stt}{\,1\!+\!\ct\,} \! \(T^a_{jk} T^b_{ki} \)
+\frac{\stt}{\,1\!-\!\ct\,}\! \(T^b_{jk} T^a_{ki} \)\!\]\!.
\end{equation}
We may further define the color-singlet states 
of the SU($N$) gauge group: 
{\small
\beqa
\hspace*{-9mm}
|0\ra_{\!q}^{} =\hsm \frac{1}{\sqrt{N\,}\,} \sum_{j=1}^{N}\!
| q_j^{} \bar{q}_j^{} \ra  , 
\quad
|0\ra_{\!\APP}^{} \!\!=\hsm 
\frac{1}{\sqrt{2(N^2 \!-\!1)\,}\,}\!\!
\sum_{a=1}^{N^2-1} \!|\AP\AP\ra,
\quad
|0\ra_{\!\ATT}^{} \!\!=\hsm 
\frac{1}{\sqrt{2(N^2 \!-\!1)\,}\,}\!\!
\sum_{a=1}^{N^2-1} \!|\AT\AT\ra.
\eeqa
}
Then, we compute the $q\bar{q}$ annihilation amplitudes
of the color-singlet states at $\mO(E^0)$:
\begin{equation}
\label{eq:Amp00-qq-AA}
\TT_{\P\P}^{(0)}
[\hspace*{0.3mm}|0\ra_{\!q}^{}\!\ito\!|0\ra_{\!\APP}^{}]
=
\fr{1}{2} \TT_{\T\T}^{(0)} 
[\hspace*{0.3mm}|0\ra_{\!q}^{}\!\ito\!|0\ra_{\!\ATT}^{}]
=\, \ii\hspace*{0.3mm}g^2 f(N)
\cot\hspace*{-0.3mm}\theta  \,,
\end{equation}
where we have defined the function 
$\,f(N)\!=\!\fr{1}{\,2\sqrt{2\,}\,}\hsm
 \sqrt{\hsm (N^2\!-\!1)\hsm /N\,}\hs$.
We note that for the color-singlet initial and final states, 
the $s$-channel contribution vanishes due to 
$\,C^{aac}\!=0$\,,\, 
and the sum of the $(t,u)$-channel contributions just differs
from the Abelian case of TMQED by an overall factor
$\,(g^2/e^2)f(N)\,$. This relation holds even without making
the high energy expansion, namely,
\beqs
\label{eq:Amp-qqPP-qqTT}
\begin{align}
\TT_{\P\P}^{}
[\hspace*{0.3mm}|0\ra_{\!q}^{}\!\ito\!|0\ra_{\!\APP}^{}] 
\,=~&\,
\frac{\,g^2\,}{e^2} f(N) \,\TT[e^-e^+ \!\ito \APP \APP ]  \,,
\\[1mm]
\TT_{\T\T}^{}
[\hspace*{0.3mm}|0\ra_{\!q}^{}\!\ito\!|0\ra_{\!\ATT}^{}] 
\,=~&\,
\frac{\,g^2\,}{e^2} f(N) \,\TT[e^-e^+ \!\ito \ATT \ATT ]  \,.
\end{align}
\eeqs
After the high energy expansion, only the $\mO(E^0)$
amplitudes survive for the TMQCD and TMQED, as shown in
Eq.\eqref{eq:Amp00-qq-AA} and Eq.\eqref{eq:Amp-ee-ApAp}
which obey the above relations \eqref{eq:Amp-qqPP-qqTT}.

\vspace*{1mm}

Finally, from 
Eqs.\eqref{eq:T[qq-AA]}\eqref{eq:T[qq-ApAp-E]-cancel} and 
Eqs.\eqref{eq:E0-qq-AA}\eqref{eq:Amp00-qq-AA},
we derive the following equivalence relation under the
high energy expansion:
\begin{equation}
\label{eq:TET-qq-AA}
\hspace*{-8mm}
\TT [q_i^{}\bar{q}_j^{}\!\ito\!\Ap^a\Ap^b] 
= \fr{1}{2}\TT [q_i^{}\bar{q}_j^{}\!\ito\!\At^a\At^b] 
+\mO\!\(\frac{m}{E} \) \!,
\end{equation}
which explicitly realizes the TET \eqref{eq:TET}.

\subsubsection{\hspace*{-3mm}Pure Gauge Boson Scattering Amplitudes}
\label{sec:4.2.2}

\begin{figure}[t]
\centering
\includegraphics[height=2.5cm, width=11.5cm]{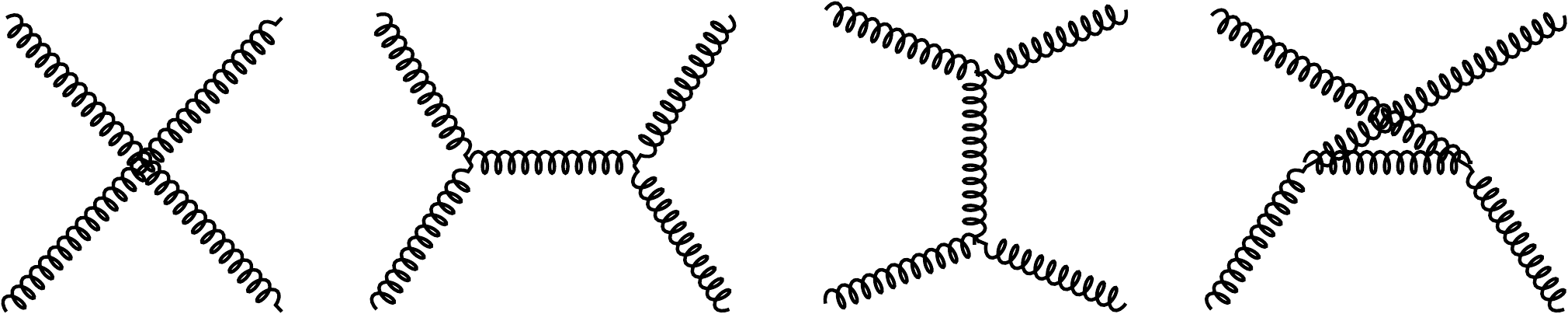}
\caption{\small\baselineskip 15pt
Feynman diagrams of the four-gauge boson scattering processes
$\Ap^a\Ap^b\!\ito\!\Ap^c\Ap^d$\, and  
$\,\At^a\At^b\ito\At^c\At^d$\,		
via the contact interaction and $(s,t,u)$-channels 
in 3d topologically massive YM theory.}
\label{fig:gg-gg}
\label{fig:4}
\vspace*{-1mm}
\end{figure}

In this subsection, we study the four-particle amplitudes
of the pure gauge boson scattering processes
$\Ap^a\Ap^b\!\ito\!\Ap^c\Ap^d$ and 
$\At^a\At^b\!\ito\!\At^c\At^d$
in the 3d non-Abelian topologically massive YM (TMYM) theory, 
where the gauge field $A_\mu^a$ belongs to 
the adjoint representation of the SU($N$) gauge group. 
The relevant Feynman diagrams are shown in Fig.\,\ref{fig:4}.

\vspace*{1mm} 

We see that
the four-gauge boson scattering amplitudes 
$\,\TT[\Ap^a\Ap^b\!\ito\!\Ap^c\Ap^d]\equiv\!\TT[4\AP]$\, and
$\,\TT[\At^a\At^b\!\ito\!\At^c\At^d]\equiv\!\TT[4\AT]$\, 
receive contributions from the contact diagram
and the pole diagrams via $(s,\,t,\,u)$-channels. 
The kinematics of such four-particle elastic scattering
processes is defined in Appendix\,\ref{app:A}.
Using the power counting rule \eqref{eq:DE},
we find that for the scattering amplitude $\,\TT[4\AP]\,$
the leading contributions of each diagram in Fig.\,\ref{fig:4}
scale as $E^4$, while for the scattering amplitude 
$\,\TT[4\AT]\,$ the individual leading contributions scale
as $E^0$. Hence, using the TET identity \eqref{eq:TET-ID2}
[or the TET \eqref{eq:TET}], we predict that the $\AP$-amplitude 
should contain exact energy cancellations
at the $\mO(E^4)$, $\mO(E^3)$, $\mO(E^2)$, and $\mO(E^1)$,
respectively. This is because the leading energy-dependence of 
the $\AP$-amplitude must match that of the corresponding  
$\AT$-amplitude of $\mO(E^0)$\, on the RHS of 
the TET identity \eqref{eq:TET-ID2} 
[or the TET \eqref{eq:TET}].

\vspace*{1mm}

Then, we compute explicitly the full scattering amplitude of 
$\AP$'s at tree level and present it in a compact form:
\begin{equation}
\label{eq:4Ap-amp}
\TT[4\AP] \,=\, g^2\! \(\! 
\frac{\,\CC_s \,\NN_s\,}{\,s\!-\!m^2\,} 
+ \frac{\,\CC_t \,\NN_t\,}{\,t\!-\!m^2\,} 
+ \frac{\,\CC_u \,\NN_u\,}{\,u\!-\!m^2\,}  \!\)\!,
\end{equation}
where the color factors are defined as usual 
$\(\CC_s,\, \CC_t,\, \CC_u\) = 
\(C^{abe}C^{cde},\, C^{ade}C^{bce},\, C^{ace}C^{dbe} \)$
with $C^{abc}$ denoting the structure constants of the
gauge group. The numerators 
$(\NN_s,\,\NN_t,\,\NN_u)$ take the following form:
\beqs
\label{eq:N-stu} 
\begin{align}
\label{eq:N-s}
\hspace*{-8mm}
\NN_s &= \frac{\,4m^2 \!-\! s\,}{\,16 m^3 s^{\frac{1}{2}}_{}\,}
\!\[\! 4 m s^{\frac{1}{2}} (5 m^2 \!+\! 4 s) \ct \!+\! \ii\hs
(4 m^4 \!+\! 29 m^2 s \!+\! 3 s^2 )\st \!\] \!,
\\[1mm]
\label{eq:N-t}
\hspace*{-8mm}
\NN_t &= -\frac{\cht }{\,16m^3\,}\!
\(s^{\frac{1}{2}} \hsm +\hsm \iii 2\hs m
\tan\!\fr{\theta}{2} \)^{\!\!2}\!\times
\nn\\
&  \hspace{9mm}
\LB\hsm 4 m  [  13 m^2\!-\!3 s \!+\! (8m^2 \!-\! s )\ct ]\cht
+\, \ii\hs s^{\frac{1}{2}}[22 m^2 \!-\!
3s\!+\!(20 m^2 \!-\! 3 s ) \ct] \sht \RB \!,
\\[1mm]
\label{eq:N-u}
\hspace*{-8mm}
\NN_u &= \frac{\sht}{\,16m^3\,}\!\(\!s^{\frac{1}{2}} 
\!-\!\ii\hs 2\hs m\cot\!\fr{\theta}{2} \!\)^{\!\!2}\!\times
\nn\\
& \hspace*{9mm}
\LB\hsm 4\hs m\hs  [  13\hs m^2\!-\!3 s \!-\! (8m^2 \!-\! s )\ct\hs ]\hs\sht
-\hs \ii \hs s^\hf  [ 22 m^2 \!-\! 3s  \!-\!  (20 m^2 \!-\! 3 s ) \ct] \cht \RB \!.
\end{align}
\eeqs
We note that in the (2+1)d spacetime
there is a kinematic exchange symmetry between the scattering 
amplitudes of $t$-channel and $u$-channel, namely, their numerators
obey the relation 
$\,\NN_u^{}(\pi\hsm +\theta) = -\NN_t^{}(\theta)\,$.
We have verified that our numerators \eqref{eq:N-t}-\eqref{eq:N-u}
indeed satisfy this kinematic exchange symmetry.

\vspace*{1mm}

We note that each term on the RHS of Eq.\eqref{eq:4Ap-amp}
scales as $E^3$ at most because summing up each contribution of
the contact diagram with the corresponding pole diagram already
cancels $\mO(E^4)$ terms, as we show in Table\,\ref{tab:4}.
For the high energy scattering with $\,E\!\gg\!m$,\,  
we expand the full amplitudes in terms of $1/\bar{s}_0^{}$\,, 
where $\,{s}_0^{}\!=\!4E^2\!\be^2\,$ and 
$\,\bar{s}_0^{}\!=\!4\bE^2\!\be^2\,$ with 
$\,\bE\!=\!E/m\,$.
Thus, we can explicitly demonstrate the exact energy cancellations
at each order of $E^n$ ($n=4,3,2,1$), respectively. 
We summarize our findings in Table\,\ref{tab:4}, from which
we prove the following exact energy cancellations: 
\beqs 
\label{eq:E-cancel}
\begin{align}
& \TT_{cj}^{(4)}\!+\!\TT_j^{(4)} =0\,, 
\label{eq:E4-cancel}
\\[2mm]
& \sum_j \! \(\TT_{cj}^{(3)}\!+\!\TT_j^{(3)}\)
= -\ii\hspace*{0.25mm}24\st\bsz^{\frac{3}{2}}
c_0^{}\,(\CC_s\!+\CC_t\!+\CC_u)
=0 \,,
\label{eq:E3-cancel}
\\ 
& \sum_j \! \( \TT_{cj}^{(2)}\!+\!\TT_j^{(2)}\)
= -128\ct\bsz\hspace*{0.3mm} 
c_0^{}\,(\CC_s\!+\CC_t\!+\CC_u) =0 \,,
\label{eq:E2-cancel}
\\
& \sum_j \! \(\TT_{cj}^{(1)}\!+\!\TT_j^{(1)}\)
= -\ii\hspace*{0.25mm}304\st\bsz^{\frac{1}{2}}
c_0^{}\,(\CC_s\!+\CC_t\!+\CC_u) =0 \,,
\label{eq:E1-cancel}
\\[-6.5mm]
\hspace*{-1mm}
&\nn 
\end{align}
\eeqs 
where $\,j\!\in\!(s,t,u)$,\, 
$c_0^{}\!=g^2/128$,\, 
and the superscript $(n)$ in
the amplitudes $(\TT_{cj}^{(n)},\,\TT_j^{(n)})$ denotes the 
contributions at the $\mO(E^n)$ 
with $\,n\!=\!1,2,3,4\,$.
The amplitudes $\,\TT_c^{}\,$ (contributed by the contact diagram) 
and $\,\TT_j^{}\,$ (contributed by 
the gauge-boson-exchange in each channel-$j$\,) 
are given by the sums:
\begin{equation}
\TT_c^{} \,=\, \sum_{j,n}\TT_{cj}^{(n)}\,,
\hspace*{10mm}
\TT_j^{} \,=\, \sum_{n}\TT_j^{(n)}\,,
\end{equation}
where $\,j\!\in\!(s,t,u)$\, and $\,n\!=\!4,3,2,\cdots$.
From Table\,\ref{tab:4} and Eq.\eqref{eq:E4-cancel},
the $\mO(E^4)$ contributions 
cancel exactly between the contact diagram and the pole diagrams 
in each channel of $\,j\!\in\!(s,t,u)$. 
Furthermore, it is striking to see that 
the sum of each $\mO(E^n)$ contributions ($n\!=\!3,2,1$)
also cancel exactly because of the Jacobi identity
$\CC_s\!+\CC_t+\CC_u=0$\,,\,
as shown in Table\,\ref{tab:4} and 
Eqs.\eqref{eq:E3-cancel}-\eqref{eq:E1-cancel}.

\linespread{1.5}
\begin{table}[t]
\centering
\begin{tabular}{c||c|c|c|c}
\hline\hline 
{\small Amplitude}
& $\times \bsz^2$
& $\times \bsz^{3/2}$
& $\times \bsz$
& $\times \bsz^{1/2}$
\\ 
\hline\hline
$\TT_{cs}$
& $8\st \,  \,\CC_s$ 
& $\ii\hspace*{0.25mm}32\st \,\CC_s$
& $64\ct\,  \CC_s$
& $\ii\hspace*{0.25mm} 64\st\,\CC_s$
\\ \hline
$\TT_{ct}$ 
& $ -(5 \!+\! 4 \ct \!-\! \ctt)\,\CC_t$ 
& $ -\ii\hspace*{0.25mm} 8(2\st\!-\!\stt) \,\CC_t$ 
& $ -32(\ct \!-\!\ctt)\, \CC_t $ 
& $ -\ii\hspace*{0.25mm} 16(2\st\!-\!5\stt)\,\CC_t$ 
\\ \hline
$\TT_{cu}$ 
& $ (5 \!-\! 4 \ct \!-\! \ctt) \,\CC_u$ 
& $ -\ii\hspace*{0.25mm}8(2\st\!+\!\stt)\,\CC_u$ 
& $ -32(\ct\!+\!\ctt)\,\CC_u$ 
&  $ -\ii\hspace*{0.25mm}16(2\st\!+\!5\stt)\,\CC_u$ 
\\ 
\hline \hline
$\TT_s$ 
& $ -8\st \,\,\CC_s$ 
& $ -\ii\hspace*{0.25mm}56 \st \,\CC_s$ 
& $ -192\ct\, \CC_s$ 
& $ -\ii\hspace*{0.25mm}368 \st\,\CC_s$ 
\\ 
\hline
$\TT_t$ 
&  $ (5 \!+\! 4 \ct \!-\! \ctt)\,\CC_t$ 
&$ -\ii\hspace*{0.25mm}8(\st\!+\!\stt)\,\CC_t$ 
& $ -32(3\ct\!+\!\ctt)\,\CC_t$ 
& $ -\ii\hspace*{0.25mm}16(17\st\!+\!5\stt)\,\CC_t$ 
\\ 
\hline
$\TT_u$ 
&  $ -(5 \!-\! 4 \ct \!-\! \ctt) \,\CC_u$ 
& $ -\ii\hspace*{0.25mm}8(\st\!-\!\stt)\,\CC_u$ 
& $ -32(3\ct\!-\!\ctt)\,\CC_u$ 
&  $ -\ii\hspace*{0.25mm}16(17\st\!-\!5\stt)\,\CC_u $
\\ 
\hline\hline
Sum 
& 0   
& 0                                      
& 0                
& 0                                    
\\
\hline\hline
\end{tabular}
\caption{\small\baselineskip 15pt
Energy cancellations in the four-gauge boson scattering amplitude 
of 3d non-Abelian TMYM theory,  
$\TT[\Ap^a\Ap^b\!\ito\!\Ap^c\Ap^d]
\!=\!\TT_c^{}\!+\!\TT_s\!+\!\TT_t\!+\!\TT_u$\,,
where the amplitude from contact diagram is 
further decomposed into three sub-amplitudes according to their 
color factors, $\TT_c=\TT_{cs}+\TT_{ct}+\TT_{cu}\hs$.\ 
The energy factor is defined as 
$\,\bar{s}_0^{}\!=\!4\bE^2\!\be^2\,$ and $\,\bE\!=\!E/m\,$.
A common overall factor $\,({g^2}\!/{128})\,$ in each amplitude is
not displayed for simplicity.}
\label{tab:4}
\vspace*{2mm}
\end{table}

\vspace*{1mm}

After all these energy cancellations, 
we systematically derive the 
remaining scattering amplitude at $\mO(E^0)$.
We also compute the amplitude $\,\TT[4\tdAT]$\, which contains
terms no more than $\mO(E^0)$ 
by the direct power counting. 
Thus, we present both scattering amplitudes
expanded to $\mO(E^0)$ as follows:
\beqs
\label{eq:4AP-4AT-E0-TMQCD}
\begin{align} 
\label{eq:4AP-E0-TMQCD}
\TT_0[4\AP] \,&=\, g^2 \!\left[  
\CC_s \!\(\frac{-9\ct\,}{4}\) \!+
\CC_t\! \(\frac{-1 \!-\! 9\ct \!-\! 4\ctt\,}{4(1 \!+\! \ct)}\) \!+
\CC_u\!\(\frac{\,1 \!-\! 9\ct \!+\! 4\ctt\,}{4(1 \!-\! \ct)} \)
\right]  \!,
\\[1.5mm]
\label{eq:4AT-E0-TMQCD}
\TT_0[4\tdAT] \,&=\, g^2 \!\left[  
\CC_s \! \(\frac{-\ct\,}{4}\) \!+  
\CC_t \! \(\frac{\,3 \!-\! \ct\,}{\,4(1\!+\! \ct)\,}\) \!+
\CC_u \! \(\frac{\,-3\!-\!\ct\,}{\,4(1\!-\!\ct)\,} \)
\right]  \!,
\end{align}
\eeqs 
where we have denoted 
$\,\tdA_{\T}^a\!=\!\fr{1}{\sqrt{2\,}\,}\AT\,$
as before. 
Comparing the two amplitudes above, 
we find that they differ by an amount: 
\begin{equation}
\TT_0[4\AP] - \TT_0[4\tdAT] \,=\,
-2g^2\ct (\CC_s\!+\CC_t\!+\CC_u) \,=\, 0\,,
\label{eq:TET-4AP-4AT} 
\end{equation}
which vanishes identically because of the Jacobi identity 
$\,\CC_s\!+\CC_t\!+\CC_u\!=0\,$.
This demonstrates explicitly the TET for
the four-gauge boson scattering in the non-Abelian TMYM theory:
\beqa
\label{eq:TET-4Ap-TMQCD} 
\TT[\Ap^a\Ap^b\!\ito\!\Ap^c\Ap^d] \,=\, 
\TT[\At^a\At^b\!\ito\!\At^c\At^d] 
+\mO\!\(\frac{m}{E} \) \!,
\eeqa 
which confirms the general TET \eqref{eq:TET}.

\vspace*{1mm} 

Before concluding this subsection, 
we stress that the present study has well understood and 
justified the structure of our gauge boson scattering amplitude
\eqref{eq:4Ap-amp} order by order 
under the high energy expansion, 
including the exact energy cancellations
at each $\mO(E^n)$ with $n=4,3,2,1$ and the proof of 
the TET relation \eqref{eq:TET-4AP-4AT} at $\mO(E^0)$.

\subsection{\hspace*{-3mm}Unitarity Bounds on TMYM and TMG Theories}
\label{sec:4.3}

In this subsection, we analyze the partial wave amplitudes of
the 3d topologically massive gauge boson scattering and
the 3d topologically massive graviton scattering 
(Sec.\,\ref{sec:5}).  
We will demonstrate that {\it the partial wave amplitudes for 
either the topologically massive gauge boson scattering 
or the topologically massive graviton scattering have
high energy behaviors no larger than $\mO(E^0)$, so they  
can respect the perturbative unitarity bound.} 

\vspace*{1mm}

For an SU($N$) gauge theory, we define the gauge-singlet 
one-particle state:
\begin{equation}
|0\ra_{\!\APP}^{} = 
\frac{1}{\sqrt{2(N^2 \!-\!1)\,}\,} 
\sum_{a=1}^{N^2-1} | \AP \AP \ra \,.
\end{equation}
Thus, we compute the scattering amplitude for the gauge-singlet 
state as follows: 
\begin{equation}
\TT[|0\ra_{\!\APP}^{} \!\!\!\ito |0\ra_{\!\APP}^{}] 
\,=\, \frac{\,g^2\hsm N\hs}{2}\!
\(\!\frac{\NN_t^\pp}{\,t\!-\!m^2\,} - 
    \frac{\NN_u^\pp}{\,u\!-\!m^2\,}\!\) \!.
\end{equation}

In $d$-dimensions, the partial wave amplitude takes the
following general form\,\cite{Soldate:1986mk}: 
\begin{equation}
\label{eq:PW-1}
a_{\ell}^{}(s)\,=\,
\frac{s^{{d}/{2}-2}}{~C_{\ell}^{\nu}(1)\lam_d^{}~}\! 
\int_{0}^{\pi} \!\! \td \theta  \[\! (\sin \theta)^{d-3}\, C_{\ell}^{\nu}(\cos \theta) \, \TT_{\rm{el}} \]  \!,
\end{equation}
where $\,C_{\ell}^{\nu}(x)\,$ is the Gegenbauer polynomial and
\begin{equation}
\nu=\fr{1}{2}(d-3) \,, \qquad
\lam_d^{}=2 \Ga\!\(\!\fr{1}{2}d\!-\!1\)\!(16 \pi)^{d/2-1} .
\end{equation}
The partial wave $a_{\ell}^{}$ should satisfy the 
unitarity condition 
$\Im(a_{\ell}^{})\!\geqq\! |a_{\ell}^{}|^2$,\, 
leading to\,\cite{Soldate:1986mk}\cite{He2005}:
\beqa
\label{eq:UCd}
|a_\ell^{}| \leqq 1 \hs ,
\hspace*{5mm}
|\Re (a_\ell^{})| \leqq \fr{1}{2} \hs , 
\hspace*{5mm}
|\Im (a_\ell^{})|\leqq 1\hs .
\eeqa 
For the present study, we have $\,d\!=\!3\hs$.\  
Thus, we can compute the real part of 
the $s$-wave amplitude \eqref{eq:PW-1} as follows:
\begin{align}
\Re (a_0^{}) &=\, 
\frac{1}{\,8\pi \sqrt{s\,}\,}\!
\int_{0}^{\pi} \!\!\!\hspace*{0.5mm}\td\theta \,
\hspace*{0.5mm}
\Re (\TT[|0\ra_{\!\APP}^{} \!\!\!\ito |0\ra_{\!\APP}^{}])
\nn\\[1mm]
&= 
-\frac{~g^2N (16m^4\!+\!24 m^2s\!+\!s^2)\,}
{32\sqrt{s\,}\hs (s\!-\!4m^2)^2\,}
\simeq - \frac{Ng^2}{\,32\sqrt{s\,}\,} \,.
\label{eq:REa0}
\end{align}
The imaginary part $\Im(a_0^{})$ has 
collinear divergences around $\,\theta\!=\!0,\pi$\,
of the integral.
After adding an angular cut on the scattering angle 
$(\,\delta\!\leqq\!\theta\!\leqq\!\pi\!-\!\delta\,)$ 
to remove the collinear divergences of the integral,
we find that $\Im(a_0^{})$ vanishes.\ 
Eq.\eqref{eq:REa0} shows that for high energy scattering 
the leading partial wave amplitude
$\,\Re(a_0^{})\hs$ scales as $E^{-1}$,\, which has good high energy 
behaviors. This is expected, because the 3d TMYM theory
is gauge-invariant and has a super-renormalizable gauge 
coupling $\,g\,$ with mass-dimension $+\hf$\,.
Applying the unitarity condition 
\eqref{eq:UCd} to $s$-wave amplitude \eqref{eq:REa0}, 
we derive the following perturbative unitarity bound: 
\begin{equation}
\sqrt{s\,} \, \geqq \, \frac{\,g^2\hsm N\,}{\,16\,} \,,
\end{equation}
which puts a lower limit on the scattering energy,
in addition to the requirement of high energy expansion
$\sqrt{s\,}\!\gg\! m$\,.

\vspace*{1mm}

Next,  we study the perturbative unitarity bound for the
TMG theory in Sec.\ref{sec:5}. Using the high-energy
graviton scattering amplitude
in Eqs.\eqref{eq:Amp-4hp-DC}-\eqref{eq:Amp0-4hp-DC}, 
we compute the partial wave amplitudes of its real and
imaginary parts as follows:  
\beqs 
\label{eq:TMG-a0}  
\begin{align}
\label{eq:TMG-Re(a0)}
\Re (a_0^{}) &\,\simeq\, 
-\frac{15 \ka^2 m^2}{\,2048\hs\pi\sqrt{s\,}\,} 
\frac{\,(3\cos\delta \!-\! \cos 3\delta)\,}{\sin^3 \delta}
\simeq\, 
-\frac{15 \ka^2 m^2}{\,1024\pi \hs\delta^3 \sqrt{s\,}\,} \,,
\\[1mm]
\label{eq:TMG-Im(a0)}
\mathfrak{Im}(a_0) &\,\simeq\, 
-\frac{\,247 \ka^2m\,}
{\,49152\hs\pi\,} 
\frac{\,(3\cos\delta \!-\! \cos 3\delta)\,}{\sin^3 \delta}
\,\simeq\, -\frac{247\ka^2 m}{\,24576\hs \pi\hs\delta^3\,}  
\,,
\end{align}
\eeqs 
where we put an angular cut on the scattering angle 
$(\hs\delta\hsm\leqq\hsm\theta\hsm\leqq\hsm\pi\hsm-\hsm\delta\hs )$ 
to remove the collinear divergences of the integral.
We see that 
both $\Re (a_0^{})$ and $\Im (a_0^{})$
exhibit good high-energy behaviors since they 
scale as $E^{-1}$ and $E^0$, respectively.
Imposing the perturbative unitarity condition on the
$s$-wave amplitude \eqref{eq:TMG-a0}, we derive
the unitarity bounds on the real and imaginary parts:
\begin{align} 
\sqrt{s} \geqq \frac{\,15\ka^2 m^2\,}{~1024\pi\hs\delta^3~}\hsx ,
\hspace*{10mm}
m \leqq \frac{~49152 \pi\delta^3~}{247\ka^2}\hsx ,
\end{align} 
where the first condition places a lower bound on the scattering energy
proportional to $\,\ka^2m^2\,$.
The second condition puts an upper bound on the graviton mass
$\hs m\hs$,\, proportional to
$\,1/\ka^2\,$ which characterizes the ultraviolet cutoff
scale of the TMG gravity since the 3d gravitational coupling
$\,\ka^2\!=\! 16\pi G \!=\! 2/\MP$ 
has a negative mass-dimension $-1\hs$, 
where $G$ and $\MP$ denote the 3d Newton constant and Planck mass
respectively.

\section{\hspace*{-3mm}Structure of Topological Graviton Amplitude from Double-Copy}
\label{sec:5}


In this section, we study the extended double-copy construction
of the massive graviton amplitude in
the 3d topologically massive gravity (TMG)
from the massive gauge boson amplitude in the 3d TMYM theory.
{\it Our focus is to analyze the structure of massive graviton 
scattering amplitudes under high energy expansion 
and newly uncover a series of striking energy-cancellations
of the graviton amplitudes in connection to the corresponding 
gauge boson amplitudes through 
the extended massive double-copy construction.} 
In section\,\ref{sec:5.1}, starting from the 3d action of the TMG 
we will analyze the equation of motion (EOM) of the 3d graviton
field and identify the physical polarization state of the graviton.
Then, in section\,\ref{sec:5.2}
we will extend the conventional double-copy method for
massless gauge/gravity theories\,\cite{BCJ}\cite{BCJ-Rev} 
to the 3d topologically massive gauge/gravity theories. 
For this we will improve the original massive four-gauge-boson 
scattering amplitude \eqref{eq:4Ap-amp}-\eqref{eq:N-stu} 
by proper choice of the gauge transformation on its kinematic numerators. With these we can reconstruct the correct four-graviton 
scattering amplitude in the TMG theory.
We stress that a key feature of the 3d TMYM and TMG theories
is that these theories can realize the topological mass-generation 
of gauge bosons and gravitons {\it in a fully gauge-invariant way}
under the path integral formulation,
which is important for the successful double-copy construction
in the 3d massive gauge/gravity theories.


\subsection{\hspace*{-3mm}3d Topologically Massive Gravity}
\label{sec:5.1}

In this subsection, we first introduce the 3d action of the TMG. 
Then, we analyze the equation of motion of the 3d graviton
field and identify the physical polarization state of the graviton.
We note that
even though the 3d massless Einstein gravity has no physical
content\,\cite{Deser:1981wh}\cite{Witten-3dTMG}\cite{Hinterbichler-Rev},
including the gravitational Chern-Simons term 
can make the TMG fully nontrivial.
The TMG action contains the conventional Einstein-Hilbert term and  
the gravitational Chern-Simons term\,\cite{Deser:1981wh}:
\begin{equation}
\label{eq:S-TMG}
\hspace*{-5mm} 
 S_{\rm{TMG}} \,=\, -\frac{2}{\,\ka^2\,} \!
\int \!\! \td^3 x \!\[\! \sqrt{-g} R
- \frac{1}{\,2\mt\,} \vep^{\mnr} \tGa{^\al_\rho _\be} \! 
\(\!\pd_{\mu} \tGa{^\be_\al_\nu} \!+\! 
\frac{2}{3} \tGa{^\be_\mu_\ga} \tGa{^\ga_\nu_ \al} \!\) \!\hsmx\] \!,
\end{equation}
where the 3d gravitational coupling constant $\ka$ is connected to
the Planck mass $\MP$ via 
$\,\ka \!=\!2/\!\sqrt{\MP}\,$
with $\,\MP\!=\!1/(8\pi G)\,$
and $\hs G\hs$ as the Newton constant.
The parameter $\,\mt\,$ in Eq.\eqref{eq:S-TMG} 
will provide the graviton mass 
$\,m=|\mt|$\,,\, 
as shown in Eq.\eqref{eq:Propa-TMG}.
Under the weak field expansion 
$\,g_{\mn}^{}\!= \eta_{\mn}^{} +\hs  
\ka\hspace*{0.3mm}h_{\mn}^{}$\,
and the linearized diffeomorphism transformation 
$\,h_{\mn} \!\to h_{\mn}'\!= h_{\mn}^{} + \pd_\mu^{}\xi_\nu^{} 
 +  \pd_\nu^{}\xi_\mu^{}\hsx$,
the change of the gravitational Chern-Simons term 
in Eq.\eqref{eq:S-TMG} gives a total derivative,
so the action is diffeomorphism invariant 
under the path integral formulation.

\vspace*{1mm}

The nonlinear EOM is derived as follows\,\cite{Hinterbichler-Rev}:
\begin{equation}
\label{eq:TMGEOM-1}
R_{\mn}-\frac{1}{2} g_{\mn}R +\frac{1}{m} C_{\mn}=0 \,,
\end{equation}
where the Cotton tensor 
$\,C_{\mn}^{}\hsm =\hsm \tensor{\vep}{_\mu^\rho^\si}\nabla_{\rho}
( R_{\si\nu}^{}-\frac{1}{4} g_{\si \nu}^{} R )$\, 
is symmetric and traceless.  
In Eq.\eqref{eq:TMGEOM-1} and the discussions hereafter,
we use the positive branch of the mass parameter 
$\,\tilde{m}\!>\!0\,$, which corresponds to the graviton with
spin $\hsx\sp \hsmx =\hsmx +2\hsx$ \cite{Deser:1981wh}.
Then, we can expand the metric tensor around the Minkowski metric 
$\,g_{\mn}^{}\hsm =\hsm \eta_{\mn}^{} \hsm +\hsm \ka h_{\mn}^{}$\, 
and impose the transverse-traceless condition 
for the graviton field. 
With these, we obtain the linearized EOM from \eqrefe{eq:TMGEOM-1}:
\begin{equation}
\label{eq:EOM-hmunu}
\[\! \eta_{\mu\al}\eta_{\nu\be} + \frac{1}{\,2m\,}
(\vep_{\mu\rho\al}^{}\eta_{\nu\be}^{} + 
\vep_{\nu\rho\be}^{}\eta_{\mu\al}^{})
\pd^{\rho}\!\]\! \pd^2 h^{\ab}_{\rm{P}} \,=\, 0 \,.
\end{equation}
We may denote the operator in the square brackets 
of Eq.\eqref{eq:EOM-hmunu} as
\begin{equation}
\widehat{O}_{\mn\ab} \,=\, 
\eta_{\mu\al}\eta_{\nu\be} + \frac{1}{\,2m\,}
(\vep_{\mu\rho\al}^{}\eta_{\nu\be}^{} + 
\vep_{\nu\rho\be}^{}\eta_{\mu\al}^{})
\pd^{\rho}\,.
\end{equation}
Then, we act the operator 
$\,\widehat{O}_{\mn\ab}^{}\,$ 
twice on the graviton field and 
impose the transverse-traceless condition 
on the physical graviton state, which leads to 
\begin{equation}
( \pd^2 - m^2 ) \pd^2 \hP^{\mn} = 0 \,.
\end{equation}
This shows that the graviton field 
obeys a Klein-Gordon-like equation 
and carries the physical mass $m$\,.

Alternatively, we can ``square'' the EOM of the TMYM theory 
\eqref{eq:PolEOM} and arrive at
\begin{align}
&\( \eta_{\mu \al} \eta_{\nu \be}
-\frac{\,\iii\vep_{\mu\rho\al}^{}\eta_{\nu\be}^{}p^{\rho}\,}{m}
-\frac{\,\iii\vep_{\nu\si\be}^{}\eta_{\mu\al}^{}p^{\si}\,}{m} -\frac{\,\vep_{\mu\rho\al}^{}\vep_{\nu\si\be}^{}
p^{\rho}p^{\si}\,}{m^2}\) \!\ep^{\al} \ep^{\be}
\nn\\
\label{eq:TMGEOM-2}
& =\, 
2\!\left[ \eta_{\mu \al} \eta_{\nu \be}
-\frac{\ii}{\,2m}
(\vep_{\mu\rho\al}^{}\eta_{\nu\be}^{}+
 \vep_{\nu\rho\be}^{}\eta_{\mu\al}^{})\hspace*{0.23mm}p^{\rho} 
\right]\! \ep^{\ab}  =\,0 \,,
\end{align}
where in the second row we have used the relations:
\begin{align}
& \ep_{\mn} \!= \ep_\mu \ep_\nu \,, \quad 
p^{\mu} \ep_{\mu}\!=0 \,, \quad 
\tensor{\ep}{_\mu^\mu}\!=0  \,,
\nn\\
&\vep_{\mu\rho\al}^{}\vep_{\nu\si\be}^{}
\hspace*{0.3mm}
p^{\rho}p^{\si}=\,
(\eta_{\mu\be}^{}\eta_{\al\nu}^{}
\!-\hsm \eta_{\mn}^{}\eta_{\ab}^{})\hs p^2\!
+\eta_{\ab}^{}\hs p_{\mu}^{} p_{\nu}^{}\! 
+\eta_{\mn}^{}\hs p_{\al}^{} p_{\be}^{}\! 
-\eta_{\nu\al}^{}\hs
p_{\mu}^{}p_{\be}^{}\!
-\eta_{\mu\be}^{}\hs
p_{\nu}^{}p_{\al}^{} \,,~~~~
\end{align}
with the momentum $p^\mu$ obeying the on-shell condition
$\,p^2 \!=\! -m^2$\,.\
Thus, we see that \eqrefe{eq:TMGEOM-2} coincides with  
\eqrefe{eq:EOM-hmunu} where the graviton field is expressed 
in the plane wave form 
$\,h^{\mn}_{\rm{P}} = \ep^{\mn}_{\rm{P}} 
 e^{-\iii p \cdot x}$\,,
with $\,\ep^{\mn}_{\P}\!=\ep^\mu_{\P}\ep^\nu_{\P}\,$.
The graviton polarization tensor $\,\ep^{\mn}_{\P}$\,
is the solution of the EOM \eqref{eq:TMGEOM-2},
where the subscript ``$_{\text{\P}}$''
indicates that it corresponds to the physical polarization
state of the graviton $\hP^{\mn}$.
Then, we impose the following gauge-fixing term:
\begin{equation}
\label{eq:LGF-TMG}
\La_{\rm{GF}}^{} \,= \fr{1}{\xi}(\FF_\mu^{})^2 \,, 
\qquad~~ 
\FF^\mu =\, \pd_\nu h^{\mn} \!-\! \fr{1}{2}\pd^\mu h\,.
\end{equation}
With the above and making the weak field expansion, 
we can derive the graviton propagator, which we present 
in Eq.\eqref{eq:Propa-TMG} of Appendix\,\ref{app:E}.


\subsection{\hspace*{-3mm}Double-Copy Construction of Graviton Amplitude and Energy Cancellations}
\label{sec:5.2}

In this subsection, we extend the conventional double-copy method for
massless gauge/gravity theories\,\cite{BCJ}\cite{BCJ-Rev} 
and construct the massive four-graviton scattering amplitude
in the 3d TMG theory from the four-gauge boson amplitude in
the 3d TMYM theory. Our focus is to analyze the structure of  
massive graviton scattering amplitudes under high energy expansion 
and newly uncover a series of striking energy-cancellations
of the graviton amplitudes in connection to the corresponding 
gauge boson amplitudes.

\vspace*{1mm}

We examine the kinematic numerators $\{\NN_j\}$
of the original massive four-gauge-boson 
scattering amplitude \eqref{eq:4Ap-amp}-\eqref{eq:N-stu}, 
and find that they violate the kinematic Jacobi identity due to
$\,\sum_j \NN_j^{}\!\neq 0\,$.\  
Then, we analyze the gauge boson amplitude 
\eqref{eq:4Ap-amp} and find that it is invariant 
under the following generalized gauge transformation:
\begin{equation}
\label{eq:GaugeTrans-N'}
\NN_j^{} \,\to \, \NN_j' = \NN_j^{} + 
\Delta\hs (s_j^{} \!-\! m^2) \,, 
\end{equation}
where $\,j\!\in\!(s,t,u)\hs$ and $\hs\Delta\hs$ is a coefficient.
Thus, by requiring the gauge-transformed numerators $\{\NN_j'\}$
to satisfy the kinematic Jacobi identity 
$\,\sum_j\NN_j'\!=\hsmx 0\,$,
we can determine the coefficient $\Delta\,$ as follows:
\begin{equation}
\Delta = 
\frac{\,-\ii\csc\hspace*{-0.25mm}\theta~}
{~32 m^3~}\!
\left[(16 m^4s^{-\frac{1}{2}}_{} \!+\! 
8 m^2 s^{\frac{1}{2}}_{} \!-\! 3 s^{\frac{3}{2}}_{}) 
-(16 m^4s^{-\frac{1}{2}}_{}  \!-\!24 m^2 s^{\frac{1}{2}}_{} 
-\! 3 s^{\frac{3}{2}}_{})
\ctt \!+\! \ii\hspace*{0.25mm} 16ms\hspace*{0.25mm}\stt 
\right] \!.
\end{equation}
Then, we apply the gauge transformation \eqref{eq:GaugeTrans-N'}
to the numerators \eqref{eq:N-stu}
and further derive the following new kinematic numerators
$(\NN_s',\,\NN_t',\,$ $\NN_u')\hs$:
\beqs
\label{eq:N'-stu}
\begin{align}
\label{eq:N'-s}
\NN_s'  =&\ \frac{\,\ii \csc\hsm\theta\,}{\,8m s^{\hf}\,}\!
\[\!8 m^4 \!+\! 26 m^2 s \!-\! 7 s^2 \!-\!
(8 m^4 \!+\! 18 m^2 s \!+\! s^2)\ctt \!-\!
\iii m s^{\hf} (20 m^2 \!+\! 7 s) \stt\!\]\!,
\\[1mm]
\label{eq:N'-t}
\NN_t' =&\, -\!
\frac{\ii \csc\hsm\theta}
{\,32m s^{\hf}\,}\!
\[\!  16 m^4 \!+\! 52 m^2 s \!-\! 14 s^2
\!+\! \(16 m^4 \!+\! 104 m^2 s \!-\! 15 s^2\) \!\ct
\!-\! 2\! \(8 m^4 \!+\! 18 m^2 s \!+\! s^2\) \!\ctt
\right.\nn\\
& \hspace*{0.8mm} 
-\! \(16 m^4   \!+\! 24 m^2 s   \!+\! s^2\)\! \cttt
+\! \iii ms^{\hf}(176 m^2 \!-\! 20 s) \st
\!-\! \iii ms^{\hf}(40 m^2 \!+\! 14 s) \stt
\nn\\
& \hspace*{0.8mm} 
\left.
- \iii ms^{\hf}(32 m^2 \!+\! 8 s)\sttt \!\]\!,
\\[1mm]
\label{eq:N'-u}
\NN_u^\pp =&\, -\! \frac{\ii \csc\theta}{\,32m s^{\hf}\,}\!\!
\[\!16 m^4 \!+\! 52 m^2 s \!-\! 14 s^2
\!-\! \(16 m^4 \!+\! 104 m^2 s \!-\! 15 s^2\) \!\ct
\!-\! 2\! \(8 m^4 \!+\! 18 m^2 s \!+\! s^2\) \!\ctt
\right.\nn\\
& \hspace*{1.4mm} 
\!+\! \(16 m^4   \!+\! 24 m^2 s   \!+\! s^2\)\! \cttt
\!-\!\iii ms^{\hf}(176 m^2 \!-\! 20 s) \st
\!-\! \iii ms^{\hf}(40 m^2 \!+\! 14 s) \stt
\nn\\
& \hspace*{0.6mm} 
\left. +\, \iii ms^{\hf}(32 m^2 \!+\! 8 s)\sttt \!\]\!,
\end{align}
\eeqs
which nicely obey the kinematic Jacobi identity 
$\,\sum_j\NN_j'\!=\hsmx 0\,$ 
by construction.
With the above, we can reexpress the gauge boson amplitude
\eqref{eq:4Ap-amp} as follows: 
\begin{equation}
\label{eq:4Ap-amp-N'}
\TT[4\AP] \,=\, g^2\! \(\! 
\frac{\,\CC_s \,\NN_s'\,}{\,s\!-\!m^2\,} 
+ \frac{\,\CC_t \,\NN_t'\,}{\,t\!-\!m^2\,} 
+ \frac{\,\CC_u \,\NN_u'\,}{\,u\!-\!m^2\,}  \!\)\!.
\end{equation}
As a consistency check, we have also verified that the gauge-transformed
numerators \eqref{eq:N'-t}-\eqref{eq:N'-u} satisfy the
kinematic exchange symmetry between the $t$-channel and $u$-channel, 
namely, 
$\,\NN_u'(\pi\hsm +\theta) = -\NN_t'(\theta)\,$.

\vspace*{1mm} 

For the four-particle scattering amplitudes of massive physical gravitons
$\,\hP\!=\!\epP^{\mn}h_{\mn}^{}$ in the 3d TMG theory, 
we can use the power counting rule
\eqref{eq:DE-TMG} or \eqref{eq:DE0-TMG} of 
section\,\ref{sec:3.2}
to count the energy-dependence
of the graviton scattering amplitudes and find that 
the leading individual contributions 
to the tree-level amplitudes scale as $E^{12}$. 
But, using the extended double-copy approach
for massive scattering amplitudes,
we will prove that such leading contributions
of $\mO(E^{12})$ in
the four-graviton scattering amplitudes 
must cancel down to $\mO(E^1)\hs$, which gives the
striking large cancellations of  $\mO(E^{12})\ito \mO(E^1)$
by an energy power of $E^{11}$. 

\vspace*{1mm}

For this purpose, we extend the conventional double-copy method 
with the color-kinematics duality for 
massless gauge/gravity theories\,\cite{BCJ}\cite{BCJ-Rev} 
to the 3d topologically massive gauge/gravity theories. 
Using our improved massive-gauge-boson amplitude   
\eqref{eq:4Ap-amp-N'}
of 3d TMYM theory and the color-kinematics duality
$\,\CC_j^{}\!\ito\NN_j'\,$,\, we construct the full 
scattering amplitude of physical gravitons, 
$\M[\hP^{}\hP\!\ito\hP\hP]\!\equiv \M[4\hP]$,\, 
in the 3d TMG theory: 
\begin{align}
\label{eq:Amp-4hp-DC}
\M[4\hP] \,=\, \frac{\,\ka^2\,}{16}\! \[\! \frac{\,(\NN_s^\pp)^2\,}{s \!-\! m^2}
\!+\! \frac{\,(\NN_t^\pp)^2\,}{t \!-\! m^2}
\!+\! \frac{\,(\NN_u^\pp)^2\,}{u \!-\! m^2}  \!\] \!,
\end{align}
where we have made the gauge-gravity coupling
conversion $\,g \ito\kappa/4\,$.
(The same conversion factor will work for the 4d double-copy
of the massless gauge/gravity theories\,\cite{BCJ}
if the same normalization of color factors is adopted.) 
Then, substituting the improved kinematic numerators 
\eqref{eq:N'-stu} into Eq.\eqref{eq:Amp-4hp-DC}, we derive
the following compact formula for the four-graviton 
scattering amplitude after significant simplifications:
\begin{equation}
\label{eq:Amp-4hp-DC2} 
\M[4\hP] \,=\,
-\frac{~\ka^2 m^2
(P_0^{}\!+\!P_2^{} \ctt \!+\! P_4^{} c_{4\theta}^{} \!+\! 
 P_6^{} c_{6\theta}^{} \!+\! \bar{P}_2^{} \stt \!+\! \bar{P}_4^{} s_{4\theta}^{} \!+\! \bar{P}_6^{} s_{6\theta}^{})\hsm\csc^2\!\theta~}
{4096\hs (3 \!+\! \bsz) (4 \!+\! \bsz)^{3/2} (2 \!+\! \bsz \!-\! 
 \bsz \ct) (2 \!+\! \bsz \!+\! \bsz \ct)} \,,
\end{equation}
where $(P_j,\, \bar{P}_j)$ are expressed as 
polynomial functions of the variable $\bsz\!=s_0^{}/m^2$,
\begin{align}
\label{eq:Amp-4hp-DC2-Pj}
P_0^{} &=\,-4\hs (7992 \bsz^2 + 4767 \bsz^3 + 692 \bsz^4) 
(4\hsm +\hsm \bsz)^\hf \,,
\nn\\[1mm]
P_2^{} &=\, (-221184 - 304128 \bsz - 114048 \bsz^2 - 10928 \bsz^3 + 505 \bsz^4) (4\hsm +\hsm\bsz)^{\hf} ,
\nn\\[1mm]
P_4 &=\, 4\hs (55296 + 45312 \bsz + 13208 \bsz^2 + 1563 \bsz^3 + 58 \bsz^4) (4\hsm +\hsm\bsz)^{\hf} ,
\nn\\[1mm]
P_6 &=\, -(98304 + 57344 \bsz + 11264 \bsz^2 + 832 \bsz^3 + 17 \bsz^4)(4\hsm +\hsm\bsz)^{\hf} ,
\\[1mm]
\bar{P}_2 &=\, \iii (-442368 - 663552 \bsz - 300672 \bsz^2 
- 46048 \bsz^3 + 540 \bsz^4 + 475 \bsz^5) \hs,  \hspace*{10mm}
\nn\\[1mm]
\bar{P}_4 &=\, \iii 4\hs (110592 + 104448 \bsz + 36880 \bsz^2 + 5828 \bsz^3 + 372 \bsz^4 + 5 \bsz^5) \hs ,
\nn\\[1mm]
\bar{P}_6 &=\,
-\iii (196608 + 139264 \bsz + 35328 \bsz^2 + 3776 \bsz^3 
+ 148 \bsz^4 + \bsz^5) \hs .
\nn
\end{align}

Next, we make high energy expansion 
for the reconstructed graviton amplitude 
\eqref{eq:Amp-4hp-DC2}-\eqref{eq:Amp-4hp-DC2-Pj}
and derive the following leading order (LO) result: 
\begin{equation}
\label{eq:Amp0-4hp-DC}
\hspace*{-4mm}	
\M_0^{}[4\hP]  \,=\, 
- \frac{\,\iii\ka^2\,}{\,2048\,}\hspace*{0.5mm} 
m\hspace*{0.3mm} s_0^{\frac{1}{2}} 
(494 \ct\!+\!19 c_{3\theta}^{} \!-\! c_{5\theta}^{}) 
\csc^3 \hspace*{-0.7mm}\theta \,.
\end{equation}
It is striking to see that the above LO graviton amplitude 
actually scales as $\,\mO(m\hspace*{0.25mm}E)\,$ 
under high energy expansion. 
Because the direct application of 
our power counting rule \eqref{eq:DE0-TMG} to the double-copy 
graviton amplitude \eqref{eq:Amp-4hp-DC} gives the 
scaling behavior $\,\M_0^{}[4\hP]\!=\!\mO(m^{-2}E^4)\hs$,\,
we can expect from Eq.\eqref{eq:Amp0-4hp-DC} 
the further nontrivial energy cancellations of 
$\,E^4\ito E^1\hs$,$\hs$ which we will analyze in the following 
paragraph.  

\vspace*{1mm}

Inspecting the expressions of $\,(\NN_j^{},\,\NN_j')$\, 
in Eqs.\eqref{eq:N-stu} and \eqref{eq:N'-stu}, 
we find that they contain individual leading terms scaling
as $\,(E^5,\,E^3)$,\, respectively. This shows that the
gauge transformation \eqref{eq:GaugeTrans-N'} leads to 
an energy cancellation of $\,E^5\ito E^3\,$
in each of the transformed numerators $\NN_j'$\,.
This has an important impact on the energy dependence
of the double-copy amplitude \eqref{eq:Amp-4hp-DC},
namely,  in each channel of 
$\,\NN_j^{\hs\prime\,2}/(s_j^{}-m^2)\,$,
it contains a leading energy term behaving as $E^4$,\, 
rather than $E^8$ from  $\,\NN_j^{\,2}/(s_j^{}\!-m^2)\,$.\,
In comparison with the leading energy-dependence of 
the individual diagrams contributing to the tree-level 
four-graviton amplitude
which scales as $E^{12}$ by the direct power counting rule
\eqref{eq:DE0-TMG}, 
our double-copy construction \eqref{eq:Amp-4hp-DC}
demonstrates that the four-graviton amplitude could have 
a leading energy dependence of $E^4$ at most in each channel. 
Hence, this double-copy construction
guarantees a nontrivial large energy cancellation in the original 
four-graviton scattering amplitude:\ $E^{12}\ito E^4$,\,
which cancels the leading energy dependence by a large
power factor of $\,E^{8}$\,.

{\small
\linespread{1.5}
\begin{table}[t]
\centering
\begin{tabular}{c||c|c|c}
\hline\hline
{\small$\!\!$Amplitude$\!\!$}
& $\times \bsz^2$ 
& $\times \bsz^{3/2}$              
& $\times \bsz$ 
\\
\hline\hline
$\M_{s}$\!\!\!\!
& \!\!$-\frac{\,99+28 \ctt +\ctf\,}{1-\ctt}$ \!\!\!
& {\footnotesize $-\iii 14(15 \ct  \!+\! \cttt)\! \csc\theta $}\!\!\!\!
& $ -\frac{\,2(75 + 326 \ctt +47 \ctf)\,}{1-\ctt} $\!\!\!\!
\\ \hline
$\M_{t}$  \!\!\!\!
& $\frac{\,99+28 \ctt +\ctf\,}{4(1-\ct)} $ \!\!\!\!
& {\footnotesize $\iii (102 \!+\! 105 \ct \!+\! 70 \ctt \!+\! 
7 \cttt \!+\! 4 \ctf)\! \csc\theta$} \!\!\!\!
& $\frac{\,75 - 107 \ct + 326 \ctt +268 \cttt + 47 \ctf 
  + 31 \ctfv}{1-\ctt} $ \!\!\!
\\ \hline
$\M_{u}$ \!\!\!\!
& $\frac{\,99+28 \ctt +\ctf\,}{4(1 + \ct)} $ \!\!\!\!
& \!{\footnotesize $\iii (-102 \!+\! 105 \ct \!-\! 70 \ctt 
  \!+\! 7 \cttt \!-\! 4 \ctf)\! \csc\theta$} \!\!\!
& \!\!$\frac{\,75 + 107 \ct + 326 \ctt -268 \cttt + 47 \ctf 
  - 31 \ctfv\,}{1-\ctt}$ \!\!\!
\\ 	\hline\hline
$\rm{Sum}$
& 0 
& 0 
& 0 
\\
\hline\hline
\end{tabular}
\caption{\small\baselineskip 15pt
Exact energy cancellations at each order of $E^4$, $E^3$, and $E^2$
in our four-graviton scattering amplitude 
\eqref{eq:Amp-4hp-DC} by double-copy construction. 
Here the notations $(\sz,\,\bsz)$ are defined by 
$\bsz\!=\!\sz/m^2$ and
$\,\sz \!=\! 4E^2\be^2$.	
A common overall factor $\,({\ka^2m^2}\!/{1024})\,$ 
in each amplitude is not displayed for simplicity.}
\label{tab:5}
\end{table}
}

\vspace*{1mm}

Furthermore, it is striking to see that {\it 
our summed full amplitude 
\eqref{eq:Amp-4hp-DC} actually scales as $E^1$
under high energy expansion} as shown
in the above Eq.\eqref{eq:Amp0-4hp-DC}.
We can demonstrate explicitly this large energy cancellation of
$\,E^{4}\!\ito E^1\hs$, 
which includes exact cancellations of
the energy power terms at each order of $E^4$, $E^3$, and $E^2$. 
We summarize our findings on these exact energy cancellations
of the full amplitude \eqref{eq:Amp-4hp-DC} 
into Table\,\ref{tab:5}.
We may understand the reason for 
such energy cancellations of 
$\,E^{4}\hsm\ito\hsm E^1\,$ as follows. 
We first note that an $S$-matrix element $\,\mathbb{S}\,$ with
$\,\EE\,$ external states and $\hs L\hs$ loops ($L\!\geqq\! 0$) 
in the (2+1)d spacetime has mass-dimension 
$\,D_\Sb^{} \hsmx =\hsm 3\hs -\fr{1}{2}\EE\,$
as given by Eq.\eqref{eq:DS}.\
Thus, the four-graviton amplitude $\M[4\hP]$ 
has mass-dimension $\,D_{\!\M}^{}\!=\!1\hs$. 
At tree level it contains
the gravitational coupling $\,\ka^2\,$ of mass-dimension $-1\hs$.
Hence, we can express the four-graviton amplitude
$\,\M[4\hP] \!=\hsm \ka^2\bM[4\hP]$, 
where $\,\bM[4\hP]\,$ 
has mass-dimension equal $+2$\,.
The tree-level amplitude $\,\bM[4\hP]\,$
contains only two parameters $(E,\,m)$, 
each of which has mass-dimension $+1\hs$.
With these we can deduce the scaling behavior  
$\,\bM[4\hP]\!\propto\! m^{n_1} E^{n_2}\,$
with $\,n_1^{}\!+n_2^{}\!=\!2\,$, under high energy expansion.
Hence, for the energy terms of $E^{n_2^{}}$
with $n_2^{}=4,3,2$, we deduce the mass-power factor
$\,n_1^{}\!=\!-2,-1,0\,$, respectively.
This means that in the massless limit $\hs m\hsm\ito 0\,$, 
the physical graviton amplitude $\,\bM[4\hP]$\,
would go to infinity (for $n_2^{}\!\geqq\! 3$) or 
remain constant (for $n_2^{}\!=2$\,).
But,  in the $\,m\ito 0\,$ limit,
the 3d graviton field becomes unphysical and 
has no physical degree of 
freedom\,\cite{Deser:1981wh}\cite{Witten-3dTMG}\cite{Hinterbichler-Rev};
so the amplitude $\hs\bM[4\hP]\,$
should vanish because the physical graviton $\hP$ no longer exists
in the massless limit.
This means that the $\,m^{n_1} E^{n_2}\,$ terms with
$\,n_1^{}\!=\!-2,-1,0\,$
should vanish, and the physical amplitude $\,\bM[4\hP]$
has to start with the leading behavior of 
$\,m^1E^1\,$ under high energy expansion.
Thus, it is expected that the energy cancellations should hold  
at each order of $E^4$, $E^3$, and $E^2$,
in agreement with what we firstly uncovered by explicit calculations 
in Table\,\ref{tab:5}.

\vspace*{1mm} 

In summary, using our double-copied graviton amplitudes
in Eqs.\eqref{eq:Amp-4hp-DC}-\eqref{eq:Amp0-4hp-DC}
and Table\,\ref{tab:5} we have uncovered 
{\it a new type of large energy cancellations}
in the original four-graviton scattering amplitude at tree level 
for the 3d TMG theory:
\begin{equation}
\label{eq:E12-E1-TMG}
\mO(E^{12}) \,\longrightarrow\, \mO(E^1)\,,~~~~~~
(\hs\text{for}~\EE_{\hP}^{}\!\!=4\hsx
 \text{~in~3d~TMG}\hs )\hs. 
\end{equation}
Furthermore, with this extended double-copy construction, 
we have established  
{\it a new correspondence between the two types of 
leading energy cancellations in the massive scattering amplitudes:
$E^4 \ito E^0\hs$ in the TMYM theory and
$\,E^{12}\!\to\hsm E^1$ in the TMG theory.}
We also note that in Eq.\eqref{eq:E12-E1-TMG} 
the exact energy cancellations in the four-graviton amplitude down 
by a large power of $\,E^{11}\,$ are even much more severe 
than the energy cancellations
($E^{10}\!\ito E^2$) in the massive four-longitudinal KK graviton
scattering amplitudes of the compactified 5d KK Einstein gravity
found before by explicit calculations\,\cite{sekhar}\cite{kurt}
and by the double-copy construction\,\cite{Hang:2021fmp}.

\vspace*{1mm}

In passing, during the finalization of the present paper in this summer, 
we became aware of a 
recent new paper\,\cite{Gonzalez:2021bes} which directly
calculated the graviton amplitude of the 3d TMG with very lengthy
expressions in its Eq.(C.1) where all the polarization tensors (vectors)
take symbolic forms. 
We have quantitatively compared our 
full graviton amplitude \eqref{eq:Amp-4hp-DC} 
(by double-copy construction) with their Eq.(C.1)
and find agreement after substituting explicitly all the 
polarization formulas into Eq.(C.1) and making
substantial simplifications of Eq.(C.1).\ 
This comparison gives an independent verification 
of our double-copy result.\ 
Our work has fully different focus from \cite{Gonzalez:2021bes}
and our analyses differ from \cite{Gonzalez:2021bes} 
in several essential ways.\ 
(i).\,The main part of our work
(Secs.\,\ref{sec:2}-\ref{sec:4}) is {\it to analyze the
mechanism of topological mass-generation of gauge bosons
and uncover nontrivial energy cancellations in the gauge boson
scattering amplitudes.}
These were not studied by \cite{Gonzalez:2021bes}.\ 
(ii).\,For this purpose, {\it we newly formulated the 
3d topological mass-generation mechanism 
at $S$-matrix level, and newly proposed and proved 
the TET for $N$-point gauge boson amplitudes} in Sec.\,\ref{sec:3}.\ 
We further verified the TET explicitly by computing the 
four-point scattering amplitudes 
of various high-energy processes for both Abelian and non-Abelian
gauge theories (with and without matter fields) 
in Secs.\,\ref{sec:4.1}-\ref{sec:4.2}. 
These were not considered by \cite{Gonzalez:2021bes}.\ 
(iii).\,Our whole study on the gauge boson amplitudes 
and the double-copied graviton amplitudes {\it has focused on 
the structure of the scattering amplitudes under high energy expansion}
and {\it on the mechanism of nontrivial large energy cancellations} 
as summarized in Tables\,\ref{tab:1}-\ref{tab:5} 
and Eq.\eqref{eq:E12-E1-TMG}.\ 
For this, we newly constructed the general 3d power counting method
in Sec.\,\ref{sec:3.2}, and used it together with the TET to prove the nontrivial energy cancellations for $N$-point 
gauge boson amplitudes in Sec.\,\ref{sec:3.3}.
These were not covered by \cite{Gonzalez:2021bes}.\ 
(iv).\,We also note that Eq.(C.2) of \cite{Gonzalez:2021bes} further 
gave more compact formula of the 4-graviton amplitude in a very 
different Briet coordinate system and cannot be directly compared to
our double-copied graviton amplitude 
\eqref{eq:Amp-4hp-DC2}-\eqref{eq:Amp-4hp-DC2-Pj}.
Their 4-gauge boson amplitude in Eq.(C.3) was also given in the
Briet coordinate system and cannot be directly compared to
our Eq.\eqref{eq:N'-stu}-\eqref{eq:4Ap-amp-N'}.\ 
The Eqs.(4.12)-(4.13) of \cite{Gonzalez:2021bes} 
gave 4-gauge boson amplitude with all polarization
vectors in symbolic forms.\  
We have further confirmed with the authors of \cite{Gonzalez:2021bes}
that our gauge boson amplitude \eqref{eq:4Ap-amp} and their Eq.(4.13) 
are in good agreement after substituting all the polarization formulas
into their Eq.(4.13) 
and after taking into account the notational difference
in defining the Mandelstam variables.$\!$\footnote{%
\baselineskip 14pt
After posting this paper to arXiv:2110.05399, we had helpful
discussions with the authors of Ref.\,\cite{Gonzalez:2021bes}.
We are glad to thank them for the comparison between their Eq.(4.13) and 
our Eq.\eqref{eq:N-stu} which confirms the good agreement between 
the independent analyses on both sides.}\,
We stress that the parts of our study for the four-gauge boson amplitudes 
and double-copied four-graviton amplitudes have focused on analyzing their
structures of energy-dependence and on uncovering the
striking large energy cancellations in these amplitudes as well as
the mechanism of such energy cancellations.  
These new findings were not covered by \cite{Gonzalez:2021bes}
whose independent study had fully different focus.
We also note that 
the structures of our non-Abelian gauge boson amplitudes 
\eqref{eq:4Ap-amp} and \eqref{eq:4Ap-amp-N'} 
are well understood and justified by nontrivial 
self-consistency checks as we explained 
at the end of Sec.\,\ref{sec:4.2.2} 
and showed in Tables\,\ref{tab:4}-\ref{tab:5}.

\section{\hspace*{-3mm}Conclusions}
\label{sec:6}

Understanding the mechanism of topological mass-generation and 
the structure of the scattering amplitudes in the 3d
Chern-Simons (CS) gauge and gravity theories is important  
for applying the modern quantum field theories to particle physics 
and condensed matter physics\,\cite{Deser:1981wh}\cite{Dunne:1998}\cite{Tong:2016kpv}.
In 3d spacetime the existence of the CS actions for gauge bosons 
and gravitons is theoretically unavoidable 
and compelling. This generates gauge-invariant topological mass-terms 
for gauge bosons and gravitons without invoking the conventional 
Higgs mechanism\,\cite{Higgs} and leads to good high energy behaviors
for the scattering amplitudes of topologically massive gauge bosons
and gravitons.

\vspace*{1mm} 

In this work, we systematically studied the mechanism of the topological 
mass-generations in 3d CS gauge theories and newly formulated it
at the $S$-matrix level. For this, we proposed and proved a new
Topological Equivalence Theorem (TET) for understanding the structure
of the scattering amplitudes of physical gauge bosons ($\AP$)  
in the topologically massive gauge theories.
We newly uncovered the nontrivial large energy cancellations
in the $N$-point gauge boson scattering amplitudes for both
the Abelian and non-Abelian CS gauge theories.
We further used an extended double-copy approach to analyze the
structure of the graviton scattering amplitudes in the 3d
topologically massive gravity (TMG) theory, with which we 
constructed the massive physical graviton scattering amplitudes
from that of the corresponding massive physical gauge bosons
in the topologically massive Yang-Mills (TMYM) theory.
From these, we newly uncovered 
a series of striking large energy cancellations 
in the four-point physical graviton scattering amplitudes, 
which ensure such massive scattering amplitudes 
to have good high energy behaviors 
and obey the perturbative unitarity bounds.\ 
We summarize these new findings more explicitly as follows.

\vspace*{1mm} 

In section\,\ref{sec:2}, we analyzed the dynamics of 
topological mass-generation in the 3d CS gauge theories.
In such dynamics, 
including the CS term \eqref{eq:L-MCS-YMCS} automatically converts 
the gauge boson's transverse polarization state $\AT$ (combined
with its longitudinal polarization state $\AL$) 
into the massive physical polarization state 
$\hs\AP \!\propto \! ( \AT +\hsm \AL)\hs$ 
as given in Eq.\eqref{eq:Ap},
while making its orthogonal combination 
$\hs\AX \!\propto\! ( \AT\hsm -\hsmx\AL)\hsx$ in Eq.\eqref{eq:Ax}
become an unphysical state.
This topological mass-generation mechanism has essential difference
from the conventional Higgs mechanism\,\cite{Higgs}, because the CS
term generates gauge-invariant mass term of $A_\mu^a\hs$ and 
no spontaneous gauge symmetry breaking and Higgs boson are invoked. 

\vspace*{1mm} 

In section\,\ref{sec:3}, 
we newly proposed and proved a TET to formulate the
topological mass-generation of gauge bosons at the $S$-matrix level,
which quantitatively connects the $N$-point 
scattering amplitudes of physical gauge bosons 
$\AP$ to the amplitudes of the corresponding 
transverse gauge bosons $\AT$
in the high energy limit. For this, we established the general TET 
identity \eqref{eq:TET-ID2}, with which we derived the TET 
\eqref{eq:TET} under high energy expansion. 
We presented a new energy power counting rule \eqref{eq:DE} 
to count the leading energy dependence of general $N$-point 
scattering amplitudes for the 3d topologically massive gauge theories
and another new power counting rule \eqref{eq:DE-TMG} 
[and \eqref{eq:DE0-TMG}] to count the 
leading energy dependence of $N$-point scattering amplitudes
for the 3d TMG theory.
(A generalized power counting method in $d$-dimensions is
given in Appendix\,\ref{app:B}.)
With these, we demonstrated that
our TET identity \eqref{eq:TET-ID2} provides a general mechanism
of nontrivial energy cancellations 
in the $N$-point $\AP$-amplitudes because
the net energy dependence of a given  
$\AP$-amplitude must match
that of the leading $\AT$-amplitude on the RHS of 
Eq.\eqref{eq:TET-ID2a}. For the high-energy 
scattering of $N$-gauge bosons $\AP$ 
(with $N\!\hsm\geqq\! 4\hsx$), 
the TET identity \eqref{eq:TET-ID2}
[or TET \eqref{eq:TET}] guarantees the nontrivial 
large energy cancellations in the $\AP$-amplitude:
$E^4 \hsm\to\hsm E^{4-N}\hs$.

\vspace*{1mm}

In section\,\ref{sec:4}, we explicitly demonstrated 
the TET \eqref{eq:TET} for the first time by using 
various high-energy four-particle scattering amplitudes  
in both the Abelian and non-Abelian topologically massive
CS gauge theories.
In sections\,\ref{sec:4.1.1}-\ref{sec:4.1.2}, 
we computed the scattering amplitudes of the annihilation processes
$\hsx \phi^-\phi^+\!\ito \APP\APP$
$(\phi^-\phi^+\!\!\ito\!\ATT\ATT)\hsx$ 
and Compton scattering 
$\hsx\phi^-\APP\!\ito\phi^-\APP$
$(\hs\phi^-\ATT\!\ito\phi^-\ATT\hs)$ 
in the topologically massive scalar QED (TMSQED), 
as shown in Fig.\,\ref{fig:1}.
In parallel, we computed the scattering amplitudes of 
annihilation processes 
$\hsx e^-e^+\!\ito \APP\APP$ $(\hs e^-e^+\!\ito\ATT\ATT\hs )$ 
and Compton scattering 
$\hsx e^-\APP\!\ito\hsx e^-\APP$
$(\hsx e^-\ATT\hsm\ito e^-\ATT)$ 
in the topologically massive spinor 
QED (TMQED),\, as shown in Fig.\,\ref{fig:2}.  
From these analyses, 
we newly uncovered the nontrivial energy cancellations
of $\hsx E^2\!\ito E^0\hsx$ in each $\APP$-amplitude,  
which are summarized in Tables\,\ref{tab:1}-\ref{tab:2} and in 
Eqs.\eqref{eq:T[2phi-AA]-cancel}\eqref{eq:T[phi-Ap]-cancel}
and Eqs.\eqref{eq:T[ee-ApAp-E2]-cancel}-\eqref{eq:T[eAp-E2]-cancel}.   
We further computed the remaining nonzero scattering amplitudes
of $\mO(E^0)$ and proved explicitly the validity of the TET as 
in Eqs.\eqref{eq:TET-SQED} and \eqref{eq:Amp-CSQED-AT}. 

\vspace*{1mm}

Next, in section\,\ref{sec:4.2}  we studied 
the structure of scattering amplitudes
in the non-Abelian topologically massive QCD (TMQCD).
In section\,\ref{sec:4.2.1}, we computed the
quark-antiquark annihilation processes 
$\hsx q_i^{}\bar{q}_j^{}\!\ito\!\Ap^a\Ap^b$
$(q_i^{}\bar{q}_j^{}\!\ito\!\At^a\At^b)$ 
for the TMQCD, which are shown in Fig.\,\ref{fig:3} and contain
additional $s$-channel diagram induced by the non-Abelian 
cubic vertex. 
We uncovered nontrivial energy cancellations of 
$\hsx E^2\!\ito E^0\hsx$ in the $\AP$-amplitude 
as summarized by Table\,\ref{tab:3} and
Eq.\eqref{eq:T[qq-ApAp-E]-cancel}.\ 
We further computed the remaining nonzero $\AP$-amplitude and
$\AT$-amplitude of $\mO(E^0)$, and proved explicitly that
the TET holds for the TMQCD as in Eq.\eqref{eq:TET-qq-AA}. 
Then, in section\,\ref{sec:4.2.2},
we systematically analyzed the four-gauge boson scattering
amplitudes of $\hs\Ap^a\Ap^b\hsmx\to\hsmx\Ap^c\Ap^d$\, and  
$\,\At^a\At^b\hsm\ito\hsm\At^c\At^d$\,	
in the TMQCD, which are shown in Fig.\,\ref{fig:4}. 
We newly uncovered the nontrivial large energy cancellations of 
$\hsx E^4\!\ito E^0\hsx$ in the $\AP$-amplitude, 
at each order of $(E^4,\,E^3,\,E^2,\,E^1)$ respectively,
which are summarized in Table\,\ref{tab:4} and
Eqs.\eqref{eq:E4-cancel}-\eqref{eq:E1-cancel}.
We further computed the remaining nonzero $\AP$-amplitude and
$\AT$-amplitude at $\mO(E^0)$ as given 
in Eqs.\eqref{eq:4AP-E0-TMQCD}-\eqref{eq:4AT-E0-TMQCD}.
With these and the Jacobi identity, we proved explicitly that
the TET indeed holds for the four-gauge boson scattering amplitudes
of the TMQCD, as shown in Eq.\eqref{eq:TET-4Ap-TMQCD}. 
Finally, in section\,\ref{sec:4.3}, we analyzed the perturbative
unitarity of the TMYM and TMG theories.\  We found that
the partial wave amplitudes \eqref{eq:REa0} and \eqref{eq:TMG-a0}
exhibit good high energy behaviors as they scale as $E^{-1}$
or $E^0$ in the high energy limit. This is expected for the
3d TMYM theory because its gauge coupling has mass-dimension
$+\hf$ and thus is super-renormalizable.
This issue becomes much more nontrivial for the 3d TMG theory 
since its Newton constant $\hs G\!\propto\!\ka^2\hs$ 
has mass-dimension $-1\hs$.
But, we should expect this theory to exhibit good high energy behavior because the massive physical graviton scattering amplitudes 
in the TMG theory can be reconstructed from the corresponding 
massive gauge boson amplitudes 
in the TMYM theory via extended double-copy approach,
as shown in section\,\ref{sec:5}.

\vspace*{1mm}

In section\,\ref{sec:5}, we extended the conventional 
double-copy approach and constructed the massive four-graviton 
scattering amplitude of the 3d TMG theory 
by using the massive four-gauge boson amplitude of the 3d TMYM theory. 
We found that the reconstructed tree-level 
four-graviton scattering amplitude could only  
scale as $E^1$ under high energy expansion.
We made the gauge transformation \eqref{eq:GaugeTrans-N'}
on the kinematic numerators $\NN_j^{}$ of Eq.\eqref{eq:N-stu}
in the four-gauge boson scattering amplitude \eqref{eq:4Ap-amp} 
such that the new numerators $\NN_j'$ in Eq.\eqref{eq:N'-stu} 
obey the Jacobi identity.  
The gauge transformation of  $\hsx\NN_j^{}\ito\NN_j'\hsx$
leads to the energy cancellations
of $\hsx E^5 \!\ito E^3\hsx$ in each kinematic numerator.\ 
This determines the individual leading energy dependence of    
the reconstructed graviton amplitude \eqref{eq:Amp-4hp-DC}
to be no more than $\hsx\mO(E^4)\hs$. 
By explicit computations, we further uncovered new energy
cancellations of $\hsx E^4 \hsm\ito E^1\hsx$ in the 
graviton scattering amplitude \eqref{eq:Amp-4hp-DC} 
which are summarized in Table\,\ref{tab:5}. 
Then, we computed the remaining nonzero
graviton amplitude as in Eq.\eqref{eq:Amp0-4hp-DC}
which scales as $\hs\mO(E^1)\hs$ only.$\hsx$
In contrast, applying the general power counting rule \eqref{eq:DE0-TMG}
we found that the individual contributions to the four-graviton
amplitude have leading energy dependence behave as $\hsx E^{12}\hs$.
From these together, we demonstrated a new type of 
striking large energy cancellations 
in the four-graviton scattering amplitude 
as in Eq.\eqref{eq:E12-E1-TMG},   
$\,\mO(E^{12}) \ito \mO(E^1)\hsx$,\,  
for the 3d TMG theory.
Furthermore, with the extended double-copy construction, we established  
{\it a new correspondence between the two types of 
leading energy cancellations in the massive scattering amplitudes:
$E^4 \!\ito\! E^0\hs$ in the TMYM theory and
$\,E^{12}\!\to\!E^1$ in the TMG theory.}

\vspace*{1mm}

Our present findings are highly nontrivial and encouraging. 
We already generally proved our TET for the $N$-point gauge boson
scattering amplitudes (section\,\ref{sec:3.1})
and uncovered the new mechanism of 
large energy cancellations for these $N$-point amplitudes
by using the TET and the general energy power counting method in 3d
(sections\,\ref{sec:3.2}-\ref{sec:3.3}). 
It would be also interesting to extend our current explicit 
calculations and analyses to the $N$-point scattering
amplitudes of gauge bosons (gravitons) in the TMYM (TMG) theories  
with more external states (such as $N\!=\!5$ and $N\!=\!6$).
Since the 3d topologically massive CS gauge theories are 
super-renormalizable and have good high energy behaviors, 
it would be desirable to extend the present
tree-level analyses up to loop levels. 
We will pursue such extended studies in future works.

\newpage 
\noindent
{\bf\Large Acknowledgements:}
\\[1mm]
We thank Stanley Deser and Henry Tye for useful discussions.
This research was supported in part by National NSF of China
(under grants Nos.\,11835005, 12175136), 
by National Key
R\,\&\,D Program of China (under grant No.\,2017YFA0402204),
and by the CAS Center for Excellence in Particle Physics (CCEPP).

\vspace*{10mm}

\appendix

\noindent
{\Large\bf Appendix:}
\vspace*{-4mm}

\section{\hspace*{-2mm}Kinematics and Feynman Rules 
of 3d CS Gauge Theories}
\renewcommand{\theequation}{\thesection.\arabic{equation}}
\label{app:A}

In this Appendix, we present definitions of relevant
kinematic variables for the four-particle scattering process and
the Feynman rules for the 3d topologically massive
Chern-Simons gauge theories.

\vspace{1mm}

For the present analysis, we choose the following
metric signature and the rank-3 anti-symmetric tensor:
\begin{equation}
\eta_{\mn} = \eta^{\mn}= \diag(-1,1,1) \,, \quad
\vep^{012} = - \vep_{012} = 1 \,.
\end{equation}
Thus, we have the momentum on-shell condition 
$\,p^2 \!=\! -m^2$.

\vspace{1mm}

For the $2 \ito 2$ elastic scattering process,
the momenta in the center-of-mass frame can be defined 
as follows: 
\begin{alignat}{3}
\label{eq:CM-Momenta}
p_{1}^\mu &=  E ( 1, 0,  \be ) , \quad
&&p_2^\mu =  E ( 1, 0,  -\be ), \quad
\nn\\
p_{3}^\mu &=   E ( 1, \be \st,  \be \ct  ), \quad
&&p_{4}^\mu =  E ( 1, -\be \st,  -\be \ct  ),
\end{alignat}
where we have defined $\,\be\!=\!\sqrt{1\!-\!m^2/E^2\,}$.\,
Thus, we further define the Mandelstam variables:
\begin{align}
s &=-\( p_{1} \!+\! p_{2} \)^{2} \!= 4E^2 , 
\nn\\
\label{eq:stu}
t &=-\( p_{1} \!-\! p_4 \)^{2} \!= 
-\fr{1}{2}\hspace*{0.3mm}s\hspace*{0.3mm}\be^2
(1\!+\!\ct)\,,
\\
u &=-\( p_{1} \!-\! p_3 \)^{2} \!=
-\fr{1}{\,2\,}\hspace*{0.3mm}s\hspace*{0.3mm}
\be^2 (1\!-\!\ct) \,.
\nn
\end{align}
For convenience of our analysis,
we can use the relation
$\,E^2 \!=\! E^2\be^2 \!+\! m^2\,$
to define another set of Mandelstam variables
$(\sz,\tz,\uz)$\hspace*{0.4mm}:
\begin{align}
\label{eq:s0-t0-u0}
\sz = 4E^2\be^2 \,,~~~~
\tz = -\fr{1}{2}\sz(1\!+\!\ct)  \,,~~~~
\uz = -\fr{1}{2}\sz(1\!-\!\ct)  \,.
\end{align}
%
The summations of $(s,t,u)$ and $(\sz,\tz,\uz)$ obey
the conditions:
\begin{equation}
s+t+u  = 4 m^{2} , ~~\qquad 
\sz +\tz + \uz =0 \,.
\end{equation}

In the rest frame with 3-momentum 
$\,p^\mu \!=(m,\,0,\,0)\equiv \bar{p}^\mu$,\,
we can solve \eqrefe{eq:PolEOM} and derive the polarization vector:
\begin{equation}
\label{eq:Pol-Rest}
\ep^\mu (\bar{p}) = \fr{1}{\sqrt{2\,}\,}
( 0 ,\, 1 ,\, -\ii \sp )  \,,
\end{equation}
where $\,\sp = {\mt}/{m}=\pm 1$\, and $\,m=|\mt|$\,.
We note that the in the rest frame the gauge boson polarization
vector has zero time-component and its two possible forms
are not independent due to the relation
$\,\ep_2^\mu = \iii\mathfrak{s}
\hspace*{0.25mm} \ep_1^\mu$\,.
Furthermore, by choosing the orthonormal basis
$e^j_1 = (1,0)$ and $e^j_2 = (0,1)$ in a plane,
we can define a polarization basis:
\begin{equation}
e^j_\pm \,=\,
\fr{1}{\sqrt{2}}(e^j_1 \pm \ii\hspace*{0.25mm} e^j_2
\hspace*{0.25mm})
\,=\,
\fr{1}{\sqrt{2\,}\,}\!\(1,\, \pm\ii\hspace*{0.25mm}\) .
\end{equation}
Thus, in the rest frame
the spatial part of the polarization vector
$\ep^\mu(\bar{p})$
can be decomposed in terms of the basis
$\{e^j_\pm\}$:
\begin{equation}
\ep^j(\bar{p}) \,=\, \ep_+^{} e^j_+ + \ep_-^{} e^j_- \,,
\end{equation}
where the coefficients $\,(\ep_+^{},\, \ep_-^{})$\, 
satisfy
$\,(\ep_+^{},\,\ep_-^{}) \!=\! (0,\,1)$ 
for $\,\sp\!=\!+1$\, and
$\,(\ep_+^{},\,\ep_-^{}) \!=\! (1,\,0)$\, 
for $\sp=\!-1$
\cite{Banerjee:2000gc}.
So, it is clear that in the 3d Chern-Simons gauge theory,
the case of $\,\sp=\!+1$ (or, $\sp\!=\!-1$)
only allows one physical polarization state
$\ep_-^{}$ (or, $\ep_+^{}$)
of the gauge boson, as expected.

\vspace*{1mm}

After taking the Lorentz boost along an arbitrary direction,
the gauge boson momentum can be generally written as
$\,p^\mu \!=\!E(1,\,\be\st,\,\be\ct)\,$. Thus, we can
Lorentz-boost the rest-frame polarization vector
\eqref{eq:Pol-Rest} to the following general polarization vector:
\begin{equation}
\label{eq:PolP-general}
\epP^\mu(p) \,=\, \frac{1}{\sqrt{2\,}\,} \!\!\(\!
\frac{\,\ii p_1^{} \!\!+\!\sp p_2^{}}{m},\
\ii\!+\!
\frac{\,p_1^{}(\ii p_1^{} \!\!+\! \sp p_2^{})}{m(m\!-\! p_0^{})},\
\sp\!+\!  \frac{\,p_2^{}(\ii p_1 \!\!+\! \sp p_2)}
{m(m\!-\! p_0^{})}\!\) \!,
\end{equation}
where  $\,\ep^\mu_+ \!=\! -(\ep^\mu_-)^*$.
Thus, by substituting \eqrefe{eq:CM-Momenta} into \eqrefe{eq:PolP-general}, 
we derive the following explicit formulas 
of the physical polarization vectors:
\begin{alignat}{3}
\ep_{1}^\mu &= \fr{\mathfrak{s}}{\sqrt{2}} ( \bE\be,\, \ii\mathfrak{s},\, \bE )\,, \quad
&&\ep_{2}^\mu = - \fr{\mathfrak{s}}{\sqrt{2}} ( \bE\be,\, -\ii\mathfrak{s} ,\, -\bE )\,,
\nn\\
\ep_{3}^\mu &= \fr{\mathfrak{s} e^{\ii \mathfrak{s} \theta}}{\sqrt{2}} ( \bE\be,\, \bE\st \!+\! \ii \mathfrak{s} \ct,\, \bE\ct \!-\! \ii \mathfrak{s} \st )
\,, \quad
&&\ep_{4}^\mu = -\fr{ \mathfrak{s} e^{\ii \mathfrak{s}\theta}}{\sqrt{2}} ( \bE\be,\, -\bE\st \!-\! \ii \mathfrak{s}\ct,\, -\bE\ct \!+\! \ii \mathfrak{s}\st )  \,,
\label{eq:CSPolarization}
\end{alignat}
where we have defined $\,\bE =\! E/m$\,.

\vspace*{1mm}

For $\mt\!>\!0$\,, 
the propagators for the Abelian and non-Abelian
topological gauge theories can be derived as follows: 
\beqs
\begin{align}
\label{eq:A-propagator}
\D_{\mn}(p) &= -\ii \! \[ \frac{1}{p^{2}+m^{2}} \! 
\( \!\eta_{\mn}^{}\! - \frac{\,p_{\mu} p_{\nu}}{p^2} -
\frac{\,\iii m \vep_{\mnr} p^{\rho}\,}{p^2} \) \!+ 
\xi \frac{\,p_\mu^{}p_\nu^{}\,}{p^4} \] \!,
\\[2mm]
\D^{ab}_{\mn} (p)&= \delta^{ab} \D_{\mn}(p) \,.
\end{align}
\eeqs
In Eq.\eqref{eq:A-propagator},
the pole  $\,p^2\!=\!0$\, is unphysical,
for which the EOM \eqref{eq:PolEOM} becomes
$\,\mt\vep^{\mu\rho\nu}p_{\rho}^{}\ep_\nu^{}=0\,$.
It can be solved as 
$\,\ep^\mu \!= f(p)\hspace*{0.3mm} p^\mu$,\, 
but it can be eliminated by the freedom of gauge transformations\,\cite{Pisarski:1985}. 
Hence, the massless mode is a pure 
gauge artifact\,\cite{Pisarski:1985}\cite{Devecchi:1994ha}.
Furthermore, we can derive the following Feynman rule of 
the cubic gauge boson vertex:
\begin{align}
V_{\mnr}^{a b c}(p_1^{},p_2^{},p_3^{}) \,=\,
g\hspace*{0.4mm} C^{abc} \! \[ \! \eta_{\mn} (p_{1} \!-\! p_{2})_{\rho}
+\eta_{\nu \rho}(p_{2}\!-\! p_{3})_{\mu}
+\eta_{\rho \mu} (p_{3} \!-\! p_{1})_{\nu}
+ \iii m\hspace*{0.4mm}\vep_{\mnr} \]\! .
\end{align}
The quartic gauge boson vertex is similar to that of the
4d QCD.

\section{\hspace*{-2mm}General Power Counting Method in 
\boldmath{$d$}-Dimensions}
\label{app:B}

In this Appendix, extending the 3d power counting method of section\,\ref{sec:3.2}, we present a general power counting
formula for the $d$-dimensional spacetime.

\vspace*{1mm}

In $d$-dimensions, we derive the mass-dimension of a given 
$S$-matrix element $\mathbb{S}$ as follows:
\begin{equation}
\label{eq:DS-d}
D_{\mathbb{S}}^{} \,=\, 
d - \frac{\,d\!-\!2\,}{2}\,\EE \,,
\end{equation}
where $\EE$ denotes the total number of external states as before.
We see that the general formula \eqref{eq:DS-d} reduces to
$\,D_{\mathbb{S}}^{}\!=\hsm 4-\EE\,$ for $\hs d\hsm =\hsm 4\hs$ and 
$\,D_{\mathbb{S}}^{}\!=\hsm 3-\fr{1}{2}\EE\,$ 
for $\hs d\hsm =\hsm 3\hs$, respectively. 
Then, we can deduce the mass-dimension of all the coupling 
constants in the $S$-matrix element $\mathbb{S}$\,:
\begin{equation}
D_C^{} \,=\, \sum_j \VV_j^{}\!
\(d-d_j^{}\!- \frac{\,d\!-\!2\,}{2} b_j^{}\!- 
\frac{\,d\!-\!1\,}{2}f_j^{}\),
\end{equation}
where $\VV_j^{}$ denotes the number of vertices of type-$j$,
and the quantities $(d_j^{},b_j^{},f_j^{})$ denote
the numbers of (partial derivatives, bosonic fields,
fermions) in each vertex of type-$j$\,,
respectively.  
We have the following general relations for each
Feynman diagram which contributes to the amplitude $\mathbb{S}$\,,
\begin{equation}
\label{eq:L-V-I-app}
L = 1\!+I\!-\VV\,,~~~~
\VV=\sum_j \VV_j^{} \,, ~~~~
\sum_j \VV_j^{}b_j^{} = 2I_B^{}+\EE_B^{}\,,~~~~
\sum_j \VV_j^{}f_j^{} = 2I_F^{}+\EE_F^{}\,, 
\end{equation}
where $L$ denotes the number of loops of a given diagram, 
$(I_B^{},\,\EE_B^{})$ denote the numbers of (internal,\,external)
bosonic lines in this diagram, and
$(I_F^{},\,\EE_F^{})$ denote the numbers of (internal,\,external)
fermionic lines in the same diagram.
With these, we derive the leading energy dependence 
of the amplitude $\mathbb{S}$:
\begin{equation}
\label{eq:DE-d-1}
D_E^{} \,=\, D_{\mathbb{S}}^{} - D_C^{} \,=\,
2(1 \!-\! \VV)+ (d \!-\! 2)L+\sum_j  \VV_j^{}\!
\(d_j^{}\!+\!\fr{1}{2}f_j^{}\) .
\end{equation}

Then, we consider the $d$-dimensional gauge theories
(including Chern-Simons term when allowed).
By imposing the relations \eqref{eq:V-1}, 
we derive the leading energy dependence of the
amplitude $\mathbb{S}$\,:
\begin{equation}
\label{eq:DE-d-YM}
D_E^{} \,=\, (\EE_{A_\rm{P}}^{} \!-\EE_{v}^{})  + (4 - \EE - \over{\VV}_3^{}) + (d-4)L  \,,
\end{equation}
where $\,\EE_{\APP}^{}\!$ is the total number of external states
of the physical gauge bosons
and $\,\EE_{v}^{}\,$ denotes number of the external states
of gauge bosons
$\,v^a\!=\!v_\mu^{}A^{a\mu}\,$
with the factor $v^\mu \!=\!\epL^\mu\!-\epS^\mu$\,.

\vspace*{1mm}

Next,  we can apply the power counting formula \eqref{eq:DE-d-1}
to the topologically massive gravity (TMG) theory. 
For this, we can derive  
the {\it leading energy-dependence} of a pure graviton
scattering amplitude $\mathbb{S}$ in the TMG theory,
which corresponds to setting 
$d_j^{}\!=\!3$ and $f_j^{}\!=\!0$ 
in \eqrefe{eq:DE-d-1}:
\begin{equation}
\label{eq:DE-d-TMG}
D_E^{} \,=\, 2\EE_{\hP} \!+ \VV_3 +2 + (d-2)L \,.
\end{equation}
where $\,\VV_3^{}$\, denotes the number of cubic vertices 
from the CS term in the TMG action.
For instance, we can check that for the TMG theory of 
$\hs d\!=\hsmx 3\hs$, 
the above Eqs.\eqref{eq:DE-d-YM} and \eqref{eq:DE-d-TMG} 
just reduce to the power counting formulas 
\eqref{eq:DE} and \eqref{eq:DE-TMG}, which we derived for
the 3d TMYM and TMG theories in section\,\ref{sec:3.2}.

\section{\hspace*{-2mm}Dirac Spinors in (2+1)d Spacetime}
\label{app:C}

The anti-symmetric and symmetric commutation relations for the gamma matrices in (2+1)d spacetime are given by
\begin{equation}
\label{eq:DiracEqPsi}
\LB \ga^{\mu}\hspace*{-0.3mm},\, \ga^{\nu} \RB 
=\, 2 \eta^{\mn} \,,  \qquad
\[\ga^\mu\hspace*{-0.3mm},\,\ga^\nu\] 
=\,  2\vep^{\mn \alpha}\ga_\al^{}\,,
\end{equation}
where we can choose the gamma matrices 
as the Pauli matrices\,\cite{Agarwal:2008pu}:
\begin{equation}
\ga^{0}\!=\iii \sigma_2^{}\!=
\begin{pmatrix}
0&1\\ -1 &0
\end{pmatrix} \!, \quad~
\ga^{1}\!= \sigma_1^{}\! =
\begin{pmatrix}
0&1\\ 1 &0
\end{pmatrix}\!, \quad~
\ga^{2}\!= \sigma_3^{}\! =
\begin{pmatrix}
1&0 \\ 0 & -1
\end{pmatrix} \!.
\end{equation}

The Dirac equation in the 3d spacetime is derived as follows\,\cite{Agarwal:2008pu}\cite{Witten:2015}:
\begin{equation}
(\slashed{\pd} - m_f^{}) \psi = 0\,,
\end{equation}
with $\,\slashed{\pd}=\ga^\mu\pd_{\mu}$\,. 
Its solution takes the plane wave form 
$\,\psi \sim u(p) e^{-\iii p\cdot x} 
\!+v(p) e^{\iii p\cdot x}$.\, 
Thus, the spinors $(u,v)$ 
satisfy the momentum-space equations:
\begin{equation}
(\slp - \iii m_f^{}) u =0  \,, \qquad
(\slp + \iii m_f^{}) v =0  \,.
\label{eq:DiracEqUandV}
\end{equation}
Then, solving \eqrefe{eq:DiracEqUandV} gives the 
spinor solutions for particle and anti-particle:
\begin{equation}
u \!=\! \frac{1}{\sqrt{-p_0 \!+\! p_1}\,} \!
\begin{pmatrix}
p_2 \!+\! \iii m_f
\\[1mm]
-p_0 \!+\! p_1
\end{pmatrix} \!, \qquad
v \!=\! \frac{1}{\sqrt{-p_0 \!+\! p_1}\,} \!
\begin{pmatrix}
p_2 \!-\! \iii m_f
\\[1mm]
-p_0 \!+\! p_1
\end{pmatrix}  \!.
\end{equation}
They obey the following spinor identities:
\begin{equation}
u\bar{u} = -\slp -  \iii m_f \,, \quad~ 
\bar{u}u = -\iii 2m_f^{} \,, \quad~
v\bar{v} =-\slp +  \iii m_f^{} \,, \quad~ 
\bar{v}v = \iii 2m_f^{} \,,
\vspace{-2mm}
\end{equation}
where $\,\bar{u}=\!u^{\dagger}\ga^0$\, and 
$\,\bar{v}=\!v^{\dagger}\ga^0$.

\section{\hspace*{-2mm}Topological Scattering Amplitudes 
with Matter Fields}
\label{app:D}

In this Appendix, we present the full amplitudes 
of the scattering processes discussed in section\,\ref{sec:4}. 
For the notational convenience of the following scattering amplitudes, 
we have defined the parameters 
$\,\be_\pm^{} \!=\! 1 \pm \be\,$ with 
$\,\be \!=\! \sqrt{1 \!-\! m^2\!/\!E^2\,}$.\, 

\vspace*{3mm}
\noindent
The scattering amplitudes of pair annihilation in the 
topologically massive scalar QED take the following forms:
\beqs
\begin{align}
\TT[\phi^-\phi^+\!\ito \APP \APP] \ &=\
-\frac{\,2e^2\!\[\!1\!-\! \bE^4 \!+\! \bE^2 (1 \!+\! \bE^2 ) \ctt 
\!+\! \iii 2\bE^3\stt\!\]\,}
{(1 \!-\! 2\bE^2)^2 \!+ 4 \bE^2(1 \!-\! \bE^2) \cct} \,,
\\[1mm]
\TT[\phi^-\phi^+\!\ito \ATT\ATT] \ &=\
-\frac{\,2e^2 \(1 \!-\! 2\bE^4 \!+\! 2\bE^4 \ctt\)\,}
{\,(1 \!-\! 2\bE^2)^2 \!+ 4 \bE^2(1 \!-\! \bE^2) \cct\,} \,.
\end{align}
\eeqs
The Compton scattering amplitudes in the 
topologically massive scalar QED are given by
\beqs
\begin{align}
\hspace*{-9mm}
\TT[\phi^-\!\APP \!\ito \phi^- \!\APP ] \,=&\
\frac{e^2}{\,1 \!-\!\be(1 \!+\!\be_+) \!-\! 2\be^2\ct\,}\times 
\\
&  
~\{(2 \!-\! \bE^2\be_-^2)\be^2 \!-\! [1\!-\!\be (1\!+\!\be_+)\!
+ \! \bE^2 \be_-^2 (1 \!-\! 2 \be^2)] \ct \!-\! \iii 2 \bE \be_-\be_+ (1 \!-\! 2 \be) \st \}\,,
\nn\\[1mm]
\hspace*{-9mm}
\TT[\phi^-\!\ATT \ito \phi^- \!\ATT ] \,=&\
\frac{~2e^2 [2 \be ^2 \!-\! (1\!-\! 2\be \!-\! \be^2) \ct ]~}
{1  \!-\! \be(1 \!+\! \be_+) \!-\! 2 \be ^2\ct }.
\end{align}
\eeqs
%
The scattering amplitudes of pair annihilation in the 
topologically massive spinor QED are derived as follows:
\beqs
\begin{align}
\TT[e^-e^+ \!\ito \APP \APP ] &\,=
\frac{~2e^2\bE \!\[\!2(1\!-\!\bE^2) \!+\! 2\bE^2 \ctt \!+\! 
\iii \bE (1\!+\!\bE^2) \stt \!\]~}
{(1 \!-\! 2\bE^2)^2 \!+\! 4 \bE^2(1 \!-\! \bE^2) \cct} \,,
\\[1mm]
\TT[e^-e^+ \!\ito \ATT\ATT ] &\,=
\frac{\iii 4e^2\bE^4\stt}
{~(1 \!-\! 2\bE^2)^2 \!+\! 4 \bE^2(1 \!-\! \bE^2) \cct~} \,.
\end{align}
\eeqs
The Compton scattering amplitudes in the 
topologically massive spinor QED:
\beqs
\begin{align}
\hspace*{-6mm}
\TT[e^-\!\APP \ito e^- \!\APP ] \,=&\
\frac{\iii e^2 \be (1 \!+\! \tan \fr{\theta}{2})}
{\,2\be_+ [1 \!-\! \be(1\!+\! \be_+) \!-\!
2 \be^2 \ct](1\!+\!\st)^\hf\,}\!
\[\!1 \!+\! \bE^2 \!-\! 4 \be\be_+
\!-\!  \bE^2 \be^2 (2\!-\!\be^2) 
\right.
\nn\\[1mm]
&\left.\hspace{2mm}
- 4 \be \be_+ \ct \!-\! (1 \!+\!
\bE^2 \be_-^2\be_+^2) \ctt \!-\!
\iii 4 \bE \be \be_-\be_+ \st \!-\!
\iii 2 \bE \be_-\be_+ \stt \]\!,
\\[2mm]
\hspace*{-6mm}
\TT[e^-\!\ATT \ito e^- \!\ATT ] \,=&\
\frac{\, \iii 2e^2 \be (1 \!-\! 2 \be\be_+ 
\!-\!\ct)(1 \!+\! \ct \!+\! \st) \,}
{\,\be_+ [1 \!-\! \be(1\!+\! \be_+) \!-\!
2\be^2 \ct](1\!+\!\st)^\hf\,}\,.
\end{align}
\eeqs
The scattering amplitudes of pair annihilation 
via the color-singlet channel in the topologically massive QCD
are connected to that of the TMQED according to 
Eq.\eqref{eq:Amp-qqPP-qqTT} in section\,\ref{sec:4.2.1}:  
\beqs
\label{eq:Amp2-qqPP-qqTT}
\begin{align}
\TT_{\P\P}^{}
[\hspace*{0.3mm}|0\ra_{\!q}^{}\!\ito\!|0\ra_{\!\APP}^{}] 
\,=~&\,
\frac{\,g^2\,}{e^2} f(N) \,\TT[e^-e^+ \!\ito \APP \APP ]  \,,
\\[1mm]
\TT_{\T\T}^{}
[\hspace*{0.3mm}|0\ra_{\!q}^{}\!\ito\!|0\ra_{\!\ATT}^{}] 
\,=~&\,
\frac{\,g^2\,}{e^2} f(N) \,\TT[e^-e^+ \!\ito \ATT \ATT ]  \,.
\end{align}
\eeqs
where the function 
$\,f(N)\!=\!\fr{1}{\,2\sqrt{2\,}\,}\hsm
 \sqrt{\hsm (N^2\!-\!1)\hsm /N\,}\hs$.

\section{\hspace*{-2mm}Graviton Propagator and Scattering Amplitude in TMG} 
\label{app:E}

From the action \eqref{eq:S-TMG} together with 
the gauge-fixing term \eqref{eq:LGF-TMG},
we derive the quadratic term of the graviton fields:
\begin{equation}
S_{\rm{TMG}} ~=\, \int \!\! \td^3 x ~
\frac{1}{2} \, h^{\mn}  \D^{-1}_{\mn \ab} \, h^{\ab} \,,
\end{equation}
where the inverse of the graviton propagator 
$\D^{-1}_{\mn \ab}$ takes the following form:
\begin{align}
\label{eq:D-1}
\D^{-1}_{\mn \ab} \,=~& \(\!1 \!-\! \frac{1}{2 \xi}\)\! \eta_{\mn} \eta_{\ab} \pd^{2}
\!-\! \frac{1}{2} \(\eta_{\mu \al} \eta_{\nu \be}
\!+\! \eta_{\mu \be} \eta_{\nu \al}\) \!\pd^{2}
\!+\! \(\frac{1}{\xi} \!-\! 1\) \! \(\eta_{\mn} \pd_{\al} \pd_{\be} + \eta_{\ab} \pd_{\mu} \pd_{\nu}\)
\nn\\[1mm]
&\, +\frac{1}{2}\!\(\!1 \!-\! \frac{1}{\xi}\)\! \(\eta_{\mu \al} \pd_{\nu} \pd_{\be} \!+\! \eta_{\mu \be} \pd_{\nu} \pd_{\al} \!+\! \eta_{\nu \al} \pd_{\mu} \pd_{\be} \!+\! \eta_{\nu \be} \pd_{\mu} \pd_{\al}\)
\nn\\[1mm]
&\, + \frac{1}{2m} \!\[ \vep_{\mu \rho \al}^{} (\pd_{\nu} \pd_{\be} \pd^{\rho} \!-\! \eta_{\nu \be} \pd^2 \pd^{\rho} )+ (\mu \leftrightarrow \nu) \!\]  \,.
\end{align}
Then, transforming \eqrefe{eq:D-1} into momentum space and 
imposing the normalization condition
\begin{equation}
\D^{-1}_{\mn \ab}\D^{\alpha \beta \rho \si} =\frac{\ii}{\,2\,} (\delta_{\mu}^{\rho} \delta_{\nu}^{\si}+\delta_{\mu}^{\si} \delta_{\nu}^{\rho} ) \,,
\end{equation}
we can derive the massive graviton propagator as follows:
\begin{equation}
\label{eq:DTMG-1}
\D_{\mn\ab}^{} \,=\, 
\frac{~\iii \hs \Delta_{\mn\ab}^{}~}{~2(p^2 \!+\! m^2)~} \,,
\end{equation}
where the numerator is given by
\begin{align}
\label{eq:DTMG-1x}
\Delta_{\mn\ab}^{} \,=& \,
-\!\eta_{\mn}^{} \eta_{\ab}^{} - \frac{\,m^2\,}{p^2}
(2\eta_{\mn}^{} \eta_{\ab}^{} \!-\! 
\eta_{\mu\al}^{}\eta_{\nu\be}^{} \!-\! 
\eta_{\mu\be}^{}\eta_{\nu\al}^{})
\!-\! \frac{1}{\,p^2\,} 
(\eta_{\mn}^{} p_{\al}^{} p_{\be}^{} \!+\! 
\eta_{\ab}^{} p_{\mu}^{} p_{\nu}^{})
\nn\\
&\ -\! 
\frac{1}{\,p^4\,} \, p_\mu^{} p_\nu^{} p_\al^{} p_\be^{}
+ \frac{\,\xi (p^2\!+\! m^2) \!-\! m^2\,}{p^4}
( \eta_{\mu\al}^{} p_{\nu}^{} p_{\be}^{} \!+\!
\eta_{\mu\be}^{} p_{\nu}^{} p_{\al}^{} \!+\!
\eta_{\nu\al}^{} p_{\mu}^{} p_{\be}^{} \!+\!
\eta_{\nu\be}^{} p_{\mu}^{} p_{\al}^{} )
\nn\\
&\ +\!  \frac{\,\iii m\hs p^\rho\,}{2 p^2}
(\vep_{\rho \mu \al}^{} \eta_{\nu \be}^{} \!+\!
\vep_{\rho \mu \be}^{} \eta_{\nu \al}^{} \!+\!
\vep_{\rho \nu \be}^{} \eta_{\mu \al}^{} \!+\!
\vep_{\rho \nu \al}^{} \eta_{\mu \be}^{} )
\nn\\
&\ - \frac{\,\iii m\hs p^\rho\,}{2p^4}
(\vep_{\rho\mu\al}^{} p_{\nu}^{} p_{\be}^{} \!+\!
\vep_{\rho\mu\be}^{} p_{\nu}^{} p_{\al}^{} \!+\!
\vep_{\rho\nu\be}^{} p_{\mu}^{} p_{\al}^{} \!+\!
\vep_{\rho\nu\al}^{} p_{\mu} p_{\be}^{} ) \,.
\end{align}
Using the notation 
$\,P_{\mn} \!= \eta_{\mn}^{} \!-\! \fr{p_{\mu} p_{\nu}}{p^2}$\,, 
we can further express the propagator \eqref{eq:DTMG-1}-\eqref{eq:DTMG-1x} 
into the following form:
\vspace*{-3mm} 
{\small 
\begin{align}
\label{eq:Propa-TMG}
\D_{\mn\ab}^{}(p) \,=&\, 
-\frac{\ii}{\,2(p^2\!+\!m^2)\,} \!
\(P_{\mu \al} P_{\nu \be} \!+\!
P_{\mu \be} P_{\nu \al} \!-\!
P_{\mu \nu} P_{\al \be} \)
\!+\! \frac{\ii}{\,2p^2\,}\! \(\!P_{\mu \al}^{}P_{\nu\be}^{} \!+\!
P_{\mu \be}^{} P_{\nu \al}^{} \!-\! 2P_{\mn}^{}P_{\al\be}^{}\)
\nn\\
&\, -\frac{\ii}{\,p^4\,}\! 
\(\eta_{\mn}^{}p_{\al}^{}p_{\be}^{}\!+\!
  \eta_{\al\be}^{} p_{\mu}^{}p_{\nu}^{}\)
+ \frac{\,\iii\xi\,}{\,2p^4\,} \! 
\(\eta_{\mu\al}^{} p_{\nu}^{} p_{\be}^{}  \!+\!  
\eta_{\mu\be}^{} p_{\nu}^{} p_{\al}^{}  \!+\!  
\eta_{\nu\al}^{} p_{\mu}^{} p_{\be}^{} \!+\! 
\eta_{\nu\be}^{} p_{\mu}^{} p_{\al}^{} \)
\nn\\
&\, -
\frac{m\hs p^{\rho}}{\,4\hs p^2(p^2\!+\!m^2)\,} \! 
\(\vep_{\rho\mu\al}^{} P_{\nu\be}^{} \!+\!
\vep_{\rho\mu\be}^{} P_{\nu\al}^{} \!+\!
\vep_{\rho\nu\al}^{} P_{\mu\be}^{} \!+\! 
\vep_{\rho\nu\be}^{} P_{\mu\al}^{} \) \!.
\end{align}
}
Under the Landau gauge $\,\xi\!=\! 0$\,, the above
propagator reduces to
{\small 
\begin{align}
\label{eq:PropaTMG-xi=0}
\hspace*{-4mm}
\D_{\mn\ab}^{\hs\xi=0}(p) =& 
-\!\frac{\ii}{\,2(p^2\!+\!m^2)\,} \!
\(P_{\mu \al} P_{\nu \be} \!+\!
P_{\mu \be} P_{\nu \al} \!-\!
P_{\mu \nu} P_{\al \be} \)
\!+\! \frac{\ii}{\,2p^2\,}\! \(\!P_{\mu \al}^{}P_{\nu\be}^{} \!+\!
P_{\mu \be}^{} P_{\nu \al}^{} \!-\! 2P_{\mn}^{}P_{\al\be}^{}\)
\nn\\
\hspace*{-4mm}
& -\!\frac{m\hs p^{\rho}}{\,4\hs p^2(p^2\!+\!m^2)\,} \! 
\(\vep_{\rho\mu\al}^{} P_{\nu\be}^{} \!+\!
\vep_{\rho\mu\be}^{} P_{\nu\al}^{} \!+\!
\vep_{\rho\nu\al}^{} P_{\mu\be}^{} \!+\! 
\vep_{\rho\nu\be}^{} P_{\mu\al}^{} \)\! 
-\frac{\ii}{\,p^4\,}\! 
\(\eta_{\mn}^{}p_{\al}^{}p_{\be}^{}\!+\!
  \eta_{\al\be}^{} p_{\mu}^{}p_{\nu}^{}\)\!.
\end{align}
}
If the last term above is removed by contracting with a conserved current or on-shell physical graviton polarization, the propagator \eqref{eq:PropaTMG-xi=0} agrees with the result of Ref.\,\cite{Deser:1981wh}.

\vspace*{1mm}

Next, we note that in section\,\ref{sec:5} we reconstructed 
the four-graviton scattering amplitude in 
Eqs.\eqref{eq:Amp-4hp-DC2}-\eqref{eq:Amp-4hp-DC2-Pj}
which is expressed in terms of the energy variable
$\bsz\!=s_0^{}/m^2$\,. For the sake of comparison, we further
reexpress the four-graviton amplitude in terms of the 
Mandelstam variable 
$\,\bs =s/m^2\hsx$,
which is connected to $\bsz\hsmx =\hsmx s_0^{} /m^2\hsx$ via
$\,\bs\hsm =\hsm \bsz\hsmx +\hsm 4\,$. 
Thus, from Eqs.\eqref{eq:Amp-4hp-DC2}-\eqref{eq:Amp-4hp-DC2-Pj},
we derive the following equivalent expressions: 
\begin{equation}
\M[4\hP] \,=\,
\frac{~\ka^2 m^2
(Q_0^{}\hsm +\hsm Q_2^{} \ctt \hsm+\hsm Q_4^{} c_{4\theta}^{} 
\hsm +\hsm Q_6^{} c_{6\theta}^{} \hsm +\hsm \bar{Q}_2^{} \stt 
\hsm +\hsm \bar{Q}_4^{} s_{4\theta}^{} \hsm +\hsm \bar{Q}_6^{} s_{6\theta}^{})\hsm\csc^2\!\theta~}
{4096\hs (1 \!-\hsm \bs)\hs \bs^{3/2}\hs 
[2 \hsm -\hsm \bs \hsm -\hsm (4 \!-\! \bs)\ct\hs ] 
[\hs 2 \hsm -\hsm \bs \hsm +\hsm (4 \!-\! \bs)\ct\hs ]} 
\,,
\end{equation}
where $(Q_j^{},\, \bar{Q}_j)$ are expressed as
polynomial functions of the variable $\bs\!=s/m^2$,
\begin{align}
Q_0^{} &\,=\hsx 
(256 + 49088 \bs - 68880 \bs^2 + 25220 \bs^3 - 2768 \bs^4) \bs^\hf \,,
\nn\\[1mm]
Q_2^{} &\,=\hsx 
(-768 - 45568 \bs + 65568 \bs^2 - 19008 \bs^3 + 505 \bs^4) \bs^\hf \,,
\nn\\[1mm]
Q_4^{} &\,=\hsx 
4 (192 - 176 \bs + 20 \bs^2 + 635 \bs^3 + 58 \bs^4)  \bs^\hf \,,
\nn\\[1mm]
Q_6^{} &\,=\hsx 
-(256 + 2816 \bs + 2912 \bs^2 + 560 \bs^3 + 17 \bs^4)  \bs^\hf\,,
\\[1mm]
\bar{Q}_2^{} &\,=\hsx \ii\hs 
(1280 - 256 \bs + 21312 \bs^2 - 8960 \bs^3 + 475 \bs^4)  \bs\,,
\nn\\[1mm]
\bar{Q}_4^{} &\,=\hsx \iii\hs 4 
(320 - 544 \bs + 676 \bs^2 + 272 \bs^3 + 5 \bs^4)  \bs\,,
\nn\\[1mm]
\bar{Q}_6^{} &\,=\hsx -\ii\hs 
(1280 + 3584 \bs + 1568 \bs^2 + 128 \bs^3 + \bs^4) \bs\,.
\nn
\end{align}


\end{document}